\documentclass[a4paper,11pt]{article}

\usepackage{jheppub} 
\usepackage{ulem}
\usepackage[T1]{fontenc} 
\usepackage{lmodern}
\usepackage{newtxtext,amssymb,amsmath,extarrows}
\usepackage{graphicx,microtype,physics,framed,empheq}
\usepackage{hyperref,slashed,graphicx,color,amsmath,amssymb}
\usepackage[numbers,sort&compress]{natbib}
\usepackage{subfigure}
\usepackage{bm}
\usepackage{soul}
\usepackage{ulem}
\usepackage{pifont}
\usepackage[dvipsnames]{xcolor}
\usepackage{float}
\usepackage{color, soul}
\sethlcolor{yellow}
\setstcolor{green}
\setulcolor{red} 
\usepackage{booktabs} 
\usepackage{multirow}
\usepackage{verbatim}
\usepackage[thicklines]{cancel}

\allowdisplaybreaks 

\graphicspath{{fig/}}

\title{\boldmath  Primordial Stochastic Gravitational Waves from Massive Higher-Spin Bosons}

\author[a]{ Haipeng An,}
\author[a,b]{ Zhehan Qin,}
\author[a,c]{ Zhong-Zhi Xianyu,}
\author[a]{ Borui Zhang}
\affiliation[a]{Department of Physics, Tsinghua University,\\Beijing 100084, China}
\affiliation[b]{Department of Applied Mathematics and Theoretical Physics, University of Cambridge,\\Wilberforce Road, Cambridge, CB3 0WA, UK}
\affiliation[c]{Peng Huanwu Center for Fundamental Theory,\\Hefei, Anhui 230026, China}

\emailAdd{anhp@mail.tsinghua.edu.cn}
\emailAdd{qzh21@mails.tsinghua.edu.cn}
\emailAdd{zxianyu@tsinghua.edu.cn}
\emailAdd{zhangbr22@mails.tsinghua.edu.cn}

\abstract{
Can a stationary stone radiate gravitational waves (GWs)? While the answer is typically ``no'' in flat spacetime, we get a ``yes'' in inflationary spacetime. In this work, we study the stationary-stone-produced GWs in inflation with a concrete model, where the role of stones is played by massive higher-spin particles. We study particles of spin-$2$ and higher produced by helical chemical potentials, and show that the induced GWs feature a scale-invariant and helicity-biased power spectrum in the slow-roll limit. Including slow-roll corrections leads to interesting backreactions from the higher-spin boson production, resulting in an intriguing scale-dependence of GWs at small scales. Given the existing observational and theoretical constraints, we identify viable parameter regions capable of generating visibly large GWs for future observations. }

\begin{document}
	\maketitle
	\flushbottom
	
	\section{Introduction}\label{Intro}
	
It is well known that massless higher-spin fields (with spin $s\geq 2$) cannot be interacting in a consistent relativistic field theory in flat spacetime \cite{Weinberg:1964ew,Coleman:1967ad,Grisaru:1976vm,Grisaru:1977kk,Aragone:1979hx,Weinberg:1980kq,Porrati:2008rm,Porrati:2012rd}. However, massive higher-spin fields are ubiquitous in particle physics.
In a free theory, they are sensible one-particle states according to the representation theory of the spacetime isometry group (e.g., the Poincaré group for Minkowski spacetime) \cite{Bekaert:2006py,Rahman:2012thy,Rahman:2015pzl,Ponomarev:2022vjb}, while introducing consistent interactions generally requires them to be composite instead of fundamental. 
The studies of higher-spin fields have a long history \cite{Fierz:1939ix,fronsdal_theory_1958,Chang:1967zzc,Singh:1974qz,deWit:1979sib},
 and are relevant to many theoretical directions, including massive gravity \cite{Arkani-Hamed:2002bjr,Hinterbichler:2011tt,Hassan:2011zd,Volkov:2011an,Hassan:2012wr,Dalmazi:2013sva,Kluson:2013jlo,deRham:2014zqa,Koenigstein:2015asa,Goon:2018fyu}, extra-dimensional models \cite{Kaluza:1921tu,Klein:1926tv,Aragone:1987dtt,Randall:1999ee,Csaki:2018muy}, strongly coupled gauge theories \cite{Witten:1979kh}, and string theory \cite{Sorokin:2004ie,Sagnotti:2011jdy,Chang:2012kt,Rahman:2012thy,Rahman:2015pzl,Ponomarev:2022vjb}.
It was also realized that massive higher-spin particles are crucial for a theory of gravity, as required by causality \cite{Buchbinder:2000ta,Camanho:2014apa}. Furthermore, higher-spin states could be essential for quantum gravity, since an infinite tower of massive higher-spin fields renders the theory renormalizable and finite as a UV completion of gravity \cite{Sorokin:2004ie,Sagnotti:2011jdy,Rahman:2015pzl,Noumi:2019ohm,Kato:2021rdz}.

The study of higher-spin fields is also well motivated in curved spacetime, particularly in de Sitter (dS) spacetime. Similar to the Minkowski case, the spacetime isometry group of dS allows arbitrary values of spin (integer or half-integer) \cite{Basile:2016aen,Sun:2021thf}.
Moreover, cosmic inflation \cite{Guth:1980zm,Linde:1981mu,Albrecht:1982wi,Baumann:2009ds,Wang:2013zva}, during which the spacetime background is nearly dS,  provides a natural testing ground for detecting these higher-spin particles.
Due to the spacetime expansion, these higher-spin particles can be produced simultaneously at the Hubble scale, $H\lesssim 10^{14}\text{ GeV}$. Through their interaction with either curvature perturbations or tensor perturbations, they can leave characteristic imprints in the late-time observables.
In particular, spectra (including mass and spin) of heavy particles produced during inflation can be identified by their imprints in the three-point (or higher-point) correlation functions of curvature perturbation, which can in principle be observed via either the Cosmic Microwave Background (CMB) \cite{Meerburg:2015zua,Planck:2018jri,Planck:2019kim,Galloni:2022mok,BICEP:2021xfz,Sohn:2024xzd} or the Large Scale Structure (LSS)  \cite{Dalal:2007cu,Slosar:2008hx,Cabass:2024wob}, a paradigm known as the Cosmological Collider physics \cite{Chen:2009zp,Baumann:2011nk,Noumi:2012vr,Gong:2013sma,Arkani-Hamed:2015bza}. 

In this work, we consider a complementary probe of higher-spin fields produced during inflation, namely gravitational waves. The basic picture is the following. During inflation, the higher-spin particles are spontaneously produced due to the spacetime expansion and the rolling inflaton. After production, these particles act as sources of tensor perturbations. These tensor perturbations are frozen outside the horizon during inflation and re-enter the horizon after inflation, becoming gravitational waves (GWs) that can be observed by current and future gravitational wave experiments.

There is a nice analogy for GW production through higher-spin particles. Normally, in Minkowski spacetime, GWs are produced if the source has a time-dependent quadrupole. Stationary objects, such as a stone, cannot radiate GWs. The situation is different in dS: Due to the absence of global energy conservation, even a stationary stone can radiate GWs.
In the inflationary spacetime, nonrelativistic higher-spin particles with mass much heavier than the Hubble scale act effectively as stationary stones, with their quadrupoles induced by the spin. Thus, with higher-spin states, we have a concrete model to realize stationary-stone-produced GWs in inflation.

For this reason, we focus on the scenario where the higher-spin particles lie in the principal series,\footnote{In dS spacetime, particles are categorized into exceptional series, discrete series, complementary series, and principal series based on their mass, serving as unitary irreducible representations of dS spacetime isometry \cite{Bekaert:2006py,Sun:2021thf,Basile:2016aen}. In our study, we focus on the scenario where the mass of the particles significantly exceeds the Hubble parameter, corresponding to the principal series. On the side of small masses below the Higuchi bound \cite{Higuchi:1986py}, there can be ``partially massless'' particles with their masses taking positive discrete values \cite{Zinoviev:2001dt,Deser:2003gw,Hinterbichler:2016fgl,Lee:2016vti,Baumann:2017jvh}, which is a special characteristic of dS spacetime. In this case, certain degrees of freedom of the field become decoupled, resulting in interesting phenomenology \cite{,Lee:2016vti,Baumann:2017jvh} that we will leave for future research endeavors.} namely $m>(s-1/2)H$ for $s\geq 1$.
However, gravitational production of heavy particles in inflation is exponentially suppressed by a Boltzmann factor $|\beta(k)|^2\propto \exp(-2\pi m/H)$, so are the resulting GWs. In recent years, it was emphasized that one can  naturally enhance the heavy particle production by including a parity-odd chemical potential generated by the rolling inflaton \cite{Turner:1987bw,Garretson:1992vt,Barnaby:2010vf,Barnaby:2011vw,Adshead:2015kza,Adshead:2018oaa,Chen:2018xck,Wang:2019gbi,Bodas:2020yho,Wang:2020ioa,Sou:2021juh,Wang:2021qez,Tong:2022cdz,Qin:2022lva,Qin:2022fbv}. Due to the additional energy injection through the inflaton rolling, the particle production rate of one helicity mode is exponentially amplified while the other suppressed. The amplified mode can thereby generate a relatively large GW signal \cite{Lue:1998mq,Alexander:2004us,Anber:2012du,Crowder:2012ik,Domcke:2016bkh,Machado:2018nqk,Machado:2019xuc,Salehian:2020dsf,Niu:2022quw}. 

For massive spinning particles with chemical potential, bosonic fields with spin $s=0,1$ have been extensively studied  \cite{Turner:1987bw,Garretson:1992vt,Barnaby:2010vf,Barnaby:2011vw,Wang:2019gbi,Bodas:2020yho,Wang:2020ioa,Sou:2021juh,Wang:2021qez,Qin:2022lva,Qin:2022fbv}, and the case of spin-$2$ has also been studied in recent years \cite{Maleknejad:2018nxz,Tong:2022cdz}. Similarly, fermionic cases with spin $s=1/2$ have been comprehensively analyzed \cite{Adshead:2015kza,Adshead:2018oaa,Chen:2018xck}, but with a notable distinction that the chemical potential could only alleviate the Boltzmann suppression to $\mathcal O(1)$ due to the Pauli blocking. 
Therefore, we aim to extend the analysis to bosonic fields with arbitrary spin in this work, and leave the study of higher-spin fermionic fields for future research.

For higher-spin particles, we focus on GWs generated primarily by spin-induced quadrupole, which is scale-invariant in the slow-roll limit. However, as the background field and the Hubble parameter evolve, the chemical potential changes accordingly. Consequently, during different periods of inflation, the particle production rate varies, resulting in varying strengths of backreaction on the background spacetime. Numerical calculations indicate that the chemical potential increases substantially during the late stage of inflation, leading to a copious amount of particle production. The particle number density becomes sufficiently large in this period that the backreaction from massive higher-spin fields cannot be ignored, causing a significant amplification of tensor perturbations. This is reflected in the shape of the gravitational wave spectrum as an increase in energy density on small scales (or high frequencies). Over a broad range of parameters, the GWs remain nearly scale-invariant and consistent with CMB observations at large scales, but become larger and potentially detectable by various current and future gravitational wave experiments at smaller scales.

The frequencies detectable by gravitational wave experiments span a wide range, from pulsar timing arrays (PTAs) such as NANOGrav \cite{NANOGrav:2023hde,NANOGrav:2023hvm,NANOGrav:2023gor}, EPTA \cite{EPTA:2023sfo,EPTA:2023fyk,EPTA:2023xxk}, IPTA \cite{Verbiest:2016vem}, and the Square Kilometre Array (SKA) telescopes \cite{Janssen:2014dka} in the nanoHz range to future planned laser interferometers such as LISA \cite{LISA:2017pwj}, Taiji \cite{Ruan:2018tsw}, Tianqin \cite{TianQin:2020hid}, BBO \cite{Kudoh:2005as,Cutler:2005qq}, and Ultimate DECIGO \cite{Kawamura:2020pcg} in the mHz to Hz range. In the frequency range of 10Hz to 100Hz, signals can be detected by the planned Einstein Telescope \cite{Hild:2008ng,Abac:2025saz}, as well as the Advanced-Laser Interferometer Gravitational-wave Observatory (aLIGO) and Advanced-Virgo (adVirgo) network \cite{KAGRA:2021kbb,LIGOScientific:2022sts,Jiang:2022uxp}.
Different parameters induce various features in the gravitational wave spectrum. Consequently, by performing joint measurements of GWs across different frequency bands, we may extract information about the mass, spin, and chemical potential of the higher-spin particles.

 The rest of this paper is organized as follows. In Sec.\;\ref{ChemPtlReview}, we briefly review the concept of chemical potential, together with its structure and properties. In Sec.\;\ref{FreeSpin2ChemPtlSect}, we present a detailed calculation of the GWs produced by the massive spin-$2$ case, derived from the explicit Lagrangian and the energy-momentum tensor of the theory. In Sec.\;\ref{spinschem}, we extend our derivations and calculations to bosons with arbitrary nonzero spins. Specifically, we derive the equations of motion (EoMs) incorporating the chemical potential, and we construct the Wronskian of spin-$s$ fields using the symplectic inner product method \cite{Lee:2016vti}, from which we obtain their mode functions. From this analysis, we reveal a discontinuity in the number of propagating degrees. for higher-spin particles as the chemical potential is turned on/off, which was observed in \cite{Tong:2022cdz} for spin-2 particles. In Sec.\;\ref{GWspins_phe}, we focus on the gravitational wave phenomenology of massive spin-$s$ fields. We derive the power spectrum of GWs and subsequently examine the effects of the slow-roll correction and backreaction. Based on experimental and theoretical constraints, we determine feasible parameter spaces and perform numerical calculations of the primordial GWs generated by massive spin-$s$ fields. We analyze the characteristics of GWs across different parameters and explore the potential for future experimental observations. Finally, we present our conclusion in Sec.\;\ref{ConclusionsSect}. More technical details are collected in the appendices. App.\;\ref{FeynmanRuleAppendix} outlines the Feynman rules of propagators and interactions, and App.\;\ref{Prop_estim} provides an estimation of the propagators. In App.\;\ref{polar_tensors}, the polarization tensors relevant to the analysis are presented. App.\;\ref{spin2EMT} discusses the energy-momentum tensor for the massive spin-2 field, followed by App.\;\ref{AppendixspinsEMT}, which extends this discussion to energy-momentum tensors for massive spin-$s$ fields. Finally, App.\;\ref{full-order_slowroll} includes the full-order solution of the slow-roll corrections.
 
\textbf{Notations and conventions:}
In most cases we use the metric $d s^2=a^2(\tau)\left[-d\tau^2+d\bm{x}^2\right]$ for the inflationary patch, where $a(\tau)$ is the scale factor and $-\infty<\tau<0$ is conformal time. In the slow-roll limit, $a(\tau)=-1/(H\tau)$ where the Hubble parameter $H$ is a constant. We also occasionally use physical time $t$, defined by $dt=a(\tau)d\tau$, and we use primes and overdots to denote derivatives with respect to conformal time \( \tau \) and physical time \( t \), respectively. We will use Greek letters for spacetime indices, \( \mu, \nu, \dots = 0, 1, 2, 3 \), and Latin letters for spatial indices, \( i, j, \dots = 1, 2, 3 \). $\epsilon^{\mu\nu\rho\sigma}$ is the 4-dimensional Levi-Civita symbol and we define the 3-dimensional Levi-Civita symbol $\epsilon^{ijk}\equiv\epsilon^{0ijk}$. 3-dimensional vectors are written in boldface, \( \bm{k} \), and unit vectors are written in hatted boldface, \( \hat{\bm{k}} \). A shorthand for the symmetrization of tensor indices is \( A_{(\mu} B_{\nu)} \equiv (A_{\mu}B_{\nu} + A_{\nu}B_{\mu})/2 \). The dimensional chemical potential parameter for the spin-$s$ field is defined as $\tilde{\kappa}^{(s)}\equiv 2\dot{\phi}_0/(s\Lambda_{c,s})$, while the dimensionless chemical potential parameter is defined as $\kappa_s\equiv \dot{\phi_0}/(H\Lambda_{c,s})$, where $H$ is the Hubble parameter, $s$ is the spin, $\Lambda_{c,s}$ is the cutoff parametrizing the scale of physics which is responsible for the chemical potential operator. And the conformal weight is defined as $\mu_s\equiv\sqrt{m_s^2/H^2-(2s-1)^2/4}$. 
We will use natural units, \( c = \hbar = 1 \), with reduced Planck mass \( M_{\text{pl}}^2 = 1/(8 \pi G) \).

\section{Review of helical chemical potential induced by inflaton rolling}\label{ChemPtlReview}
The chemical potential is a core concept that originates from statistic mechanics, where it acts as the Lagrange multiplier $\kappa$ controlling the total particle number in a grand canonical ensemble. The partition function of a grand canonical ensemble is well-known:
\begin{align}
    \mathcal{Z}=e^{-\frac{(H-\kappa N)}{T}}.
\end{align}
In a field theory, the partition function can be conveniently expressed as a phase-space path integral:
\begin{align}
\mathcal{Z}=\int~\mathcal{D}\Phi\mathcal{D}\Pi e^{-i\int dt\left[H[\Phi,\Pi]-\kappa N[\Phi,\Pi]-\int d^3\bm x\Pi\dot{\Phi}\right]}, 
\end{align}
where the Hamiltonian $H[\Phi,\Pi]$ is a function of the field variable $\Phi$ and its canonical conjugate momentum $\Pi$. As in statistic mechanics, the chemical potential $\kappa$ is defined to be the coefficient of the particle number operator $N[\Phi,\Pi]$. To make full use of the spacetime symmetry, we integrate out the canonical momentum and adopt the Lagrangian formalism, that is,
\begin{align}
    \mathcal{Z}=\int~\mathcal{D}\Phi e^{i\int d^4x\sqrt{-g}\left[\mathcal{L}(\Phi,\partial\Phi)+\kappa_\mu(x)J^\mu(\Phi,\partial\Phi)\right]},
\end{align}
where $\kappa_\mu J^\mu$ represents the general form of a background field coupled to the current density which is identified as a chemical potential, and the particle number is given by $N=\int_\Sigma d^3x_\mu J^\mu$.

However, the chemical potential defined in this form should satisfy some conditions to influence the particle production in a nontrivial way \cite{Chen:2018xck,Wang:2019gbi,Sou:2021juh}. 
More specifically, we can consider an example of a free complex scalar field $\Phi$ with the following Lagrangian \cite{Wang:2019gbi}:
\begin{align}
\mathcal{L}=\left[\left(\partial_t+i\kappa\right)\Phi^{\star}\right]\left[\left(\partial_t-i\kappa\right)\Phi\right]-|\partial_i\Phi|^2-m^2|\Phi|^2.
\end{align}
The scalar field $\Phi$ is charged under a global $\mathrm{U(1)}$ symmetry with the conserved charge $\mathcal{Q}=-i(\dot{\Phi}^\star\Phi-\Phi^\star\dot{\Phi})$.
If we do a field redefinition $\Phi\to e^{i\kappa t}\Phi$, the chemical potential in the above formula will be eliminated since its effect is just shifting the frequency $\omega\to\omega-\kappa$ in the mode expansion $\Phi\sim \int d^3\bm k\,\exp(-i\omega t+i\bm{k}\cdot \bm{x})(a_{\bm{k}}+b^{\dagger}_{-\bm{k}})$. Therefore, the chemical potential defined in this way is merely a reclassification of positive and negative modes, which does not have the physical effect of enhancing the particle production during the inflation.

The above argument suggests that, to have a physical effect, the chemical potential term should not be fully eliminated by a field redefinition. This can be realized in at least two ways: 1) The symmetry associated with the current $J^\mu$ is explicitly or spontaneously broken. 2) The symmetry is gauged and there exists a gauge field background that is not a pure gauge. The latter possibility more resembles the ordinary Schwinger pair production \cite{Lozanov:2018kpk,Maleknejad:2019hdr,Chua:2018dqh}, while the former option is more directly realized during inflation and has been more extensively explored \cite{Chen:2018xck,Wang:2019gbi,Wang:2020ioa,Bodas:2020yho,Sou:2021juh,Tong:2022cdz,Bodas:2024hih}. This is because the gradient of the rolling inflaton background $\phi_0(t)$ can be naturally viewed as an $1$-form albeit pure-gauge field:
\begin{align}
\kappa_\mu=\frac{\nabla_\mu\phi_0}{\Lambda_{c,s}} = (a\kappa,0,0,0).
\end{align}
Since $d\kappa=0$, we need a nonconserved current. Nonconserved currents made of particles of spin-0 \cite{Bodas:2020yho}, spin-$1/2$ \cite{Chen:2018xck,Wang:2019gbi}, spin-$1$ \cite{Wang:2020ioa,Bodas:2024hih}, and spin-2 \cite{Tong:2022cdz} have been studied in the literature. For higher-spin cases, the allowed chemical potential term consistent with the theory is fixed uniquely if we impose some reasonable assumptions and constraints, similar to what was done to spin-2 case in \cite{Tong:2022cdz},  and this will be discussed in Sec.\;\ref{spinschem}.
	\section{Massive spin-$2$ field with chemical potential}\label{FreeSpin2ChemPtlSect}
	\subsection{The effective Lagrangian and equations of motion}\label{IntroducingSin2ChemPtl}
	We begin with the effective Lagrangian of a massive spin-$2$ field with chemical potential:
\begin{align}\label{effect_Lag_spin2}
S_\text{spin-2} = S_{\rm EH}+S_{\rm FP}+S_{c}.
\end{align}
Here $S_{\rm EH}$ denotes the Einstein-Hilber action:
\begin{align}\label{effect_Lag_EH}
S_{\text{EH}}
	&=\int d^4x\sqrt{-g}\left[-\frac{1}{4}\nabla_{\mu}\Sigma_{\nu\lambda}\nabla^\mu\Sigma^{\nu\lambda}+\frac{1}{2}\nabla_{\mu}\Sigma^{\mu\lambda}\nabla^\nu\Sigma_{\nu\lambda}+\frac{1}{4}\nabla_{\mu}\Sigma\nabla^{\mu}\Sigma-\frac{1}{2}\nabla^{\nu}\Sigma^{\mu}_{~\nu}\nabla_{\mu}\Sigma\right.\notag\\
	&\left.-\frac{1}{2}H^2\left(\Sigma_{\mu\nu}\Sigma^{\mu\nu}+\frac{1}{2}\Sigma^2\right)\right],
\end{align}
where $\Sigma\equiv g^{\mu\nu}\Sigma_{\mu\nu}$.
The mass term takes the  Fierz-Pauli form \cite{maggiore2007gravitational,Koenigstein:2015asa,Tong:2022cdz}:
\begin{align}
    S_{\rm FP}=-\int d^4x\sqrt{-g}\frac{1}{4}m^2\left(\Sigma_{\mu\nu}\Sigma^{\mu\nu}-\Sigma^2\right),
\end{align}
which ensures that no ghost is introduced at the level of free theory. Furthermore, to avoid negative-norm states, the mass is generally required to satisfy the Higuchi bound, which is given by $m^2 \geq s(s-1)H^2$ \cite{Higuchi:1986py}. As noted earlier, we have chosen the principal series, and the Higuchi bound is automatically satisfied.
Finally, we have introduced the following chemical potential term: 
\begin{align}
    S_c=-\int d^4x\sqrt{-g}\frac{\nabla_{\mu}\phi}{2\Lambda_{c,2}}\varepsilon^{\mu\nu\rho\sigma}\Sigma_{\nu\lambda}\nabla_{\rho}\Sigma_{\sigma}^{~\lambda},
\end{align}
where $\varepsilon^{\mu\nu\rho\sigma} \equiv \epsilon^{\mu\nu\rho\sigma}/{\sqrt{-g}}$ is the Levi-Civita tensor. For the case of a massive spin-$2$ field, the chemical potential of this form is the unique choice to ensure the theory's self-consistency \cite{Tong:2022cdz}, namely:
(i) the chemical potential comes from a local operator with the lowest possible mass dimension;
(ii) the operator must take the form $\kappa_\mu J^\mu$ with the current $J^\mu$ non-conserved;
(iii) the operator is quadratic in $h_{\mu\nu}$ and gives rise to a consistent linear theory without ghost;
(iv) the chemical potential term should break de Sitter (dS) boosts to relax Boltzmann suppression while preserving all other dS isometries.

The EoM of the massive spin-2 field is given by the variation of the action
\begin{align}
    \frac{\delta S_{\text{spin-$2$}}}{\sqrt{-g}\delta\Sigma^{\mu\nu}}=0,
\end{align}
which yields:
\begin{align}\label{eq_spin2EOM}
0=&\nabla_\rho\nabla^\rho\Sigma_{\mu\nu}+\nabla_{\mu}\nabla_{\nu}\Sigma-\nabla_{\mu}\nabla_{\rho}\Sigma^{\rho}_{~\nu}-\nabla_{\nu}\nabla_{\rho}\Sigma^{\rho}_{~\mu}+g_{\mu\nu}\left(\nabla_{\rho}\nabla_{\sigma}\Sigma^{\rho\sigma}-\nabla_\rho\nabla^\rho\Sigma\right)-2H^2\left(\Sigma_{\mu\nu}+\frac{1}{2}g_{\mu\nu}\Sigma\right)\notag\\
&-m^2(\Sigma_{\mu\nu}-g_{\mu\nu}\Sigma)-\frac{\nabla_\alpha\phi}{\Lambda_{c,2}}\varepsilon^{\alpha\kappa\rho\sigma}\left(g_{\kappa\nu}\nabla_\rho\Sigma_{\mu\sigma}+g_{\kappa\mu}\nabla_\rho\Sigma_{\nu\sigma}\right).
\end{align}
Below we try to count the number of degrees of freedom for the spin-2 field $\Sigma_{\mu\nu}$. If we turn off the chemical potential, we can establish the Fierz-Pauli constraints on the field components by analyzing the EoM Eq.~\eqref{eq_spin2EOM}:
\begin{align}\label{eq_ttconst}
\nabla^{\mu}\Sigma_{\mu\nu}=0,\quad \Sigma^{\mu}_{~\mu}=0.
\end{align}
Thus, the field possesses $10-4-1=5$ degrees of freedom. Indeed, the number of degrees of freedom can also be obtained through group theory methods by analyzing the $\mathrm{SO(3)}$ group representation corresponding to the spin-$2$ field.

Now we turn on the chemical potential. Under some reasonable assumptions \cite{Tong:2022cdz}, the transverse traceless constraint still holds, with which the original EoM is simplified to the following:
\begin{align}\label{EOMspin2}
    \left(\nabla_{\rho}\nabla^\rho-(m^2+2H^2)\right)\Sigma_{\mu\nu}-\frac{\nabla_\alpha\phi}{\Lambda_{c}}\varepsilon^{\alpha\kappa\rho\sigma}\left(g_{\kappa\nu}\nabla_\rho\Sigma_{\mu\sigma}+g_{\kappa\mu}\nabla_\rho\Sigma_{\nu\sigma}\right)=0.
\end{align}
Surprisingly, as we will explicitly show later, the EoM Eq.~\eqref{EOMspin2} requires the vector modes with helicity $\lambda=\pm1$ actually disappear. This result indicates that, in the presence of the chemical potential, only $5-2=3$ degrees of freedom are able to propagate. In fact, this is a common phenomenon in the theory of massive higher-spin fields with chemical potential, as will be detailed in Sec.\;\ref{spinschem}.

\subsection{Wronskian and mode functions}\label{spin2wrosk}
The massive spin-$2$ field $\Sigma_{\mu\nu}$ can be decomposed in terms of helicity eigenstates in the Fourier space:
\begin{align}\label{helidecomspin2}
    \Sigma_{\mu\nu}(\tau,\bm{k})=\sum_{\lambda=0,\pm1,\pm2}\Sigma^{\lambda}_{\mu\nu}(\tau,\bm{k}),
\end{align} 
where 
\begin{equation}
    \Sigma_{\mu\nu}(\tau,\bm{k})=\int \frac{d^3\bm k}{(2\pi)^3}\Sigma_{\mu\nu}(\tau,\bm x)e^{i\bm{k}\cdot\bm{x}}
\end{equation}
is the Fourier transformation of spin-$2$ field.
We can further write these eigenstates in the component form explicitly \cite{Lee:2016vti,Tong:2022cdz}:
\begin{align}
	\begin{array}{l l l l}
		& \Sigma_{00}^0 = \Sigma_0^0, \qquad& \Sigma_{00}^{\pm1} = 0, \qquad& \Sigma_{00}^{\pm2} = 0, \\
		& \Sigma_{0i}^0 = \Sigma_1^0 \epsilon^0_i,\qquad & \Sigma_{0i}^{\pm1} = \Sigma_1^{\pm1} \epsilon^{\pm1}_i, \qquad& \Sigma_{0i}^{\pm2} = 0, \\
		& \Sigma_{ij}^0 = \Sigma_2^0 \epsilon^0_{ij}+\frac{1}{3}\Sigma_0^0\delta_{ij},\qquad & \Sigma_{ij}^{\pm1} = \Sigma_2^{\pm1} \epsilon^{\pm1}_{ij}, \qquad& \Sigma_{ij}^{\pm2} = \Sigma_2^{\pm2} \epsilon^{\pm2}_{ij},
	\end{array}
\end{align}
from which we can observe that the components with helicity $|\lambda|> n$ vanish, where $n$ is the number of the spatial indices that takes values in $0,1,2$. We define the polarization tensors to satisfy the following conditions:
\begin{align}
\begin{array}{l l l l l}
&\epsilon_i^0(\hat{\bm{k}})=\hat{\bm{k}}_i,\qquad &\bm{k}_i\epsilon_i^{\pm1}(\hat{\bm{k}})=0,\qquad &\bm{k}_i\epsilon_{ij}^{0}(\hat{\bm{k}})=\epsilon_j^0(\hat{\bm{k}}),\qquad&\bm{k}_i\epsilon_{ij}^{\pm1}(\hat{\bm{k}})=\frac{3}{2}\epsilon_j^{\pm1}(\hat{\bm{k}})\\\\
&\bm{k}_i\epsilon_{ij}^{\pm2}(\hat{\bm{k}})=0,\qquad &\epsilon_i^{\pm1}(\hat{\bm{k}})\epsilon_i^{\pm1\star}(\hat{\bm{k}})=2,\qquad &\epsilon_{ij}^{\pm2}(\hat{\bm{k}})\epsilon_{ij}^{\pm2\star}(\hat{\bm{k}})=4,\qquad&\epsilon_{ij}^{\pm2}(\hat{\bm{k}})=\epsilon_{ij}^{\mp2\star}(\hat{\bm{k}})=\epsilon_{ij}^{\pm2}(-\hat{\bm{k}}),
\end{array}
\end{align}
and the general form of the rank-$n$ polarization tensor $\epsilon^\lambda_{i_1\cdots i_n}$ can be found in App.\;\ref{polar_tensor}.

Then, we aim to compute the contraction of rank-1 polarization tensors corresponding to different momentum directions. In particular, we are concerned with $\epsilon^{\lambda}_{i}(\hat{\bm{k}})\epsilon^{\lambda^\prime}_{i}(\hat{\bm{p}})$ with $\lambda,\lambda'=\pm1$. 
 We start by performing a spatial rotation to align $\hat{\bm{p}}$ with the direction of $\hat{\bm{k}}$:
\begin{align}
	\epsilon^{\lambda}_{i}(\hat{\bm{k}})\epsilon^{\lambda^\prime}_{i}(\hat{\bm{p}})&=\epsilon^{\lambda}_{i}(\hat{\bm{k}})R^T_{ij}\epsilon^{\lambda^\prime}_{j}(\hat{\bm{k}}),
\end{align}
where the rotation matrix and the polarization tensor can be written explicitly
\begin{align}
	R_{ij}=\left(\begin{array}{lll}
		1 &0 &0\\
		0 &\mathrm{cos}\theta &-\mathrm{sin}\theta \\
		0 &\mathrm{sin}\theta &\mathrm{cos}\theta
	\end{array}
		\right),\quad
		\epsilon^\lambda_i(\bm{k})=\left(\begin{array}{l}
			1 \\
			\lambda i \\
			0 
		\end{array}
			\right),
 \end{align}
thus the contraction becomes
\begin{align}\label{contact_didd_direct}
	\epsilon^{\lambda}_{i}(\hat{\bm{k}})\epsilon^{\lambda^\prime}_{i}(\hat{\bm{p}})&=1-\lambda\lambda^\prime \mathrm{cos}\theta=1-\lambda\lambda^\prime\frac{{\bm{k}}\cdot\bm{p}}{|\bm{k}||\bm{p}|}.
\end{align}

We first consider the EoM of helicity $\lambda=0$ mode derived from Eq.~(\ref{EOMspin2}):
\begin{align}\label{spin2_scalarmode}
    \partial^2_\tau \Sigma_0^0+2aH\partial_\tau \Sigma_0^0+\left(k^2+m^2a^2\right)\Sigma_0^0=0,
\end{align}
with $a(\tau)=-1/(H\tau)$ the scale factor, and we can see that the evolution of the helicity-0 mode is not affected by the chemical potential, and thus suffers from the Boltzmann suppression.

The vector modes that have been mentioned previously vanish due to the presence of the chemical potential \cite{Tong:2022cdz}. More specifically, we can write down the EoMs for modes $\Sigma_{n}^{\pm 1}$ from Eq.~\eqref{EOMspin2} as well as the constraints Eq.~\eqref{eq_ttconst} as the following:
\begin{align}
\partial_{\tau}^2\Sigma^{\pm1}_{1}+\left[k^2\pm k\tilde{\kappa}a+\left(m^2-2H^2\right)a^2\right]\Sigma^{\pm1}_{1}&=0,\label{spin2_vectormode1}\\
\partial_{\tau}^2\Sigma^{\pm1}_{2}-2aH\partial_\tau \Sigma^{\pm1}_{2}+\left[k^2\pm k\tilde{\kappa}a+\left(m^2-2H^2\right)a^2\right]\Sigma^{\pm1}_{2}&=-i\frac{4kaH}{3}\Sigma^{\pm1}_{1},\label{spin2_vectormode2}\\
-i\frac{2}{3k}\left(\partial_\tau\Sigma^{\pm1}_{1}+2aH\Sigma^{\pm1}_{1} \right)&=\Sigma^{\pm1}_{2},\label{spin2_vectormodeconstraint}
\end{align}
where $\tilde{\kappa}\equiv \dot{\phi}_0/\Lambda_{c,2}$. These equations together impose an algebraic constraint on $\Sigma_{1}^{\pm1}$:
\begin{align}\label{chemforbiddenspin2}
    iHk\tilde{\kappa} a^2 \Sigma_{1}^{\pm1}=0,
\end{align}
and thus $\Sigma_{1}^{\pm 1}$ vanish, so do $\Sigma_{2}^{\pm 1}$. This implies that the vector mode cannot have propagating degrees of freedom, which is a common phenomenon in the theory of higher-spin fields with chemical potential, as will be shown in Sec.\;\ref{spinschemforbid}.
 
Finally, the EoMs of helicity $\lambda=\pm2$ modes are:
\begin{align}\label{spin2_tensormode}
\partial_{\tau}^2\Sigma_2^{\pm2}-2aH\partial_\tau\Sigma_2^{\pm2}+\left[k^2\pm2 k\tilde{\kappa}a+{\left(m^2-2H^2\right)}a^2\right]\Sigma_2^{\pm2}=0,
\end{align}
and the mode functions can be solved as
\begin{align}\label{mode_func_lam2}
	\Sigma^{\pm2}_2(\tau,k)=-\frac{\mathcal{N}_{\pm2}}{\tau}W_{\pm i\kappa,i\mu}(2ik\tau),
\end{align}
where $\mu\equiv\sqrt{m^2/H^2-9/4},~\kappa\equiv\tilde{\kappa}/{H}$ and $W_{\nu,\mu}(z)$ is the Whittaker W function.
The normalization constant $\mathcal{N}_{\pm2}$ can be determined by solving the Wronskian (or normalization condition) which is from canonical commutation relations for different components of the field \cite{Tong:2022cdz}. Explicitly, we make use of the following inner product defined on a 3-dimensional spatial slice $\Sigma$:
\begin{align}
\left(A_{\mu\nu},B_{\rho\sigma}\right)\equiv \int_{\Sigma}\sqrt{\hat{g}}d^3x_\rho M^\rho (A,B),
\end{align}
where $\hat{g}$ is the induced metric and $M^\rho$ is a current defined as:
\begin{align}
    M^\rho(A,B)\equiv A^\star_{\mu\nu}\nabla^\rho B^{\mu\nu} -B^{\mu\nu}\nabla^\rho A^\star_{\mu\nu}+\frac{\phi}{\Lambda_{2,c}\sqrt{-g}}\epsilon^{\alpha\nu\rho\sigma}\nabla_\alpha \left(A^\star_{\mu\nu} B^{\mu}_{~\sigma} -B^{\mu\nu} A^\star_{\mu\sigma}\right).
\end{align}
This current is conserved, $\nabla_\mu M^\mu=0$ due to the EoM. With the above inner product, the normalization condition can be expressed as 
\begin{align}
    \left(\Sigma^\alpha_{\mu\nu}e^{i\bm{k}\cdot\bm{x}},\Sigma^\beta_{\mu\nu}e^{i\bm{k^\prime}\cdot\bm{x}}\right)=2\delta_{\alpha\beta}\delta^{(3)}(\bm{k}-\bm{k^\prime}),
\end{align}
Solving this normalization condition, we get:
\begin{align}
\mathcal{N}_{\pm2}=\frac{e^{\mp\pi\kappa/2}}{2H\sqrt{k}}.
\end{align}	

Now, by combining Eq.~(\ref{spin2_scalarmode}), Eq.~(\ref{spin2_vectormode1}), Eq.~(\ref{spin2_vectormode2}), and Eq.~(\ref{spin2_tensormode}), we can see that the dispersion relation takes the form
\begin{align}\label{dispersion_rel}
\omega^2 = k^2 +\lambda\kappa k + m^2, 
\end{align}
and the term linear in $k$ represents the effect of the chemical potential. When $\kappa > 0$, we observe that the mode with negative helicity $\lambda < 0$ is exponentially enhanced, while the mode with positive $\lambda >0 $ is exponentially suppressed, and vice versa. This is the key feature of this type of chemical potential. The helicity-dependent behavior is a manifestation of the spontaneous parity breaking due to the chemical potential coupling, and it results in a parity-violating power spectrum of GWs.  

To conclude, for positive $\kappa$, the state with the negative helicity ($\lambda = -2$) is most enhanced with respect to others. Therefore, in the following analysis of induced GWs, we will only keep the $\lambda=-2$ component as a good approximation.

\subsection{Gravitational wave calculations}\label{GWspin2}
In this subsection, we study the power spectrum of GWs generated by the massive spin-$2$ field with a chemical potential enhancement. The GWs come from the tensor perturbation of the metric during inflation, which can be written as
 \begin{align}
    d s^2=a^2(\tau)\left[-d\tau^2+\left(\delta_{ij}+h^{TT}_{ij}\right)dx^idx^j\right],
\end{align}
where the tensor perturbations $h^{TT}_{ij}$ is traceless and transverse, $a(\tau)=-1/(H\tau)$ is the scale factor and $\tau$ is the conformal time which goes from $-\infty$ to $0$. 
As a massless spin-2 field, $h^{TT}_{ij}$ can be decomposed into two modes
    \begin{align}
    h_{ij}^{TT}(\tau,\bm{k})&=\sum_{\lambda=\pm 2}\epsilon_{ij}^{\lambda}(\bm{k})h^{\lambda}(\tau,\bm{k})=\sum_{\lambda=\pm 2}\epsilon_{ij}^{\lambda}(\bm{k})\left(h^{\lambda}(\tau,k)b_{\bm{k},\lambda}+h^{\lambda\star}(\tau,k)b^{\dagger}_{-\bm{k},\lambda}\right).
    \end{align}
  The EoMs of the tensor perturbations yields
    \begin{align}\label{EoMtensorperfree}
        \partial_\tau^2 h^{TT}_{ij}(\tau,\bm{k})+k^2 h^{TT}_{ij}(\tau,\bm{k})+2aH\partial_\tau h^{TT}_{ij}(\tau,\bm{k})=0.
    \end{align}
 The mode function is obtained by directly solving the above equation with Bunch-Davies initial condition:
\begin{align}
	h^\lambda(\tau,k)&=\frac{2H}{M_{\rm pl}\sqrt{2k^3}}(1+ik\tau)e^{-ik\tau}.
\end{align}
The presence of matter introduces a source term to the EoMs of the tensor perturbations becomes \cite{maggiore2007gravitational}
\begin{align}\label{EoMtensorper}
        \partial_\tau^2 h_{ij}^{TT}(\tau,\bm{k})+k^2 h_{ij}^{TT}(\tau,\bm{k})+2aH\partial_\tau h_{ij}^{TT}(\tau,\bm{k})=16\pi GT_{ij}^{TT}(\tau,\bm{k}),
    \end{align}
where $G$ is the Newton constant. The solution of Eq.~(\ref{EoMtensorper}) can be obtained by the  standard Green function method:
\begin{align}
    h_{ij}^{TT}(\tau_0,\bm{k})=\int d{\tau}^\prime G(\tau_0,\tau^\prime,k) 16\pi G{T}_{ij}^{TT}(\tau^\prime,\bm{k}),
\end{align}
where $\tau_0=0$ is evaluated at the end of the inflation and the retarded Green’s function $G(\tau_0,\tau^\prime,k)$ satisfies the equation
\begin{align}
     \partial^2_{\tau} G(\tau,\tau^\prime,{k})+k^2 G(\tau,\tau^\prime,{k})+2aH\partial_{\tau} G(\tau,\tau^\prime,{k})=\delta(\tau-\tau^\prime).
\end{align}
The inhomogeneous particular solution to this equation is \cite{Cook:2011hg}:
\begin{align}\label{eq_retardedpropagator}
G(\tau,\tau^\prime,k)=a^2(\tau^\prime)\frac{H^2}{k^3}\tilde{G}(\tau,\tau^\prime,k),
\end{align}
where we have introduced the reduced retarded Green's function $\tilde{G}(\tau,\tau^\prime,k)$, defined as
\begin{align}
    \tilde{G}(\tau,\tau^\prime,k)\equiv \left[\left(1+k^2\tau\tau^\prime\right)\mathrm{sin}k(\tau-\tau^\prime)+k(\tau^\prime-\tau)\mathrm{cos}k(\tau-\tau^\prime)\right]\theta(\tau-\tau^\prime).
\end{align}
Evaluating the conformal time at the end of the inflation, the reduced retarded Green's function becomes
\begin{align}
    \tilde{G}(\tau_0,\tau^\prime,k)=-\mathrm{sin}k\tau^\prime+k\tau^\prime\mathrm{cos}k\tau^\prime.
\end{align}

As we have shown previously, the massive spin-$2$ field with chemical potential can be decomposed into different helicity eigenstates:
\begin{align}\label{expandmassivespin2}
		\Sigma_{ij}(\tau,\bm{k})&=\sum_{\lambda=0,\pm1,\pm2}\epsilon_{ij}^\lambda(\hat{\bm{k}})\Sigma^\lambda_2(\tau,\bm{k})\supset\epsilon_{ij}^{-2}(\hat{\bm{k}})\left(\Sigma^{-2}_2(\tau,k)a_{\bm{k}}+\Sigma^{{-2},\star}_2(\tau,k)a^\dagger_{-\bm{k}}\right),
\end{align}
 where we only focus on the (spatial components of the) highest helicity modes with $\lambda=-2$, which is maximally enhanced by the chemical potential, and its mode function is given in Eq.~(\ref{mode_func_lam2}).

To calculate the GWs induced by the massive spin-$2$ field $\Sigma_{\mu\nu}$, we need the energy-momentum tensor of the latter. The energy-momentum tensor of a massive spin-$2$ field can be derived by varying its Lagrangian with respect to the spacetime metric, whose complete form is collected in App.\;\ref{spin2EMT}. From the energy-momentum tensor Eqs.~\eqref{spin2EMTorbit} and (\ref{spin2EMTspin}), it can be observed that there are various forms of interactions between the graviton and the massive spin-$2$ field. Among these terms, we choose the pure spin term\footnote{We divide the energy-momentum tensor into two parts: one generated by orbital angular momentum and the other by spin angular momentum. The precise definitions and explicit forms are provided in App.\;\ref{spin2EMT}. Our classification is based on the explicit spatial indices $i,j$ of the energy-momentum tensor $T_{ij}$. If the spatial indices are associated with spatial derivative terms, they are attributed to orbital angular momentum; if they are carried by the field itself, they are attributed to spin angular momentum.} with two time derivatives, which is the leading term in pure spin interactions containing two modes with the highest helicity:\footnote{Contributions from other terms such as $a^{-2}m^2\Sigma_{il}\Sigma_{jl}$ and $a^{-2}H^2\Sigma_{il}\Sigma_{jl}$ are either of the same order or suppressed by powers of $H/m\ll 1$}
\begin{align}\label{EMTspin2chosen}	
	\tilde{T}_{ij}\supset
-2a^{-4}\partial_{\tau}\Sigma_{il}\partial_{\tau}\Sigma_{jl},
\end{align} 
where the factor of $2$ preceding $a^{-4}$ is the spin-dependent factor of the energy-momentum tensor. For the spin-$2$ field, this factor corresponds to $s = 2$. A general and detailed derivation of this factor for arbitrary spin bosons will be given in App.\;\ref{AppendixspinsEMT}.

Next, we calculate the transverse and traceless part of the term in Eq.~(\ref{EMTspin2chosen}). The transverse and traceless part of the energy-momentum tensor can be written explicitly as
\begin{align}
    \tilde{T}_{ij}^{TT}(\tau,\bm{k})&=\Lambda_{ij,kl}(\hat{\bm{k}})T_{kl}(\tau,\bm{k})=-2a^{-4}(\tau)\int \frac{d^3\bm{p}}{(2\pi)^3}\Lambda_{ij,kl}(\hat{\bm{k}})\partial_{\tau}\Sigma_{km}(\tau,\bm{k-p})\partial_{\tau}\Sigma_{lm}(\tau,\bm{p})\notag\\
    &=-2a^{-4}(\tau)\sum_{\lambda,\lambda^\prime}\int \frac{d^3\bm{p}}{(2\pi)^3} \partial_{\tau}\Sigma^{\lambda}(\tau,\bm{p})\partial_{\tau}\Sigma^{\lambda^{\prime}}(\tau,\bm{k}-\bm{p})\Lambda_{ij,kl}(\hat{\bm{k}})\epsilon_{lm}^{\lambda}(\hat{\bm{p}})\epsilon_{km}^{\lambda^\prime}(\widehat{\bm{k}-\bm{p}}),
\end{align}
where $\Lambda_{ij,kl}(\hat{\bm{k}})$ is a projection operator which projects a rank-2 tensor into its symmetric, traceless and transverse component:
\begin{align}
    \Lambda_{ij,kl}(\hat{\bm{k}})=\Lambda_{ij,kl}^{+}(\hat{\bm{k}})+\Lambda_{ij,kl}^{-}(\hat{\bm{k}}),
\end{align}
where
\begin{equation}
\Lambda_{ij,kl}^{\pm}(\hat{\bm{k}})=\frac{1}{4}\epsilon_{ij}^{\pm}(\hat{\bm{k}})\epsilon_{kl}^{\pm}(-\hat{\bm{k}})    
\end{equation}
are the projection tensors corresponding to different helicities of gravitons. 
 
We then compute the power spectrum of GWs by first evaluating the two-point function of tensor perturbations, as:
\begin{align}\label{tensor2pinot}
    &\langle  h_{ij}^{TT}(\tau_0,\bm{k}) h_{ij}^{TT\star}(\tau_0,\bm{k})\rangle\notag\\
    &=(\frac{2}{M_{\rm pl}^2})^2\int d\tau^\prime \int d\tau^{\prime\prime}G(\tau_0,\tau^\prime,k)G(\tau_0,\tau^{\prime\prime},k)\tilde{T}_{ij}^{TT}(\tau^\prime ,\bm{k})\tilde{T}_{ij}^{TT}(\tau^{\prime\prime},\bm{k})\notag\\
    &=16\frac{H^4}{M_{\rm pl}^4}\frac{1}{k^6}\int d\tau^\prime a^{-2}(\tau^\prime)\tilde{G}(\tau_0,\tau^\prime,k)\int d\tau^{\prime\prime}a^{-2}(\tau^{\prime\prime})\tilde{G}(\tau_0,\tau^{\prime\prime},k)\int \frac{d^3\bm{p}}{(2\pi)^3}\int \frac{d^3\bm{q}}{(2\pi)^3}\notag\\
    &\times\Lambda_{kl,k^\prime l^\prime}(\hat{\bm{k}})\langle\partial_{\tau^\prime}\Sigma_{km}(\tau^\prime,\bm{k-p})\partial_{\tau^\prime}\Sigma_{lm}(\tau^\prime,\bm{p})\partial_{\tau^{\prime\prime}}\Sigma_{k^\prime n}^\star(\tau^{\prime\prime},\bm{k-q})\partial_{\tau^{\prime\prime}}\Sigma^\star_{l^\prime n}(\tau^{\prime\prime},\bm{q})\rangle.
\end{align}
Applying Wick’s theorem and neglecting the disconnected terms, the two-point function of tensor perturbations can be constructed from the two-point function of the massive spin-$2$ field as:\footnote{Here we only consider the Gaussian part of $\Sigma_{ij}$ and neglect its interaction with the inflaton fluctuation.}
\begin{align}
    &\langle\partial_{\tau^\prime}\Sigma_{km}(\tau^\prime,\bm{k-p})\partial_{\tau^\prime}\Sigma_{lm}(\tau^\prime,\bm{p})\partial_{\tau^{\prime\prime}}\Sigma_{k^\prime n}^\star(\tau^{\prime\prime},\bm{k-q})\partial_{\tau^{\prime\prime}}\Sigma^\star_{l^\prime n}(\tau^{\prime\prime},\bm{q})\rangle\notag\\
    &=\langle \partial_{\tau^\prime}\Sigma_{lm}(\tau^\prime,\bm{p})\partial_{\tau^{\prime\prime}}\Sigma^\star_{l^\prime n}(\tau^{\prime\prime},\bm{q}) \rangle\langle \partial_{\tau^\prime}\Sigma_{km}(\tau^\prime,\bm{k-p})\partial_{\tau^{\prime\prime}}\Sigma_{k^\prime n}^\star(\tau^{\prime\prime},\bm{k-q}) \rangle.
\end{align}
We can see the above expectation value is non-vanishing only when $\bm q=\bm p$.

We can write down the explicit expression for the momenta
\begin{align}\label{momenta}
	\bm{k}&=(0,0,k)\notag\\
	\bm{p}&=kl(0,-\mathrm{sin}\theta,\mathrm{cos}\theta)\notag\\
	\bm{k}-\bm{p}&=k(0,l\mathrm{sin}\theta,1-l\mathrm{cos}\theta)
\end{align}
whose inner products and modules then yield
\begin{align}\label{momen_cont}
 \begin{array}{lll}
\bm{k}\cdot\bm{p}=k^2l\mathrm{cos}\theta, & \bm{k}\cdot(\bm{k}-\bm{p})=k^2(1-l\mathrm{cos}\theta),&\bm{p}\cdot(\bm{k}-\bm{p})=-k^2l^2+k^2l\mathrm{cos}\theta\\
	|\bm{k}|=k,&|\bm{p}|=kl,&|\bm{k}-\bm{p}|=k\sqrt{l^2+1-2l\mathrm{cos}\theta}
 \end{array},
\end{align}
and further expand the massive spin-$2$ field in Eq.~(\ref{tensor2pinot}) according to Eq.~(\ref{expandmassivespin2}) and contract the polarization tensors using Eq.~(\ref{contact_didd_direct}) to obtain the corresponding angular-dependent factor:
\begin{align}\label{angular_spin2}
&\sum_{\lambda,\lambda',\lambda''}\Lambda_{ij,kl}(\hat{\bm{k}})\epsilon_{il^\prime}^{\lambda^\prime}(\hat{\bm{p}})\epsilon_{jl^\prime}^{\lambda^{\prime\prime}}(\widehat{\bm{k}-\bm{p}})\epsilon_{km}^{\lambda^\prime,\star}(\hat{\bm{p}})\epsilon_{lm}^{\lambda^{\prime\prime},\star}(\widehat{\bm{k}-\bm{p}})\notag\\
&=\sum_{\lambda,\lambda',\lambda''}\Lambda_{ij,kl}(\hat{\bm{k}})\epsilon_{il^\prime}^{\lambda^\prime}(\hat{\bm{p}})\epsilon_{jl^\prime}^{\lambda^{\prime\prime}}(\widehat{\bm{k}-\bm{p}})\epsilon_{km}^{\lambda^\prime}(-\hat{\bm{p}})\epsilon_{lm}^{\lambda^{\prime\prime}}(\widehat{\bm{p}-\bm{k}})\notag\\
&\supset\frac{1}{4}\sum_{\lambda=\pm2}\epsilon_{ij}^{\lambda}(\hat{\bm{k}})\epsilon_{kl}^{\lambda}(-\hat{\bm{k}})\epsilon_{il'}^{-2}(\bm{\bm{p}})\epsilon_{jl'}^{-2}(\widehat{\bm{k}-\bm{p}})\epsilon_{km}^{-2}(-\hat{\bm{p}})\epsilon_{lm}^{-2}(\widehat{\bm{p}-\bm{k}})\notag\\
&=\frac{1}{4}\sum_{\lambda=\pm2}\left[(1-\frac{\lambda}{2}\mathrm{cos}\theta)\left(1-\frac{\lambda}{2}\frac{1-l\mathrm{cos}\theta}{\sqrt{1+l^2-2l\mathrm{cos}\theta}}\right)\left(1-\frac{\mathrm{cos}\theta-l}{\sqrt{1+l^2-2l\mathrm{cos}\theta}}\right)\right]^2,
\end{align}
where in the third line we only keep the chemical-potential-enhanced mode $\lambda^\prime,\lambda^{\prime\prime}=-2$ (we consider the case that $\kappa>0$). 
The time derivative of the $\lambda=-2$ mode is
 \begin{align}
	\partial_{\tau}\Sigma^{-2}_2(\tau,k)=-\mathcal{N}_{-2}\left(-\frac{1}{\tau^2}W_{-i\kappa,i\mu}(2ik\tau)+\frac{1}{\tau}\partial_{\tau}W_{-i\kappa,i\mu}(2ik\tau)\right).
\end{align}
Putting everything into Eq.~(\ref{tensor2pinot}) and averaging over the spatial direction of momentum, 
the power spectrum of the tensor perturbations induced by the massive spin-$2$ field then yields
\begin{align}\label{GWpowerspectrum_spin2}
	\mathcal{P}_h(k)
 &=\frac{2k^3}{(2\pi)^2}\frac{1}{V}\int \frac{d\hat{\bm{k}}}{4\pi}\langle h^{TT}_{ij}(\bm{k},\tau_0),h^{\star TT}_{ij}(\bm{k},\tau_0)\rangle\notag\\
	&=\frac{H^4}{M_{\rm pl}^4}\frac{e^{2\pi\kappa}}{16\pi^4}\int^{\infty}_0dl l\notag\\
	&\times\int^{\pi}_0d\theta\frac{\mathrm{sin}\theta}{\sqrt{l^2-2l\mathrm{cos}\theta+1}}\sum_{\lambda=\pm}\left[(1-\lambda\mathrm{cos}\theta)\left(1-\lambda\frac{1-l\mathrm{cos}\theta}{\sqrt{1+l^2-2l\mathrm{cos}\theta}}\right)\left(1-\frac{\mathrm{cos}\theta-l}{\sqrt{1+l^2-2l\mathrm{cos}\theta}}\right)\right]^2\notag\\
	&\times\bigg|\int^{x_{\rm max}}_{x_{0}}dx_1x_1^2\tilde{G}(\tau_f,x_1)\left(-\frac{1}{x_1^2}W_{-i\kappa,i\mu}(-2ilx_1)+\frac{1}{x_1}\partial_{x_1}W_{-i\kappa,i\mu}(-2ilx_1)\right)\notag\\
	& \times\left(-\frac{1}{x_1^2}W_{-i\kappa,i\mu}(-2i\sqrt{l^2-2l\mathrm{cos}\theta+1}x_1)+\frac{1}{x_1}\partial_{x_1}W_{-i\kappa,i\mu}(-2i\sqrt{l^2-2l\mathrm{cos}\theta+1}x_1)\right) \bigg|^2,
	\end{align}
 where we redefine the variable $-k\tau\equiv x$ to investigate the scale dependence of the power spectrum, and $\tilde{G}(\tau_f,x)\equiv x\mathrm{cos}x-\mathrm{sin}x$.\footnote{
One can also do the calculations via the ``in-in'' formalism (see e.g. \cite{Chen:2017ryl}), and our calculation corresponds to the ``classical loop'' \cite{Cespedes:2023aal}. More specifically, the retarded propagator of graviton} \eqref{eq_retardedpropagator} is the imaginary part of its bulk-to-boundary propagator, while we only use the real part of the bulk-to-bulk propagator of $\Sigma$. Therefore, the two time integrals are factorized, and the loop integral is free from UV divergence. We also assume the correction from the quantum loop will not change the order of magnitude and thus negligible.

 Notice that we have set an upper bound $x_{\rm max}$ for the conformal time integrals. Nevertheless, the result would still be scale-invariant as $x_{\rm max}$ is an intrinsic cut-off determined in a scale-invariant way.
 The intrinsic cut-off is determined by the scale of particle generation, at which particles are abundantly produced. This scale is defined by the tachyonic point of the EoM (a detailed discussion of the tachyonic point will be provided in Sec.\;\ref{phenoconstraints}). We focus solely on the region where the mode function exhibits tachyonic instability, using this energy scale as the hard cut-off for the physical momentum of the modes. This choice excludes the contribution of vacuum modes and captures the main physics of particle production \cite{Niu:2022quw}. At the tachyonic point, the mass term in the EoM becomes negative, indicating that the energy density begins to be dominated by the massive spin-$2$ particles. The EoM for the helicity $\lambda=-2$ mode of the massive spin-$2$ field yields
	\begin{align}
\partial_\tau^2\Sigma^{-2}-2aH\partial_\tau\Sigma^{-2}+\left(p^2- 2\frac{p\kappa}{H\tau}+\frac{\left(m^2-2H^2\right)}{H^2\tau^2} \right)\Sigma^{-2}=0.
	\end{align}
 We perform the redefinition $\chi^{-2}\equiv a^{-1}\Sigma^{-2} $ to eliminate the first derivative term. With this redefinition, it becomes straightforward to determine the tachyonic point
\begin{align}
		(-p\tau)_{\text{tach}}=l(-k\tau)_{\text{tach}}=\kappa+\sqrt{\kappa^2-(\frac{m^2}{H^2}-2)}=\kappa+\sqrt{\kappa^2-\mu^2-\frac{1}{4}}.
	\end{align}
	where $p\equiv kl$, as previously defined. To ensure that the solution remains real, the chemical potential must satisfy the following condition: 
	\begin{align}
		\kappa^2\geq \frac{m^2}{H^2}-2.
	\end{align}
 Therefore, for the integral of $x\equiv-k\tau$ in Eq.~(\ref{GWpowerspectrum_spin2}), the upper limit is 
    \begin{align}
		x_{\rm max}=\frac{\kappa+\sqrt{\kappa^2-\mu^2-\frac{1}{4}}}{l}.
	\end{align}
 For the $l$ integral, we find that the integrand decreases rapidly for large values of $l$. Therefore, in numerical calculations, we can choose a suitable value $ l$, beyond which the contribution of integrals can be considered negligible.
 
 With the tensor power spectrum computed in Eq.~(\ref{GWpowerspectrum_spin2}), we can find the amplitude of the GWs observable today as \cite{Niu:2022quw}.
\begin{align}\label{amplitude}
    \Omega_{h}^{\text{GW}}(k)=\frac{1}{24}\Omega_{0}\mathcal{P}_h(k),
\end{align}
where $\Omega_0\simeq 8.6\times 10^{-5}$ and $\mathcal{P}_{h}(k)$ is the primordial power spectrum when the mode with comoving momentum $k$ exits the horizon.

\section{Massive spin-$s$ fields with chemical potential}\label{spinschem}
In the previous section, we studied primordial GWs generated by massive spin-$2$ fields enhanced by chemical potential and found that they are scale invariant, similar to the case of massive spin-1 fields studied in the literature \cite{Wang:2020uic,Niu:2022quw}. Naturally, we expect this result can be generalized to higher-spin bosonic fields. 

Below, we first introduce the chemical potential term for massive spin-$s$ fields with the lowest dimension satisfying requirements such as unitarity. We then derive the EoMs and solve the mode functions. Our analysis shows that some of the modes with intermediate helicities are constrained to be zero in the presence of the chemical potential, similar to what happens for massive spin-$2$ fields.

Then, we consider the slow-roll correction and backreaction. We show that these two effects, in particular the latter, make the resulting GW spectrum dependent on the spin through the scaling behavior. We then determine feasible parameter space by including a range of experimental and theoretical constraints, and numerically compute the power spectra of GWs for different spins.

\subsection{Massive spin-$s$ fields in dS}
Bosonic states of higher spins can be described by higher-rank symmetric tensors.
Similar to the spin-2 case, a spin-$ s $ field with $ n $ spatial indices can be decomposed into helicity eigenstates as 
 \begin{align}\label{helicity_decomposition}
	I_{\tau\cdots \tau i_1\cdots i_{n}}(\tau,\bm{k})=\sum_{\lambda}I^{\lambda}_{n,s}(\tau,\bm{k})\varepsilon^{\lambda}_{i_1\cdots i_n}(\hat{\bm{k}},\epsilon^{\pm})
  \end{align} 
  where $\varepsilon^{\lambda}_{i_1\cdots i_n}(\hat{\bm{k}},\epsilon^{\pm})$ is a rank-$n$ polarization tensor which satisfies\cite{Lee:2016vti}
  \begin{align}
   &\text{Totally symmeric:}\quad\varepsilon^{\lambda}_{i_1\cdots i_n}(\hat{\bm{k}},\epsilon^{\pm})=\varepsilon^{\lambda}_{(i_1\cdots i_n)}(\hat{\bm{k}},\epsilon^{\pm})\\
   &\text{Traceless:}\quad \varepsilon^{\lambda}_{i_1\cdots ii\cdots i_n}(\hat{\bm{k}},\epsilon^{\pm})=0\\
   &\text{Transverse:}\quad\hat{\bm{k}}_{i_1}\cdots \hat{\bm{k}}_{i_m}\varepsilon^{\lambda}_{i_1\cdots i_n}(\hat{\bm{k}},\epsilon^{\pm})=0,\quad \text{for } m>n-|\lambda|
  \end{align}
  where $\epsilon^\pm_i(\hat{\bm{k}})$ is the rank-1 polarization tensor satisfying:
 \begin{align}
     \hat{\bm{k}}_i\epsilon^\pm_i(\hat{\bm{k}})=0,\quad \epsilon^\pm_i(\hat{\bm{k}}) \epsilon^{\pm,\star}_i(\hat{\bm{k}})=2,
 \end{align}
 and more mathematical properties and explicit expressions of $\varepsilon^{\lambda}_{i_1\cdots i_n}(\hat{\bm{k}},\epsilon^{\pm})$ can be found in App.\;\ref{polar_tensor}.
The construction of the Lagrangian for higher-spin fields is a nontrivial task due to the increasing mismatch between the number of states and the number of field components. In comparison, writing down the EoM is easier \cite{Deser:2003gw,deser_massive_2004,Sun:2021thf}. Therefore, we will directly work with the EoMs.

For a free spin-$s$ field in (3+1)-dim dS spacetime, the EoM reads \cite{Deser:2003gw,deser_massive_2004,Sun:2021thf}:
\begin{align}\label{freespinsEoM}
	\left(\nabla^2-H^2[(2-s(s-2))]-m_s^2\right)I_{\mu_1\cdots  \mu_s}=0
\end{align}
with the Fierz-Pauli constraints
\begin{align}\label{PFconstraints}
     I^\nu_{~\nu\mu_2\cdots \mu_s}=0,\quad \nabla^\nu I_{\nu\mu_2\cdots \mu_s}=0.
\end{align}
It is then straightforward to count the number of degrees of freedom \cite{Bekaert:2006py}:
\begin{align}
	N_{\text{dof}}&=
    \binom{s+3}{s}-\binom{s+1}{s-2}-\left[\binom{s+2}{s-1}-\binom{s-4}{s-3}\right] = 2s+1.
 \end{align}

Now we include the chemical potential, which is introduced by the following term in the action:\footnote{The coefficient $1/2$ in front of the chemical potential is to match the coefficient of the kinetic term in Eq.~(\ref{kineticspins})}
\begin{align}\label{chempspins}
	S_c=-\int d^4x\epsilon^{\mu\nu\rho\sigma}\frac{\nabla_\mu\phi}{2\Lambda_{c,s}}I_{\nu\lambda_1\cdots \lambda_{s-1}}\nabla_{\rho}I_\sigma^{~\lambda_1\cdots \lambda_{s-1}}.
\end{align}
Again, this is a unique choice for chemical potential that satisfies a few conditions, including: the operator should take the form of $\kappa_\mu J^\mu$, where current should be non-conserved; the operator should be local and quadratic since we focus on the linear theory; and the operator should have the lowest mass dimension that breaks dS boosts to relax the Boltzmann suppression, but preserves all the other dS isometries.

Since this operator does not break the symmetry of spatial rotations, the decomposition Eq.~\eqref{helicity_decomposition} still applies. Also, the chemical potential Eq.~\eqref{chempspins} only has one spatial derivative acting on fields and thus will not introduce ghost modes. Furthermore, such a chemical potential is consistent with the transverse traceless constraint Eq.~\eqref{PFconstraints} since the chemical potential term manifests as $i\hat{\bm{k}}_r \epsilon^{nrm} I_{n\lambda_1 \dots \lambda_{s-1}} I_m^{~\lambda_1 \dots \lambda_{s-1}}$,\footnote{Only this term is left after expanding the covariant derivative, since other terms will vanish due to the contraction of the antisymmetric indices of $\epsilon^{nrm}$ with the symmetric indices of the field.} effectively constituting a $k$-dependent mass term. This effective mass exhibits strict helicity dependence which is the same for all the components of a fixed helicity eigenstate. Crucially, since components of different helicities decouple at the linear level, each constraint equation involves only fields of a single helicity. Furthermore, the chemical potential induces identical mass shifts for all components of a given helicity, so it enters the constraint equations only as an overall mass shift relative to the zero–chemical–potential case. Moreover, Fierz–Pauli constraint analysis reveals a mass-independent structure, reflected in its mass-free final form. Consequently, introducing the chemical potential does not change the constraint equations.

 \subsection{EoMs and Wronskian}\label{spinswroks}
After introducing the chemical potential, the EoM and constraints of the massive spin-$s$ field yield
\begin{align}\label{comspin}
&\left(\nabla^2-H^2[(2-s(s-2))]-m_s^2\right)I_{\mu_1\cdots  \mu_s}\notag\\
&-2\times\frac{\varepsilon^{\alpha\beta\gamma\sigma}\nabla_\alpha\phi}{s\Lambda_{c,s}}\left(\sum_{k=1}^sg_{\mu_k \beta}\nabla_\gamma I_{\mu_1\cdots \mu_{k-1}\sigma\mu_{k+1}\cdots \mu_s}\right)=0,\notag\\
&I^\nu_{~\nu\mu_2\cdots \mu_s}=0,\quad \nabla^\nu I_{\nu\mu_2\cdots \mu_s}=0.
\end{align}
We classify the components of the spin-$s$ field into the following three categories based on the number of spatial indices
\begin{align}
I_{\tau\cdots \tau},\quad I_{\tau\cdots \tau i_1\cdots i_n}~(n\leq s-1),\quad I_{i_1\cdots i_s},
\end{align}
which respectively correspond to the 0-component without any spatial indices, components with mixed indices, and those with purely spatial indices. In Eq.~(\ref{comspin}), we expand all covariant derivatives and express all contractions of indices as metric products. After rearranging terms, we derive the EoMs corresponding to three distinct types of components characterized by arbitrary spin and spacetime dimension. For $n=0$, the EoM is given by:
\begin{align}
	\partial^{2}_\tau I_{\tau\cdots \tau}+2aH\partial_{\tau}I_{\tau\cdots \tau}+\left[k^2+\left(m_s^2-(s+1)(s-2)H^2\right)a^2\right]I_{\tau\cdots \tau}=0.
\end{align}
Notably, the EoM of $I_{\tau\cdots \tau}$ does not contain the chemical potential term, implying that the evolution of $I_{\tau\cdots \tau}$ remains unaffected by the chemical potential. Consequently, the gravitational wave signal generated by the 0-component of the spin-$s$ field remains unaffected by the chemical potential. Hence, the contribution of this component is disregarded in the phenomenological calculation. For components with spatial indices satisfying $1\leq n'\leq s-1$,  
\begin{align}
	0&=\partial_{\tau}^2I_{\tau\cdots \tau i_1\cdots i_{n'}}+k^2I_{\tau\cdots \tau i_1\cdots i_{n'}}+m_s^2a^2I_{\tau\cdots \tau i_1\cdots i_{n'}}\notag\\
	&-\left[n'(3-n')+(s+1)(s-2)\right]a^2H^2I_{\tau\cdots \tau i_1\cdots i_{n'}}-(2n'-2)aH\partial_\tau I_{\tau\cdots \tau i_1\cdots i_{n'}}\notag\\
	&+2aH\sum_{k=1}^{n'}ik_{i_k} I_{\tau\cdots  \tau i_1\cdots i_{k-1}i_{k+1}\cdots i_{n'}}-a^2H^2\sum_{k=1}^{n'}\sum_{p\neq k}^{n'}\delta_{i_pi_k}I_{\tau\cdots  \tau i_1\cdots i_{k-1}i_{k+1}\cdots i_{p-1}i_{p+1}\cdots i_{n'}}\notag\\
	&+iak\tilde{\kappa}^{(s)}\sum_{k=1}^{n'}\epsilon^{i_krm}\hat{\bm{k}}_rI_{\tau\cdots \tau i_1\cdots m\cdots i_{n'}}-H\tilde{\kappa}^{(s)}\sum_{k=1}^{n'}\sum_{p\neq k}^{n'}\epsilon^{i_ki_pm}I_{\tau\cdots  \tau i_1\cdots m\cdots \tau\cdots  i_{n'}},
\end{align}
wherein $\hat{\bm{k}}_r$ denotes the $r$-th component of the unit vector in the direction of the comoving momentum $k$. Furthermore, for components with spatial indices satisfying $n=s$, the EoM yields
\begin{align}
	0&=\partial_{\tau}^2I_{i_1\cdots i_s}-aH(2s-2)\partial_\tau I_{i_1\cdots i_s}+\left[k^2+\left(m_s^2-(2s-2)H^2\right)a^2\right]I_{i_1\cdots i_s}\notag\\
    &+2aH\sum^{s}_{k=1}ik_{i_k}I_{i_1\cdots \tau\cdots i_s}-a^2H^2\sum_{k=1}^s\sum_{p\neq k}^s\delta_{i_ki_p}I_{i_1\cdots \tau\cdots \tau\cdots i_s}\notag\\
	&+iak\tilde{\kappa}^{(s)}\sum^s_{k=1}\epsilon^{i_krm}\hat{\bm{k}}_rI_{i_1\cdots m\cdots i_s}-H\tilde{\kappa}^{(s)}\sum^s_{k=1}\sum^s_{p\neq k}\epsilon^{i_ki_pm}I_{i_1\cdots m\cdots \tau\cdots i_s}.
\end{align}
We find that the above two cases can be combined and written in a unified form ($1\leq n\leq s$)
\begin{align}\label{EoMnspatial}
	0&=\partial_{\tau}^2I_{\tau\cdots \tau i_1\cdots i_n}+k^2I_{\tau\cdots \tau i_1\cdots i_n}+m_s^2a^2I_{\tau\cdots \tau i_1\cdots i_n}\notag\\
	&-\left[n(3-n)+(s+1)(s-2)\right]a^2H^2I_{\tau\cdots \tau i_1\cdots i_n}-(2n-2)aH\partial_\tau I_{\tau\cdots \tau i_1\cdots i_n}\notag\\
	&+2aH\sum_{k=1}^{n}ik_{i_k} I_{\tau\cdots \tau i_1\cdots i_{k-1}i_{k+1}\cdots i_n}-a^2H^2\sum_{k=1}^{n}\sum_{p\neq k}^{n}\delta_{i_pi_k}I_{\tau\cdots \tau i_1\cdots i_{k-1}i_{k+1}\cdots i_{p-1}i_{p+1}\cdots i_n}\notag\\
	&+iak\tilde{\kappa}^{(s)}\sum_{k=1}^{n}\epsilon^{i_krm}\hat{\bm{k}}_rI_{\tau\cdots \tau i_1\cdots m\cdots i_n}-H\tilde{\kappa}^{(s)}\sum_{k=1}^{n}\sum_{p\neq k}^{n}\epsilon^{i_ki_pm}I_{\tau\cdots  \tau i_1\cdots m\cdots \tau\cdots i_n}.
\end{align}
It follows from the above equation that for $I_{\tau\cdots \tau i_1\cdots i_n}$, the EoM includes terms with spatial indices satisfying $m<n$. Consequently, the EoMs governing the spin-$s$ field components form a system of coupled equations. According to Eq.~(\ref{helicity_decomposition}), one can perform the helicity decomposition and represent these components in terms of polarization tensors: 
\begin{align}
	I_{\tau\cdots \tau i_1\cdots i_{n}}(\tau,\bm{k})=\sum_{\lambda}I^{\lambda}_{n,s}(\tau,\bm{k})\varepsilon^{\lambda}_{i_1\cdots i_n}(\hat{\bm{k}},\epsilon^{\pm}),\quad
	\varepsilon^{\lambda}_{ii\cdots i_n}=0,\quad \bm{k}_{i_1}\cdots \bm{k}_{i_m}\varepsilon^{\lambda}_{i_1\cdots i_n}=0.
  \end{align} 
We contract Eq.~(\ref{EoMnspatial}) with the polarization tensors defined in Eq.~(\ref{polarization_tensor}), and we find that for the component with $n$ spatial indices, the EoMs for its highest helicity states with $|\lambda|=n$ decouples from other components, which can generally be expressed as
\begin{align}\label{EoMhelicn}
	&\partial_\tau^2I_{n}^{\pm n}-(2n-2)aH\partial_\tau I_{n}^{\pm n}+\left[k^2+\left[m_s^2-\left(n(3-n)+(s+1)(s-2)\right)H^2\right]a^2\right]I_{n}^{\pm n}\notag\\
&\pm nak\tilde{\kappa}^{(s)}I_{n}^{\pm n}=0,
\end{align}
and the EoM of $I_{n}^{\pm (n-1)}$ also has a relatively simple form, involving only coupling with $I_{n-1}^{\pm (n-1)}$ 
\begin{align}\label{EoMhelin-1}
	&\partial_\tau^2I_n^{\pm(n-1)}-(2n-2)aH\partial_\tau I_n^{\pm(n-1)}+\left[k^2+\left[m_s^2-\left(n(3-n)+(s+1)(s-2)\right)H^2\right]a^2\right]I_n^{\pm(n-1)}\notag\\
&+aH\frac{2n}{2n-1}ikI_{n-1}^{\pm(n-1)}\pm(n-1)ak\tilde{\kappa}^{(s)}I_n^{\pm(n-1)}=0.
\end{align}
Having obtained the EoMs, we need to consider the constraints satisfied by the massive spin-$s$ field
\begin{align}
	\nabla^\mu I_{\mu\mu_1\cdots \mu_{s-1}}=0,\quad I^{\mu}_{~\mu\mu_1\cdots \mu_{s-2}}=0.
\end{align}
We expand the constraints into components $I_{\tau\cdots  \tau i_1\cdots i_n}$ and the transverse condition becomes
\begin{align}
ik\hat{\bm{k}}_mI_{\tau\cdots \tau mi_1\cdots i_{n}}=\partial_\tau I_{\tau\cdots \tau i_1\cdots i_{n}}+2aHI_{\tau\cdots \tau i_1\cdots i_{n}}.
\end{align}
We then do the helicity decomposition and write the components with polarization tensors explicitly
\begin{align}
ik\hat{\bm{k}}_m\sum_{|\lambda|}^{n+1}I^{\lambda}_{n+1} \epsilon^{\lambda}_{(i_1\cdots m\cdots i_{|\lambda|-1}}f_{i_{|\lambda|}\cdots i_{n})}=\sum_{|\lambda^\prime|}^{n}\left(\partial_\tau I^{\lambda^\prime}_{n}+2aHI^{\lambda^\prime}_{n}\right)\epsilon^{\lambda^\prime}_{(i_1\cdots i_{|\lambda^\prime|}}f_{i_{|\lambda^\prime|+1}\cdots i_{n})}.
\end{align}
It is straightforward to obtain the result of $|\lambda|=n$:
\begin{align}
	I_{n+1}^{\pm n}=-\frac{i}{k}\frac{n+1}{2n+1}\left(\partial_{\tau}I_{n}^{\pm n}+2aHI_{n}^{\pm n}\right)
\end{align}
for $n\leq s-1$.
We notice that terms that do not vanish after contracting the polarization tensors are those with equal number of $\epsilon^{\pm}_i$. For $|\lambda|\leq n$, the polarization tensor is  
\begin{align}
	\varepsilon^{\lambda}_{i_1\cdots i_n}(\bm{k},\epsilon^\pm)=\frac{1}{(2|\lambda|-1)!!}\sum_{m=0}^{n-|\lambda|}B_m\frac{1}{n}\epsilon^\lambda_{(i_1\cdots i_{|\lambda|}}\bm{k}_{i_{|\lambda|+1}}\cdots \bm{k}_{i_{|\lambda|+m}}\delta_{i_{|\lambda|+m+1}\cdots i_n)}.
  \end{align}
 Next, we focus on the traceless condition, which can be derived similarly and it becomes
  \begin{align}
	&I_{\tau\cdots \tau mmi_1\cdots  i_n}=I_{\tau\cdots \tau i_1\cdots  i_n}\notag\\
	\Rightarrow&\sum_{|\lambda|}^{n+2}I^{\lambda}_{n+2}\epsilon^{\lambda}_{(i_1\cdots m\cdots m\cdots i_{|\lambda|-2}}f_{i_{|\lambda|-1}\cdots i_{n})}=\sum_{|\lambda^\prime|}^{n}I^{\lambda^\prime}_{n}\epsilon^{\lambda^\prime}_{(i_1\cdots i_{|\lambda^\prime|}}f_{i_{|\lambda^\prime|+1}\cdots i_{n})},\quad n\leq s-2.
  \end{align}
  We now compute the mode functions associated with the maximal helicity states $\lambda = \pm s$. Our emphasis on these modes is motivated by several factors. First, as demonstrated in prior analyses, EoMs and constraints for $\lambda = \pm s$ helicity modes decouple from lower-helicity modes, thereby simplifying the computational process. Phenomenologically, as evidenced by the dispersion relation (Eq.~(\ref{dispersion_rel})) and supported by derivations in Eqs.~(\ref{EoMnspatial}), (\ref{EoMhelicn}), and (\ref{EoMhelin-1}), the effective chemical potential for a helicity eigenstate exhibits a linear dependence on the helicity. Given the exponential dependence of the wavefunction on the chemical potential parameter, the $\lambda = -s$ mode (for $\tilde{\kappa}^{(s)} > 0$) experiences significant enhancement relative to lower-helicity counterparts. Consequently, this mode produces the most pronounced phenomenological signal. Furthermore, as detailed in Sec.\;\ref{spinschemforbid}, modes with $|\lambda| < s$ exhibit a so-called "chemical-potential discontinuity" mechanism. Consequently, we prioritize the amplified $\lambda = -s$ mode due to its maximal enhancement and empirical detectability. The EoMs for the highest helicity modes are:
\begin{align}\label{EoMhs}
	&\partial_\tau^2I_{s}^{\pm(s)}-(2s-2)aH\partial_\tau I_{s}^{\pm(s)}+\left[k^2+\left[m_s^2-2\left(s-1\right)H^2\right]a^2\right]I_{s}^{\pm(s)}\pm sak\tilde{\kappa}^{(s)}I_{s}^{\pm(s)}=0.
\end{align}
The mode function can be directly solved, and the solution can be expressed as the Whittaker function: 
\begin{align}
	I^{(\pm s)}_s(\tau,k)=\mathcal{N}_{\pm s}\tau^{1-s}W_{\pm i\kappa_s,i\mu_s}(2ik\tau).
\end{align}
Here we adopt the symplectic inner product method proposed in \cite{Lee:2016vti,Tong:2022cdz} solving mode function for massive spin-$2$ field and generalize it to arbitrary spin cases.
First, we can construct a current in terms of two solutions $A_{\mu_1\cdots \mu_s}$ and $B_{\mu_1\cdots \mu_s}$ of the EoM of the massive spin-$s$ field Eq.~(\ref{comspin}). The current becomes
\begin{align}
	K^\rho(A,B)&=A^\star_{\mu_1\cdots \mu_s}\nabla^\rho B^{\mu_1\cdots \mu_s}-B_{\mu_1\cdots \mu_s}\nabla^\rho A^{\star\mu_1\cdots \mu_s}\notag\\
	&+\frac{1}{2}\frac{2\phi}{s\Lambda_{c,s}}\sum_{k=1}^s\varepsilon^{\alpha\nu\rho\sigma}\nabla_\alpha\left(g_{\mu_k\sigma}A^\star_{\mu_1\cdots \mu_{k-1}\mu_{k+1}\cdots \mu_s\nu}B^{\mu_1\cdots \mu_k\cdots \mu_s}-g_{\mu_k\nu}B^{\mu_1\cdots \mu_k\cdots \mu_s}A^\star_{\mu_1\cdots \mu_{k-1}\mu_{k+1}\cdots \mu_s\sigma}\right)\notag\\
	&=A^\star_{\mu_1\cdots \mu_s}\nabla^\rho B^{\mu_1\cdots \mu_s}-B_{\mu_1\cdots \mu_s}\nabla^\rho A^{\star\mu_1\cdots \mu_s}+\frac{2\phi}{\Lambda_{c,s}}\varepsilon^{\alpha\nu\rho\sigma}\nabla_\alpha\left(A^\star_{\mu_1\cdots \mu_{s-1}\nu}B^{\mu_1\cdots \mu_{s-1}}_\sigma\right).
\end{align}
 We can prove that the current is conserved on-shell, which yields
\begin{align}
	\nabla_\rho K^\rho=0
\end{align}
utilizing the EoM given in Eq.~(\ref{comspin}). Then, we define the inner product for the spin-$s$ field by integrating the current over a spatial hypersurface $\Sigma(\tau)$
\begin{align}\label{wronskian}
	&\langle A_{\mu_1\cdots \mu_s},B_{\mu_1\cdots \mu_s}\rangle_\tau \notag\\
	&\equiv \int_{\Sigma(\tau)}\sqrt{\hat{g}}d^3x_\rho K^\rho(A,B)\notag\\
    &=\int d\Sigma ~n_\rho \sqrt{\hat{g}}\left[A^\star_{\mu_1\cdots \mu_s}\nabla^\rho B^{\mu_1\cdots \mu_s}-B_{\mu_1\cdots \mu_s}\nabla^\rho A^{\star\mu_1\cdots \mu_s}+\frac{2\phi}{\Lambda_{c,s}}\varepsilon^{\alpha\nu\rho\sigma}\nabla_\alpha\left(A^\star_{\mu_1\cdots \mu_{s-1}\nu}B^{\mu_1\cdots \mu_{s-1}}_\sigma\right)\right]\notag\\
	&=a^{-2(s-1)}\eta^{\mu_1\nu_1}\cdots \eta^{\mu_s\nu_s}\int d^3x\left[A^\star_{\mu_1\cdots \mu_s}\nabla_\tau B_{\nu_1\cdots \nu_s}-B_{\nu_1\cdots \nu_s}\nabla_\tau A^{\star}_{\mu_1\cdots \mu_s}\right]\notag\\
	&+2a^{-2(s-1)}a\frac{\phi}{\Lambda_{c,s}}\eta^{\mu_1\nu_1}\cdots \eta^{\mu_{s-1}\nu_{s-1}}\int d^3x\epsilon^{anm}\nabla_a\left(A^\star_{\mu_1\cdots \mu_{s-1}n}B_{\nu_1\cdots \nu_{s-1}m}\right),
\end{align}
where $\sqrt{\hat{g}}$ is the induced metric. 
Deriving the Wronskian for the modes of the highest helicity is straightforward: perform the Fourier transformation and replace all indices of the fields in Eq.~(\ref{wronskian}) with spatial indices.
\begin{align}\label{wronsforhs}
	&\langle A^{(\pm s)}_{i_1\cdots i_s}(\tau,\bm{k})e^{i\bm{k}\cdot\bm{x}},B^{(\pm s)}_{i_1\cdots i_s}(\tau,\bm{k^\prime})e^{i\bm{k^\prime}\cdot\bm{x}}\rangle \notag\\
	&=a^{-2(s-1)}\int d^3x\left[A^\star_{i_1\cdots i_s}(\tau,\bm{k})\partial_\tau B_{i_1\cdots i_s}(\tau,\bm{k^\prime})-B_{i_1\cdots i_s}(\tau,\bm{k^\prime})\partial_\tau A^{\star}_{i_1\cdots i_s}(\tau,\bm{k})\right]e^{i(\bm{k^\prime}-\bm{k})\cdot \bm{x}}\notag\\
	&+2a^{-2(s-1)}a\frac{\phi}{\Lambda_{c,s}}\int d^3x\epsilon^{anm}\left(\nabla_a\left(A^\star_{i_1\cdots i_{s-1}n}(\tau,\bm{k})e^{-i\bm{k}\cdot\bm{x}}\right)B_{i_1\cdots i_{s-1}m}(\tau,\bm{k^\prime})e^{i\bm{k^\prime}\cdot\bm{x}}\right.\notag\\
	&\left.+\nabla_a\left(B_{i_1\cdots i_{s-1}m}(\tau,\bm{k^\prime})e^{i\bm{k^\prime}\cdot\bm{x}}\right)A^\star_{i_1\cdots i_{s-1}n}(\tau,\bm{k})e^{i\bm{k}\cdot\bm{x}}\right).
\end{align}
The second term in Eq.~(\ref{wronsforhs}) vanishes due to the cancellation and the contractions between symmetric rank-s tensors and anti-symmetric tensor $\epsilon^{anm}$.
After simplification, the Wronskian or the normalization condition of $I^{(\pm s)}_s$ becomes
\begin{align}\label{Wronskian_spins}
	&\langle A^{(\pm s)}_{i_1\cdots i_s}(\tau,\bm{k})e^{i\bm{k}\cdot\bm{x}},B^{(\pm s)}_{i_1\cdots i_s}(\tau,\bm{k^\prime})e^{i\bm{k^\prime}\cdot\bm{x}}\rangle=2i\delta^{(3)}(\bm{k}-\bm{k^\prime})\notag\\
	\Rightarrow&-ia^{-2(s-1)} 2^s\left(I^{\star(\pm s)}_s\partial_\tau I^{(\pm s)}_s-I^{(\pm s)}_s\partial_\tau I^{\star(\pm s)}_s\right)=2.
\end{align}
Since the Wronskian is independent of conformal time, we calculate the above equation at $\tau\to 0$ and the normalization constant of the highest helicity modes becomes
\begin{align}
	|&\mathcal{N}_{\pm s}|^2=\frac{e^{\mp\pi\kappa_s}}{2^sH^{2(s-1)}k}(-1)^{2(s-1)}\notag\\
\Rightarrow&\mathcal{N}_{\pm s}=\frac{e^{\mp\frac{\pi\kappa_s}{2}}}{2^{\frac{s}{2}}H^{(s-1)}\sqrt{k}}(-1)^{(s-1)}\quad\text{up to a phase}.
\end{align}
Finally, we obtain the mode functions of helicity-$\pm s$ fields
\begin{align}  
    I^{(\pm s)}_s(\tau,k)&=\frac{e^{\mp\frac{\pi\kappa_s}{2}}}{2^{\frac{s}{2}}H^{(s-1)}\sqrt{k}}(-\tau)^{1-s}W_{\pm i\kappa_s,i\mu_s}(2ik\tau).  
\end{align}  
Strikingly, the solution reveals a hallmark structural asymmetry: across spin-$s$ fields, the highest helicity modes exhibit marked suppression ($\lambda = +s$) or exponential enhancement ($\lambda = -s$) for positive $\kappa_s$, a definitive signature of parity-violating.  
\subsection{The chemical-potential discontinuity}
\label{spinschemforbid}

As established in Section~\ref{spin2wrosk}, a critical phenomenon emerges for spin-2 fields: helicity-$\lambda = \pm1$ modes vanish (Eq.~(\ref{chemforbiddenspin2})), reducing the degrees of freedom from $5$ to $3$ under chemical potential. We term this mechanism the \textit{chemical-potential discontinuity}, as our analysis reveals its universal manifestation in massive higher-spin systems. In the following subsection, we will delve deeper into the exploration of this phenomenon.  

We first observe dynamical coupling in EoMs and constraints between: (i) submaximal helicity $\lambda = \pm(n-1)$ modes with $n$ spatial indices, and (ii) the highest helicity modes possessing $n-1$ spatial indices. Accordingly, we explicitly construct EoMs and constraints for the coupled systems $I_{n}^{\pm(n-1)}$ and $I_{n-1}^{\pm(n-1)}$ ($n \leq s$):  
\begin{equation}\label{chemfrEoM}
\left\{\begin{aligned}&\partial_\tau^2I_n^{\pm(n-1)}-(2n-2)aH\partial_\tau I_n^{\pm(n-1)}+\left[k^2+\left[m_s^2-\left(n(3-n)+(s+1)(s-2)\right)H^2\right]a^2\right]I_n^{\pm(n-1)}\\
&+aH\frac{2n}{2n-1}ikI_{n-1}^{\pm(n-1)}\pm(n-1)ak\tilde{\kappa}^{(s)}I_n^{\pm(n-1)}=0\\
\\
&\partial_\tau^2I_{n-1}^{\pm(n-1)}-(2n-4)aH\partial_\tau I_{n-1}^{\pm(n-1)}+\left[k^2+\left[m_s^2-\left((n-1)(4-n)+(s+1)(s-2)\right)H^2\right]a^2\right]I_{n-1}^{\pm(n-1)}\\
&\pm(n-1)ak\tilde{\kappa}^{(s)}I_{n-1}^{\pm(n-1)}=0\\
\\
&I_{n}^{\pm(n-1)}=-\frac{i}{k}\frac{n}{2n-1}\left(\partial_{\tau}I_{n-1}^{\pm(n-1)}+2aHI_{n-1}^{\pm(n-1)}\right)
\end{aligned}\right.
\end{equation}
We first take the time derivative of the third equation
\begin{align}\label{Innm11der}
	\partial_\tau I_{n}^{\pm(n-1)}=-\frac{i}{k}\frac{n}{2n-1}\left(\partial^2_\tau I_{n-1}^{\pm(n-1)}+2a^2H^2\partial_\tau I_{n-1}^{\pm(n-1)}+2aH\partial_\tau I_{n-1}^{\pm(n-1)}\right).
\end{align}
Then we take a further time derivative of the third equation of Eq.~(\ref{chemfrEoM})
\begin{align}\label{Innm12der}
	\partial_\tau^2I_{n}^{\pm(n-1)}=-\frac{i}{k}\frac{n}{2n-1}\left(\partial^3_\tau I_{n-1}^{\pm(n-1)}+4a^2H^2\partial_\tau I_{n-1}^{\pm(n-1)}+4a^3H^3I_{n-1}^{\pm(n-1)}+2aH\partial^2_\tau I_{n-1}^{\pm(n-1)}\right).
\end{align}
From the second equation of Eq.~(\ref{chemfrEoM}), the second derivative of $I_{n-1}^{\pm(n-1)}$ can be expressed as
\begin{align}
	\partial_\tau^2I_{n-1}^{\pm(n-1)}&=(2n-4)aH\partial_\tau I_{n-1}^{\pm(n-1)}-\left[k^2+\left[m_s^2-\left((n-1)(4-n)+(s+1)(s-2)\right)H^2\right]a^2\right]I_{n-1}^{\pm(n-1)}\notag\\
	&\mp(n-1)ak\tilde{\kappa}^{(s)}I_{n-1}^{\pm(n-1)}.
\end{align}
Then we substitute the above equation into Eq.~(\ref{Innm12der}), eliminating the third-order derivatives of $I_{n-1}^{\pm(n-1)}$ and the result becomes
\begin{align}\label{partial2Inn-1}
	&\partial_\tau^2I_{n}^{\pm(n-1)}\times \frac{k}{-i}\frac{2n-1}{n}\notag\\
	&=\mp a^2Hk\tilde{\kappa}^{(s)}(n-1)I_{n-1}^{\pm(n-1)}\mp ak\tilde{\kappa}^{(s)}(n-1)\partial_\tau I_{n-1}^{\pm(n-1)}\notag\\
	&+\left(-n^2-4+7n+(s+1)(s-2)\right)a^2H^2\partial_\tau I_{n-1}^{\pm(n-1)}+(2n-2)aH\partial^2_\tau I_{n-1}^{\pm(n-1)}\notag\\
	&+\left(-n^2-2+5n+(s+1)(s-2)\right)a^3H^3I_{n-1}^{\pm(n-1)}-2m_s^2a^3HI_{n-1}^{\pm(n-1)}-m_s^2a^2\partial_\tau I_{n-1}^{\pm(n-1)}.
\end{align}
From the first equation of Eq.~(\ref{chemfrEoM}), the second derivative of $I_{n}^{\pm(n-1)}$ can also be expressed as
\begin{align}\label{EoMInn-1}
	&\partial_\tau^2I_{n}^{\pm(n-1)}\times\frac{k}{-i}\frac{2n-1}{n}\notag\\
    &=aH(2n-2)\left(\partial^2_\tau I_{n-1}^{\pm(n-1)}+2a^2H^2\partial_\tau I_{n-1}^{\pm(n-1)}+2aH\partial_\tau I_{n-1}^{\pm(n-1)}\right)\notag\\
	&-\left[k^2+\left[m_s^2-\left(n(3-n)+(s+1)(s-2)\right)H^2\right]a^2\right]\left(\partial_{\tau}I_{n-1}^{\pm(n-1)}+2aHI_{n-1}^{\pm(n-1)}\right)I_{n-1}^{\pm(n-1)}\notag\\
	&-aH\frac{2n}{2n-1}ik\frac{k}{-i}\frac{2n-1}{n}I_{n-1}^{\pm(n-1)}\mp (n-1)ak\tilde{\kappa}^{(s)}\left(\partial_{\tau}I_{n-1}^{\pm(n-1)}+2aHI_{n-1}^{\pm(n-1)}\right)\notag\\
	&=\mp(n-1) ak\tilde{\kappa}^{(s)}\partial_\tau I_{n-1}^{\pm(n-1)}\mp(n-1)k\tilde{\kappa}^{(s)}2a^2HI_{n-1}^{\pm(n-1)}\notag\\
	&+a^2H^2\partial_\tau I_{n-1}^{\pm(n-1)}\left(7n-4-n^2+(s+1)(s-2)\right)\notag\\
	&+2a^3H^3I_{n-1}^{\pm(n-1)}\left(5n-2-n^2+(s+1)(s-2)\right)\notag\\
	&+aH(2n-2)\partial_{\tau}^2I_{n-1}^{\pm(n-1)}\notag\\
	&-2a^3Hm_s^2I_{n-1}^{\pm(n-1)}-m_s^2a^2\partial_\tau I_{n-1}^{\pm(n-1)},
\end{align}
\noindent Utilizing Eq.~(\ref{chemfrEoM}), Eq.~(\ref{Innm11der}), and Eq.~(\ref{Innm12der}), we express all $I_{n}^{\pm(n-1)}$ on the right-hand side in terms of $I_{n-1}^{\pm(n-1)}$. Comparing Eq.~(\ref{EoMInn-1}) with Eq.~(\ref{partial2Inn-1}) yields  
\begin{align}\label{chemforspins}  
    \mp(n-1)k\tilde{\kappa}^{(s)}2a^2HI_{n-1}^{\pm(n-1)}=0.  
\end{align}  
For $n\neq 1$ and $\tilde{\kappa}^{(s)}\neq 0$, this forces $I_{n-1}^{\pm(n-1)} = 0$ . The constraints consequently nullify the sub-highest helicity modes $I_{n}^{\pm(n-1)}$ with $n$ spatial indices.  

Focusing on spin-2 (with higher-spin cases analogous), Eqs.~(\ref{chemforbiddenspin2}) and~(\ref{chemforspins}) show that, when the comoving momentum $k$, the chemical potential $\tilde{\kappa}$, and the Hubble parameter $H$ are simultaneously non-zero, vector modes must vanish to maintain the self-consistency of the EoMs and their associated constraints. Conversely, if any parameter vanishes, vector modes propagate normally, restoring the five degrees of freedom. Thus the disappearance of degrees of freedom hinges on three factors:   

We begin by analyzing the effects of the comoving momentum $k$ and the chemical potential $\tilde\kappa$. The chemical potential explicitly breaks dS boosts, preventing rest-frame transitions, while non-zero $k$ reduces the little group from $\mathrm{SO(3)}$ to $\mathrm{SO(2)}$, with helicity decomposition becoming mandatory. Consequently, non-rest-frame dispersion relations diverge across helicities, manifesting mode splitting. Eq.~(\ref{spin2_tensormode}), Eq.~(\ref{spin2_vectormode1}), and Eq.~(\ref{spin2_scalarmode}) confirm this through helicity-mass dependencies: $ 2k\tilde{\kappa} $ (tensor), $ k\tilde{\kappa} $ (vector), and $ 0 $ (scalar). Thus, in the presence of a chemical potential, the physics degrees of freedom in rest and non-rest frames can differ—including their degrees of freedom for dynamical evolution. As expected, in the rest frame ($ k = 0 $), Eq.~(\ref{chemforbiddenspin2}) vanishes, restoring all five degrees of freedom with the same dispersion—consistent with unbroken $ \mathrm{SO(3)} $ symmetry. Alternatively, setting $\tilde{\kappa}$ to zero restores dS boosts, allowing rest-frame transitions, wherein the propagating degrees of freedom naturally return to five.

We next analyze the role of the Hubble parameter $ H $. Introducing the chemical potential in flat spacetime ($ H = 0 $, $ \tilde{\kappa} \neq 0 $, $ k \neq 0 $), the vector mode equations decouple, yielding identical dynamics for $ \Sigma^{\pm1}_1 $ and $ \Sigma^{\pm1}_2 $. Substituting the constraint into Eq.~(\ref{spin2_vectormode1}) reproduces Eq.~(\ref{spin2_vectormode2}), confirming self-consistency and the non-vanishing of $ \Sigma^{\pm1}_2 $, as anticipated. This confirms that the phenomenon of the disappearance of degrees of freedom is related to the curvature of spacetime, which may be a characteristic of dS spacetime.

As proposed in \cite{Tong:2022cdz}, this phenomenon may originate from linear-theory approximations, inducing a van Dam-Veltman-Zakharov (vDVZ)-type discontinuity \cite{Zakharov:1970cc,vanDam:1970vg,maggiore2007gravitational} characterized by non-smooth $\kappa \to 0$ limits,  in which the number of degrees of freedom undergoes a sudden change. The vDVZ discontinuity arises when graviton massless limits fail to recover general relativity predictions. Similarly, for spin-2 fields with chemical potentials, helicity-mode decoupling at $\kappa \to 0$ reveals a structural discontinuity in the theory's dynamics, marked by a discontinuous shift in degrees of freedom.  

If we draw parallels to the resolution method of the vDVZ discontinuity, this issue of chemical-potential discontinuity might be addressed in a non-linear theory by employing the Vainshtein mechanism \cite{Vainshtein:1972sx,Deffayet:2001uk,Babichev:2013usa}, which involves resumming higher-order terms to restore continuity in the theory. Such nonlinear corrections may regularize degree-of-freedom transitions, enabling consistent descriptions of massive spin-2 fields in dS spacetime with chemical potential.

A complementary resolution emerges within Effective Field Theory (EFT): systematically incorporating all admissible higher-order Lagrangian terms. These terms will not only modify the EoMs, but will also alter the form of the algebraic constraint, specifically Eq.~(\ref{chemforspins}).

\subsection{Particle production}\label{spinsparpro}
In this subsection, we attempt to derive the average particle number density of massive spin-$s$ fields. The mode function of the helicity-$\pm s$ field yields
\begin{align}
	I^{(\pm s)}_s&=\frac{e^{\mp\frac{\pi\kappa_s}{2}}}{2^{\frac{s}{2}}H^{(s-1)}\sqrt{k}}(-1)^{(s-1)}\tau^{1-s}W_{\pm i\kappa_s,i\mu_s}(2ik\tau).
\end{align}
The IR limit of the mode functions can be written as
\begin{align}
	I^{(\pm s)}_s\stackrel{\tau\to0}{\longrightarrow}\alpha_{\pm s}\frac{1}{2^\frac{s}{2}H^{(s-1)}\sqrt{\mu_s}}(-\tau)^{\frac{3}{2}-s+i\mu_s}+\beta_{\pm s}\frac{1}{2^\frac{s}{2}H^{(s-1)}\sqrt{\mu_s}}(-\tau)^{\frac{3}{2}-s-i\mu_s},
\end{align}
where $\alpha_{\pm s}$ and $\beta_{\pm s}$ are the Bogoliubov coefficients. To obtain the Bogoliubov coefficients we take the late-time limit and the mode functions yield
\begin{align}
	&I^{(\pm s)}_s\stackrel{\tau\to0}{\longrightarrow}\notag\\
	&\frac{e^{\mp\frac{\pi\kappa_s}{2}}}{2^{\frac{s}{2}}H^{(s-1)}\sqrt{\mu_s}}\frac{(1-i)}{\sqrt{2}}\left[\frac{\sqrt{2\mu_s}(2k)^{i\mu_s}e^{\frac{\pi\mu_s}{2}}\Gamma\left[-2i\mu_s\right]}{\Gamma[\frac{1}{2}\mp i\kappa_s-i\mu_s]}(-\tau)^{\frac{3}{2}-s+i\mu_s}+\frac{\sqrt{2\mu_s}(2k)^{-i\mu_s}e^{-\frac{\pi\mu_s}{2}}\Gamma[2i\mu_s]}{\Gamma[\frac{1}{2}\mp i\kappa_s+i\mu_s]}(-\tau)^{\frac{3}{2}-s-i\mu_s}\right].
\end{align}
Do the matching and we can read the Bogoliubov coefficient
\begin{align}
	\beta_{\pm s}=(1-i)\frac{\sqrt{\mu_s}(2k)^{-i\mu_s}e^{\frac{\pi(\mp\kappa_s-\mu_s)}{2}}\Gamma[2i\mu_s]}{\Gamma[\frac{1}{2}\mp i\kappa_s+i\mu_s]}.
\end{align}
We finally obtain the average particle number density
\begin{align}\label{num_density}
	\langle n_{\pm s}(\bm{k})\rangle^\prime =|\beta_{\pm s}|^2=\frac{1+e^{2\pi(\mp\kappa_s+\mu_s)}}{e^{4\pi\mu_s}-1}.
\end{align}
However, since we take the late-time limit, the history of the particle production is still unknown. To obtain more accurate information about particle production, including the evolution of particle number with time, the energy scale of particle production, and the width of particle production, one needs to adopt the method of Stokes line \cite{Barry:1989zz,Enomoto:2020xlf,Hashiba:2021npn,Sou:2021juh}.

\section{Gravitational wave phenomenology of massive spin-$s$ fields}\label{GWspins_phe}
\subsection{Gravitational waves}\label{GWspins}
For computing the GWs generated by the massive spin-$s$ field enhanced by a chemical potential, similar to the calculations for GWs produced by the massive spin-$2$ field in Sec.\;\ref{GWspin2}, we first extract the leading term from the energy-momentum tensor calculated in App.\;\ref{AppendixspinsEMT} which yields
\begin{align}
	\tilde{T}_{ij}^{\rm s}= -sa^{-2s}(\tau)\partial_\tau I_{ii_1\cdots i_{s-1}}\partial_\tau I_{ji_1\cdots i_{s-1}}.
\end{align}
Then we use the projection operator $\Lambda_{ij,kl}$ to project the energy-momentum into its transverse and traceless part as $\tilde{T}^{sTT}_{ij}=\Lambda_{ij,kl}\tilde{T}^{s}_{kl}$. By convolving the resulting transverse and traceless energy-momentum tensor with the Green's function, we obtain the tensor perturbations 
\begin{align}
    h_{ij}^{TT}(\tau_0,\bm{k})=\int d{\tau}^\prime G(\tau_0,\tau^\prime,k) 16\pi G\Lambda_{ij,kl}(\hat{\bm{k}})\tilde{T}^{s}_{kl}(\tau^{\prime},\bm{k}).
\end{align}

In order to obtain a general expression for the GWs produced by massive fields with arbitrary spin, we need to calculate the angular dependence in detail. Similar to Eq.~(\ref{angular_spin2}), the angular term yields (we only care about the highest helicity modes, here we take $\lambda=-s$ considering $\kappa_s>0$)
\begin{align}\label{angularspins}
&\Lambda_{kl,mn}(\hat{\bm{k}})\epsilon^{-s}_{ki_1\cdots i_{s-1}}(\hat{\bm{p}})\epsilon^{-s}_{li_1\cdots i_{s-1}}(\widehat{\bm{k}-\bm{p}})\epsilon^{-s}_{mi^\prime_1\cdots i^\prime_{s-1}}(-\hat{\bm{p}})\epsilon^{-s}_{ni^\prime_1\cdots i^\prime_{s-1}}(\widehat{\bm{p}-\bm{k}})\notag\\
	&=\frac{1}{4}\sum_{\lambda=\pm2}\epsilon^\lambda_{kl}(-\bm{\hat{k}})\epsilon^\lambda_{mn}(\bm{\hat{k}})\epsilon^{-s}_{ki_1\cdots i_{s-1}}(\hat{\bm{p}})\epsilon^{-s}_{li_1\cdots i_{s-1}}(\widehat{\bm{k}-\bm{p}})\epsilon^{-s}_{mi^\prime_1\cdots i^\prime_{s-1}}(-\hat{\bm{p}})\epsilon^{-s}_{ni^\prime_1\cdots i^\prime_{s-1}}(\widehat{\bm{p}-\bm{k}})\notag\\
	&=\frac{1}{4}\sum_{\lambda=\pm2}\bigg|\epsilon_k^{-1}(\hat{\bm{p}})\epsilon^{-\lambda}_{kl}(\hat{\bm{k}})\epsilon_l^{-1}(\widehat{\bm{k}-\bm{p}})\times\prod_{k=1}^{s-1}\epsilon^{-1}_{i_k}(\hat{\bm{p}})\epsilon^{-1}_{i_k}(\widehat{\bm{k}-\bm{p}})\bigg|^2.
 \end{align}
In the third line of the equation,  we decompose the polarization tensors of rank-$s$ into a product of $\epsilon_i^{-1}$ and contract all indices.  

Utilizing the momentum expressions provided above and the results of the contraction of polarization tensors~Eq.~(\ref{contact_didd_direct}), we can compute Eq.~(\ref{angularspins}). We divide it into two parts, where the first part corresponds to the contraction between projection operators and the rank-2 tensor ($\lambda=\pm2$)
\begin{align}
	\epsilon_k^{-1}(\hat{\bm{p}})\epsilon^{-\lambda}_{kl}(\hat{\bm{k}})\epsilon_l^{-1}(\widehat{\bm{k}-\bm{p}})&=\left(\epsilon_k^{-1}(\hat{\bm{p}})\epsilon_k^{-\frac{\lambda}{2}}(\hat{\bm{k}})\right)\times\left(\epsilon_l^{-\frac{\lambda}{2}}(\bm{k})\epsilon_l^{-1}(\widehat{\bm{k}-\bm{p}})\right)\notag\\
	&=\left(1-\frac{\lambda}{2}\frac{\bm{k}\cdot\bm{p}}{|\bm{k}||\bm{p}|}\right)\left(1-\frac{\lambda}{2}\frac{\bm{k}\cdot(\bm{k}-\bm{p})}{|\bm{k}||\bm{k}-\bm{p}|}\right)\notag\\
	&=\left(1-\frac{\lambda}{2}\mathrm{cos}\theta\right)\left(1-\frac{\lambda}{2}\frac{1-l\mathrm{cos}\theta}{\sqrt{l^2+1-2l\mathrm{cos}\theta}}\right),
\end{align}
and the second part corresponds to the contraction of indices between rank-s tensors 
\begin{align}
	\prod_{k=1}^{s-1}\epsilon^{-1}_{i_k}(\hat{\bm{p}})\epsilon^{-1}_{i_k}(\widehat{\bm{k}-\bm{p}})=\left(1-\frac{\mathrm{cos}\theta-l}{\sqrt{l^2+1-2l\mathrm{cos}\theta}}\right)^{s-1}.
\end{align}
Finally, we obtain the angular part of the 2-point function of tensor perturbations induced by the highest helicity modes of the massive spin-$s$ field
\begin{align}
	&\frac{1}{4}\sum_{\lambda=\pm2}\bigg|\epsilon_k^{-1}(\hat{\bm{p}})\epsilon^{-\lambda}_{kl}(\hat{\bm{k}})\epsilon_l^{-1}(\widehat{\bm{k}-\bm{p}})\times\prod_{k=1}^{s-1}\epsilon^{-1}_{i_k}(\hat{\bm{p}})\epsilon^{-1}_{i_k}(\widehat{\bm{k}-\bm{p}})\bigg|^2\notag\\
	&=\frac{1}{4}\sum_{\lambda=\pm2}\bigg|\left(1-\frac{\lambda}{2}\mathrm{cos}\theta\right)\left(1-\frac{\lambda}{2}\frac{1-l\mathrm{cos}\theta}{\sqrt{l^2+1-2l\mathrm{cos}\theta}}\right)\left(1-\frac{\mathrm{cos}\theta-l}{\sqrt{l^2+1-2l\mathrm{cos}\theta}}\right)^{s-1}\bigg|^2.
\end{align}
Subsequently, we use a similar method for computing the massive spin-$2$ field to calculate the power spectrum of tensor perturbations and it yields 
\begin{align}\label{Eq:GWspins}
\mathcal{P}_{h(s)}(k)&=\mathcal{P}^{+}_{h(s)}(k)+\mathcal{P}^{-}_{h(s)}(k)\notag\\
&=\frac{2k^3}{(2\pi)^2}\frac{1}{V}\int \frac{d\hat{\bm{k}}}{4\pi}\langle h^{TT}_{ij}(\bm{k},\tau_0),h^{\star TT}_{ij}(\bm{k},\tau_0)\rangle\notag\\
 &=\frac{H^4}{M_{\rm pl}^4}\frac{e^{2\pi{\kappa}_s}}{4\pi^4}\frac{s^2}{2^{2s}}\int^{\infty}_0dl l\notag\\
	&\times\int^{\pi}_0d\theta\frac{\mathrm{sin}\theta}{\sqrt{l^2-2l\mathrm{cos}\theta+1}}\sum_{\lambda=\pm2}\bigg|\left(1-\frac{\lambda}{2}\mathrm{cos}\theta\right)\left(1-\frac{\lambda}{2}\frac{1-l\mathrm{cos}\theta}{\sqrt{l^2+1-2l\mathrm{cos}\theta}}\right)\left(1-\frac{\mathrm{cos}\theta-l}{\sqrt{l^2+1-2l\mathrm{cos}\theta}}\right)^{s-1}\bigg|^2\notag\\
	&\times\bigg|\int^{x_{\rm max}}_{0}dx_1\left(x_1\mathrm{cos}x_1-\mathrm{sin}x_1\right)\left((1-s)x_1^{-1}W_{-i\kappa_s,i\mu_s}(-2ilx_1)+\partial_{x_1}W_{-i\kappa_s,i\mu_s}(-2ilx_1)\right)\notag\\
	& \times\left((1-s)x_1^{-1}W_{-i\kappa_s,i\mu_s}(-2i\sqrt{l^2-2l\mathrm{cos}\theta+1}x_1)+\partial_{x_1}W_{-i\kappa_s,i\mu_s}(-2i\sqrt{l^2-2l\mathrm{cos}\theta+1}x_1)\right) \bigg|^2,
\end{align}
where $\mathcal{P}^{\pm}_{h(s)}(k)$ represents the helicity-dependent power spectrum for GW polarization states. The $k$-independence confirms a scale-invariant spectrum. For the massive spin-2 case, the result aligns with Sec.\;\ref{GWspin2}. Explicit spin dependence originates from four principal sources: (i) the $s^2$ prefactor, (ii) the $2^{-2s}$ normalization factor, (iii) angular momentum contributions, and (iv) Whittaker-function coefficients in the $x$-integration. Angular momentum terms originate from polarization tensor contractions, while $2^{-2s}$ reflects their normalization. In the $l \to \infty$ limit, the spin-dependent angular term asymptotically approaches 2, canceling out the $2^{-2s}$ factor. Thus, spin effects are dominated by the $s^2$ prefactor and $x$-integration.  Preliminarily, since the expression is proportional to the power of spin, suggesting that higher spin results in stronger gravitational wave intensity. Full numerical GW computations and their phenomenological implications are detailed in Sec.\;\ref{numresult}.  

 \subsection{The slow-roll correction at CMB scales}\label{slowCMB}
In the preceding analysis, we derived the GW spectrum generated by a massive spin-$s$ field, exhibiting scale invariance. This scale invariance originates from the dS isometry of the inflationary spacetime. However, realistic inflationary backgrounds deviate from exact dS symmetry, necessitating incorporation of slow-roll corrections. It is typical to quantify the deviation of inflation from an exact exponential expansion through the following slow-roll parameters \cite{maggiore2007gravitational}:  
\begin{align}
    \epsilon_H\equiv -\frac{\dot{H}}{H^2}\ll 1,\quad\quad \eta_H\equiv \frac{\dot{\epsilon_H}}{H\epsilon_H}\ll 1,
\end{align}
where $\epsilon_H$ and $\eta_H$ are termed the first and the second slow-roll parameters, respectively. It is also convenient to define the slow-roll parameters in terms of the inflaton potential rather than the Hubble parameter
\begin{align}
     \epsilon_V\equiv \frac{M_{\rm pl}}{2}\left(\frac{\partial_\phi V(\phi)}{V(\phi)}\right)^2\ll 1,\quad\quad \eta_V\equiv M_{\rm pl}\left(\frac{\partial^2_\phi V(\phi)}{V(\phi)}\right)\ll 1.
\end{align}
The slow-roll parameters defined in the aforementioned methods are not equivalent; however, they differ only by a quadratic term in $\epsilon_H$ and $\eta_H$, namely
\begin{align}
    \epsilon_V=\epsilon_H+\mathcal{O}(\epsilon^2),\quad \eta_V=2\epsilon_H-\frac{1}{2}\eta_H+\mathcal{O}(\epsilon^2).
\end{align}
We first qualitatively assess slow-roll corrections to the GW spectrum. Subhorizon modes ($ |k\tau| \gg 1 $) remain unaffected by spacetime curvature evolution or Hubble parameter time dependence. Superhorizon modes ($ |k\tau| \ll 1 $) remain constant and are, therefore, insensitive to the evolution of spacetime. Only near-horizon modes ($ |k\tau| \sim 1 $) exhibit sensitivity to $ H(k) $. The $ k $-dependence enters through the Hubble parameter $ H(k) $, modifying the power spectrum as  
\begin{align}  
    \mathcal{P}_h(k)=\mathcal{P}_h(H(k)),  
\end{align}  
and the tilt of the tensor power spectrum can be calculated directly  
\begin{align}  
    n_t=\frac{d\log\mathcal{P}_h(H(k))}{d\log k}=\frac{d\log\mathcal{P}_h(H(k))}{d\log H(k)}\times\frac{d\log H(k)}{d\log k}.  
\end{align}  

A more rigorous alternative incorporates first-order slow-roll corrections into the Hubble parameter’s time dependence within the EoMs, followed by resolving the EoMs and the corresponding Wronskian. We employ this latter methodology for more precise results.  

Focusing on tensor perturbations, their dynamics are governed by  
\begin{align}  
    \partial^2_\tau \chi(\tau,\bm{k})+\left(k^2-\frac{\partial^2_\tau a}{a}\right)\chi(\tau,\bm{k})=0,  
\end{align}  
where $ \chi \equiv a(\tau)h $. Incorporating slow-roll corrections ($ a \propto \tau^{-(1+\epsilon_H)} $) modifies this to  
\begin{align}  
    \partial^2_\tau \chi(\tau,\bm{k})+\left(k^2-\frac{2+3\epsilon_H}{\tau^2}\right)\chi(\tau,\bm{k})=0,  
\end{align}  
yielding the mode function of tensor perturbations  
\begin{align}  
    h_k(\tau)=\frac{\chi_k(\tau)}{a(\tau)}\simeq (1-\epsilon_H)\frac{\sqrt{\pi}}{2}H (-\tau)^{\frac{3}{2}}\mathrm{H}^{(1)}_\nu (-k\tau),  
\end{align}  
with $ \mathrm{H}^{(1)}_\nu(z) $ denoting the first-kind Hankel function and $ \nu \approx \frac{3}{2} + \epsilon_H $.   

Next, we calculate the slow-roll correction of the highest helicity modes of the spin-$s$ field. The modified EoM of the spin-$s$ field is Eq.~(\ref{EoMhs}) which becomes
\begin{align}\label{slow_eom}
   0= &\partial_\tau^2I_{s}^{\pm(s)}-(2s-2)\left(-\frac{1+\epsilon_H}{\tau}\right)\partial_\tau I_{s}^{\pm(s)}+\left[k^2+m_s^2(1+2\epsilon_H)\frac{k^{-2\epsilon}}{H^2_k}\frac{1}{\tau^{2(1+\epsilon_H)}}-\left(\frac{1+2\epsilon_H}{\tau^2}\right)2\left(s-1\right)\right]I_{s}^{\pm(s)}\notag\\
    &\pm sk\left(1+\epsilon_H\right)\frac{k^{-\epsilon_H}}{H_k}(-\tau)^{-(1+\epsilon_H)}{\tilde{\kappa}}^{(s)}I_{s}^{\pm(s)}.
\end{align}
In the above equation, the slow-roll corrections 
\begin{align}
    aH\simeq -\frac{1+\epsilon_H}{\tau},\quad a(\tau)\simeq (1+\epsilon_H)\frac{k^{-\epsilon_H}}{H_k}(-\tau)^{-(1+\epsilon_H)}
\end{align}
are taken into account, where $H_k$ is the value of the Hubble parameter when the mode with comoving momentum $k$ exits the horizon. Since the above equation cannot be solved analytically, we expand the solution to the first order of the slow-roll parameter $\epsilon_H$
\begin{align}\label{1odersolut}
I_s^{\pm(s)}(\tau,k)=I_0^{\pm(s)}(\tau,k)+\epsilon_H I_1^{\pm(s)}(\tau,k),
\end{align}
where the zeroth-order solution is known and satisfies the equation
\begin{align}\label{zeroeom}
     0= &\partial_\tau^2I_{0}^{\pm(s)}-(2s-2)\left(-\frac{1}{\tau}\right)\partial_\tau I_{0}^{\pm(s)}+\left[k^2+m_s^2\frac{1}{H^2_k\tau^2}-\left(\frac{1}{\tau^2}\right)2\left(s-1\right)\right]I_{0}^{\pm(s)}\pm sk\frac{-1}{H_k\tau}{\tilde{\kappa}}^{(s)}I_{0}^{\pm(s)}.
\end{align}
We also expand the comoving momentum $k^{-\epsilon_H}$ and comoving time $(-\tau)^{-(1+\epsilon_H)}$ to the first order of the slow-roll parameter
\begin{align}\label{ktauexp}
    k^{-\epsilon_H}&\simeq \left(1-\epsilon_H\mathrm{log}\frac{k}{H}\right)H^{-\epsilon},\\
    (-\tau)^{-(1+\epsilon_H)}&\simeq (-\tau)^{-1}-H^{\epsilon_H}(-\tau)^{-1}\mathrm{log}(-H\tau),
\end{align}
 Expanding equation Eq.~(\ref{slow_eom}) according to the above expressions and substituting the solution Eq.~(\ref{1odersolut}) into the equation, retaining terms up to first order of $\epsilon_H$, we obtain
\begin{align}
     0&= \partial_\tau^2I_{0}^{\pm(s)}-(2s-2)\left(-\frac{1}{\tau}\right)\partial_\tau I_{0}^{\pm(s)}+\left[k^2+m_s^2\frac{1}{H^2_k\tau^2}-\left(\frac{1}{\tau^2}\right)2\left(s-1\right)\right]I_{0}^{\pm(s)}\pm sk\frac{-1}{H_k\tau}{\tilde{\kappa}}^{(s)}I_{0}^{\pm(s)}\notag\\
     &+\epsilon_H\left\{\partial_\tau^2I_{1}^{\pm(s)}-(2s-2)\left(-\frac{1}{\tau}\right)\partial_\tau I_{1}^{\pm(s)}+\left[k^2+m_s^2\frac{1}{H^2_k\tau^2}-\left(\frac{1}{\tau^2}\right)2\left(s-1\right)\right]I_{1}^{\pm(s)}\right.\notag\\
     &\quad\quad\quad \pm sk\frac{-1}{H_k\tau}{\tilde{\kappa}}^{(s)}I_{1}^{\pm(s)}+\frac{2(s-1)}{\tau}\partial_\tau I_0^{\pm(s)}\notag\\
     &\left.\quad\quad\quad+\frac{1}{\tau^2}\left(2\frac{m_s^2}{H_k^2}-4(s-1)\mp(-\tau)sk\frac{\tilde{\kappa}^{(s)}}{H_k}-\frac{m_s^2}{H_k^2}\mathrm{log}(k^2\tau^2)\pm sk\frac{\tilde{\kappa}^{(s)}}{H_k}\tau\mathrm{log}(-k\tau)\right)I_0^{\pm(s)}\right\},
\end{align}
where the first line vanishes and the first-order solution satisfies a non-homogeneous second-order differential equation. The homogeneous part of this equation is identical to that of the zeroth-order solution (actually, we can obtain equations of arbitrary order, and we have placed the result in the App.\;\ref{full-order_slowroll})
\begin{align}\label{firsteom}
    &\partial_\tau^2I_{1}^{\pm(s)}-(2s-2)\left(-\frac{1}{\tau}\right)\partial_\tau I_{1}^{\pm(s)}+\left[k^2+m_s^2\frac{1}{H^2_k\tau^2}-\left(\frac{1}{\tau^2}\right)2\left(s-1\right)\right]I_{1}^{\pm(s)}
    \pm sk\frac{-1}{H_k\tau}{\tilde{\kappa}}^{(s)}I_{1}^{\pm(s)}\notag\\
    &=-\frac{2(s-1)}{\tau}\partial_\tau I_0^{\pm(s)}-\frac{1}{\tau^2}\left(2\frac{m_s^2}{H_k^2}-4(s-1)\mp(-\tau)sk\frac{\tilde{\kappa}^{(s)}}{H_k}-\frac{m_s^2}{H_k^2}\mathrm{log}(k^2\tau^2)\pm sk\frac{\tilde{\kappa}^{(s)}}{H_k}\tau\mathrm{log}(-k\tau)\right)I_0^{\pm(s)}.
\end{align}
Redefining the dimensionless variable $ x \equiv -k\tau $ in Eqs.~(\ref{zeroeom}) and (\ref{firsteom}) reveals that the zeroth- and first-order equations are $ k $-independent. Thus, slow-roll corrections to the power spectrum arise exclusively via $ k $-dependent terms in the Wronskian and interaction term through $ a(\tau) $. However, the $ a(\tau) $-dependence in these terms precisely cancels. Consequently, slow-roll corrections to the spin-$ s $ field’s tensor power spectrum lack explicit $ k $-scaling (e.g., terms like $ k^{\epsilon_H f(s)} $, where $ f(s) $ is spin-dependent) and instead imprint solely through $ H_k $. Spin-explicit dependencies in the power spectrum emerge solely through the $ {m}/{H_k} $ ratio within the spin-$ s $ field’s mode function.
\begin{align}
	I^{(\pm s)}_0&=\frac{e^{\mp\frac{\pi\kappa_{s,k}}{2}}}{2^{\frac{s}{2}}H_k^{(s-1)}\sqrt{k}}(-1)^{(s-1)}\tau^{(1-s)}W_{\pm i\kappa_{s,k},i\mu_{s,k}}(2ik\tau),
\end{align}
where $\kappa_{s,k}=\dot{\phi_0}/(\Lambda_{c,s}H_{k})$ and $\mu_{s,k}=\sqrt{{m_s^2}/{H_k^2}-{\left(1-2s\right)^2}/{4}}$ both depend on $k$. 
The tilt of the spin-$s$ field-induced tensor power spectrum becomes
\begin{align}
    n_t^{(1)}&=\frac{d\mathrm{log}\mathcal{P}_h}{d\mathrm{log}k}\bigg |_{k=k_\star}\notag\\
    &=\left(\frac{d\mathrm{log}H_k^4}{d\mathrm{log}k}+\frac{d2\pi\kappa_{s,k}}{d\mathrm{log}k}+\frac{d\mathrm{log}{\mathcal{I}_s}}{d\mathrm{log}\mu_{s,k}}\frac{d\mathrm{log}\mu_{s,k}}{d\mathrm{log}H_k}\frac{d\mathrm{log}H_k}{d\mathrm{log}k}+\frac{d\mathrm{log}{\mathcal{I}_s}}{d\mathrm{log}\kappa_{s,k}}\frac{d\mathrm{log}\kappa_{s,k}}{d\mathrm{log}H_k}\frac{d\mathrm{log}H_k}{d\mathrm{log}k}\right)\bigg|_{\mu_s=\mu_{s}^\star,\kappa_s=\kappa_{s}^\star,k=k_\star},
\end{align}
where $\kappa_{s}^\star={\dot{\phi}_0}/({\Lambda_{c,s}H_\star})$ and $\mu_{s}^\star=\sqrt{{m_s^2}/{H_\star^2}-{\left(1-2s\right)^2}/{4}}$ are the chemical potential and the conformal weight expressed in terms of $H_\star$, which is the value of the Hubble parameter when the pivot comoving momentum $k_\star=k_{\text{CMB}}\simeq0.002~\text{Mpc}^{-1}$ crosses the horizon. The form of the integral $\mathcal{I}_s$ is
\begin{align}
  \mathcal{I}_s &=\int^{\infty}_0dl l\notag\\
	&\times\int^{\pi}_0d\theta\frac{\mathrm{sin}\theta}{\sqrt{l^2-2l\mathrm{cos}\theta+1}}\sum_{\lambda=\pm2}\bigg|\left(1-\frac{\lambda}{2}\mathrm{cos}\theta\right)\left(1-\frac{\lambda}{2}\frac{1-l\mathrm{cos}\theta}{\sqrt{l^2+1-2l\mathrm{cos}\theta}}\right)\left(1-\frac{\mathrm{cos}\theta-l}{\sqrt{l^2+1-2l\mathrm{cos}\theta}}\right)^{s-1}\bigg|^2\notag\\
	&\times\bigg|\int^{x_{\rm max}}_{0}dx_1\left(x_1\mathrm{cos}x_1-\mathrm{sin}x_1\right)\left((1-s)x_1^{-1}W_{-i\kappa_s^\star,i\mu_s^\star}(-2ilx_1)+\partial_{x_1}W_{-i\kappa_s^\star,i\mu_s^\star}(-2ilx_1)\right)\notag\\
	& \times\left((1-s)x_1^{-1}W_{-i\kappa_s^\star,i\mu_s^\star}(-2i\sqrt{l^2-2l\mathrm{cos}\theta+1}x_1)+\partial_{x_1}W_{-i\kappa_s^\star,i\mu_s^\star}(-2i\sqrt{l^2-2l\mathrm{cos}\theta+1}x_1)\right) \bigg|^2,
\end{align}
 and the dependence of $H_k$ on $k$ is (the detailed derivation can be found in the literature \cite{maggiore2007gravitational})
\begin{align}
    \frac{d\mathrm{log}H_k}{d\mathrm{log}k}=-\epsilon_V.
\end{align}
The dependence of the conformal weight $\mu_s^\star$ on $k$ manifests in $H_k$, specifically, the derivative of $\mu_s^\star$ with respect to $k$ is
\begin{align}
    \frac{d\mathrm{log}\mu_s^\star}{d\mathrm{log}k}=\frac{d\mathrm{log}\mu_s^\star}{d\mathrm{log}H_k}\frac{d\mathrm{log}H_k}{d\mathrm{log}k}=-\epsilon_V\times\frac{1}{\frac{(2s-1)^2}{4}\frac{H_\star^2}{m_s^2}-1},
\end{align}
and the dependence of the chemical potential $\kappa_s^\star$ on $H_k$ is
\begin{align}
    \frac{d\mathrm{log}\kappa_{s}^\star}{d\mathrm{log}H_k}=-1.
\end{align}
We thus derive the tilt of the power spectrum of the tensor perturbations induced by the massive spin-$s$ field:  
\begin{align}  
    n_t^{(1)}=-4\epsilon_V+\frac{1}{\mathcal{I}_s}\frac{d{\mathcal{I}_s}}{d\mu_{s}^\star}\left(\mu^\star_{s}+\frac{(2s-1)^2}{4\mu^\star_s}\right)\epsilon_V+\left(\frac{1}{\mathcal{I}_s}\frac{d{\mathcal{I}_s}}{d\kappa_{s}^\star}+2\pi\right)\kappa_s^\star\epsilon_V,  
\end{align}  
where $m_s$ is expressed via conformal weight $\mu_s$ and spin $s$. This result explicitly demonstrates spin dependence in the spectral tilt, however, slow-roll corrections induce only minor modifications to the GW spectrum, particularly for low-spin fields. Spectral tilt variations across spins are insubstantial, and parameter degeneracy among $s$, $\mu_s$, and $\kappa_s$ precludes direct spin determination. Experimental distinction of spins via slow-roll corrections thus remains a formidable challenge.  
\subsection{The backreaction effect}\label{slowback}
Previous analyses primarily focused on CMB scales, where temporal evolution of the chemical potential $\kappa_s$ and Hubble parameter $H$ is negligible. A detailed examination reveals that $\kappa_s$ grows during inflation as a result of slow-roll corrections —specifically, background field evolution. An increasing $\kappa_s$ amplifies spin-$s$ particle production, thereby increasing their energy density. Moreover, the backreaction on the spacetime background, arising from the inverse decay of massive spin-$s$ fields, becomes significant and modifies inflaton dynamics, particularly at late times (extensively analyzed for spin-1 in \cite{Wang:2020uic,Niu:2022quw}). Specifically, the backreaction decelerates the background field’s rolling velocity $\dot{\phi}_0$. Meanwhile, the cumulative evolution of $\dot{\phi}_0$ and $H$ in turn modulates $\kappa_s ={\dot{\phi}_0}/({\Lambda_{c,s} H})$.

Our analysis identifies three principal motivations for investigating backreaction effects: (i) its physical inevitability at late inflationary stages, where backreaction becomes non-negligible; (ii) copious late-time particle production, which predicts substantial GW signals—these signals evade CMB observational limits and lie within the sensitivity bands of next-generation GW detectors; (iii) backreaction’s spin-dependent energy density, thus enabling spin differentiation via backreaction-induced GW signatures. To quantify this, we solve Einstein equations to derive the inflaton field $\phi$ and scale factor $a(\tau)$:
 \begin{equation}
	\left\{\begin{aligned}
		&\dot{\rho}(t)+3H(t)\left[\rho(t)+p(t)\right],\\
        &3H^2=8\pi G\rho(t).
	\end{aligned}\right.
\end{equation}
If we only consider the time evolution of the inflaton background, the EoM takes the following form:
\begin{equation}\label{br1o}
	\left\{\begin{aligned}
		& \ddot{\phi}_0+3H\dot{\phi}_0+\frac{dV}{d\phi_0}=\langle\frac{1}{2\Lambda_{c,s}}\frac{\epsilon^{\mu\nu\rho\sigma}}{\sqrt{-g}}\nabla_{\mu}I_{\nu\lambda_1\cdots \lambda_{s-1}}\nabla_{\rho}I_{\sigma}^{\lambda_1\cdots \lambda_{s-1}}\rangle,\\
		&3H^2M_{\rm pl}^2-\frac{1}{2}\dot{\phi_0}^2-V(\phi_0)=\langle T_{00}\rangle.
	\end{aligned}\right.
\end{equation}
Due to the complexity of directly deriving the complete Lagrangian for massive spin-$s$ fields, directly obtaining the energy-momentum tensor is intractable. Instead, we employ a particle-based approximation: summing individual particle energies weighted by their number density and integrating over momentum space. The upper limit of the momentum integral is determined by the tachyonic instability scale $k_{\mathrm{pmax}} \approx \kappa_s H$ (Eq.~(\ref{pscaleparpro})), yielding the spin-$s$ energy density approximation:   
\begin{align}  
    \langle T_{00}\rangle \simeq \rho_{s} = \int^{k_{\rm pmax}}_0 \frac{d^3\bm{k_p}}{(2\pi)^3} \omega_{k_p} \times n(k_p,\tau) = \frac{1}{2\pi^2} \int^{k_{\rm pmax}}_0 dk k_p^2 \sqrt{m_s^2 + k_p^2} \, n(k_p,\tau),  
\end{align}  
where $k_p$ denotes physical momentum, $k_{\mathrm{pmax}} \equiv \kappa_s H + \sqrt{\kappa_s^2 H^2 - \left(m_s^2 - s(s-1)H^2\right)} \approx \kappa_s H$, and $n(k_p,\tau)$ can be calculated for any $\tau$ via the Stokes line method. Here, we consider the late-time limit, where the particle number density simplifies to $n_k = \left({1 + e^{2\pi(\mp\kappa_s + \mu_s)}}\right)/\left({e^{4\pi\mu_s} - 1}\right)$, exhibiting no $k_p$- or $\tau$-dependence. Thus, the energy density of the spin-$s$ field is given by  
\begin{align}\label{enegydensity}  
    \rho_s &= \frac{1}{2\pi^2} \int^{k_{\rm pmax}}_0 dk k_p^2 \sqrt{m_s^2 + k_p^2} \, n(k_p,\tau) \notag \\  
    &= \frac{m_s^4}{16\pi^2} \left[ \left( \frac{\kappa_s H}{m_s} + 2\left( \frac{\kappa_s H}{m_s} \right)^3 \right) \sqrt{1 + \left( \frac{\kappa_s H}{m_s} \right)^2 } - \ln \left[ \frac{\kappa_s H}{m_s} + \sqrt{1 + \left( \frac{\kappa_s H}{m_s} \right)^2 } \right] \right] e^{2\pi(\kappa_s - \mu_s)}.  
\end{align}  
The value of this energy density is mainly determined by the exponential factor $\exp[2\pi(\kappa_s-\mu_s)]$. Within the parameter range satisfying the phenomenological requirements outlined in Sec.\;\ref{phenoconstraints}, this value is much smaller than $3H^2M_{\rm pl}^2$ within a relatively large parameter range. Therefore, in the calculation of backreaction, we neglect the contribution of this term.
 To calculate the evolution of $\phi_0$, as we have discussed previously, the chemical potential primarily enhances one of the modes with the highest helicity and the mode also contributes the most to the backreaction. Thus the right-hand side of the first line of Eq.~(\ref{br1o}) can be calculated directly by substituting the highest helicity component $I_{i_1\cdots i_s}$ with $s$ spatial indices (and we change time variable $t\to \tau$):
\begin{align}\label{xxxx}
	&\frac{a^{-2(s-1)}}{\sqrt{-g}}\frac{1}{2\Lambda_{c,s}}\epsilon^{\mu n\rho s }\nabla_{\mu}I_{ni_1\cdots i_{s-1}}\nabla_{\rho}I_{si_1\cdots i_{s-1}}\notag\\
	&=\frac{a^{-2(s+1)}}{2\Lambda_{c,s}}a\left[2\epsilon^{nrs}\nabla_{0}I_{ni_1\cdots i_{s-1}}\nabla_{r}I_{si_1\cdots i_{s-1}}+2\epsilon^{mrs}\nabla_{m}I_{0i_1\cdots i_{s-1}}\nabla_{r}I_{si_1\cdots i_{s-1}}\right].
\end{align} 

After doing the contraction of polarization tensors and finishing the average $\frac{1}{v}\langle\cdots \rangle$, the result for the source term contributed by the highest helicity modes yields
\begin{align}\label{brsecond}
	&\langle\frac{1}{2\Lambda_{c,s}}\frac{\epsilon^{\mu\nu\rho\sigma}}{\sqrt{-g}}\nabla_{\mu}I_{\nu\lambda_1\cdots \lambda_{s-1}}\nabla_{\rho}I_{\sigma}^{\lambda_1\cdots \lambda_{s-1}}\rangle\notag\\
	&=\frac{a^{-2(s+1)}}{\Lambda_{c,s}}\frac{2^s}{2\pi^2}\left(\frac{1}{2}\int^{k_{max}}_0dk k^3\partial_\tau|I^{\pm s}_s(\tau,k)|^2+a(s-1)H\int^{k_{max}}_0dk k^3|I^{\pm s}_s(\tau,k)|^2\right).
\end{align}
Building on this, substituting the mode function $I^{\pm s}_{s}(\tau,k)$ and redefining $x \equiv -k\tau$, Eq.~(\ref{brsecond}) transforms into:  
\begin{align}  
    &\left\langle\frac{1}{2\Lambda_{c,s}}\frac{\epsilon^{\mu\nu\rho\sigma}}{\sqrt{-g}}\nabla_{\mu}I_{\nu\lambda_1\cdots \lambda_{s-1}}\nabla_{\rho}I_{\sigma}^{\lambda_1\cdots \lambda_{s-1}}\right\rangle \notag\\  
    &=-\frac{H^4e^{\pi\kappa_s}}{4\pi\Lambda_{c,s}}\left[\int^{x_{\text{max}}}_0dx\left(x^3\partial_x |W_{-i\kappa_s,i\mu_s}(-2ix)|^2+4(1-s)x^2|W_{-i\kappa_s,i\mu_s}(-2ix)|^2\right)\right],  
\end{align}  
which is spin-dependence and this implies that the strength of the backreaction on spacetime geometry varies with spin, which also gives us insight that, in the presence of backreaction, we may be able to distinguish different spins through GWs at small scales.
The coupled evolution of $\phi_0$ and $a(\tau)$ is governed by:  
\begin{equation}\label{br1}  
    \left\{\begin{aligned}  
        & \ddot{\phi}_0+3H\dot{\phi}_0+\frac{dV(\phi_0)}{d\phi_0} \\  
        &=-\frac{H^4e^{\pi\kappa_s}}{4\pi\Lambda_{c,s}}\left[\int^{x_{\text{max}}}_0dx\left(x^3\partial_x |W_{-i\kappa_s,i\mu_s}(-2ix)|^2+4(1-s)x^2|W_{-i\kappa_s,i\mu_s}(-2ix)|^2\right)\right] \\  
        &3H^2M_{\mathrm{pl}}^2-\frac{1}{2}\dot{\phi}_0^2-V(\phi_0)\simeq 0  
    \end{aligned}\right..  
\end{equation}  

For computational efficiency, we adopt e-folds $N$ (remaining before inflation’s end) as the independent variable, with $dN = -H dt$. Eq.~(\ref{br1}) then becomes:  
\begin{equation}\label{br2}  
    \left\{\begin{aligned}  
        & \frac{d^2{\phi}_0}{dN^2}+\left(-3+\frac{d\log H}{dN}\right)\frac{d{\phi}_0}{dN}+\frac{1}{H^2}\frac{dV(\phi_0)}{d\phi_0} \\  
        &=-\frac{H^2e^{\pi\kappa_s}}{4\pi\Lambda_{c,s}}\left[\int^{x_{\text{max}}}_0dx\left(x^3\partial_x |W_{-i\kappa_s,i\mu_s}(-2ix)|^2+4(1-s)x^2|W_{-i\kappa_s,i\mu_s}(-2ix)|^2\right)\right] \\  
        &H^2=\frac{V(\phi_0)}{3-\frac{1}{2}\left(\frac{d\phi_0}{dN}\right)^2}  
    \end{aligned}\right.,  
\end{equation}  
where the $M_{\rm pl}$ has been extracted from all parameters with the dimension of mass.

We assume the inflaton potential takes the form of the Starobinsky model \cite{Starobinsky:1980te} which is one of the most favored by the recent experimental observations, given by  
\begin{align}  
    V(\phi_0)=V_0\left(1-e^{-\sqrt{\frac{2}{3}}{\phi_0}}\right)^2,  
\end{align}  
with $V_0^{1/4} \simeq 0.003M_{\mathrm{pl}}$. A generalized Starobinsky potential \cite{Domcke:2016bkh} is also considered:\footnote{In this model, $\dot{\phi}_0 < 0$, and therefore, in the subsequent calculations, we redefine $\kappa_s \equiv \frac{|\dot{\phi}_0|}{\Lambda_{c,s}H}$ to ensure that it is positive.}
\begin{align}  
     V(\phi_0)=\frac{3}{4}V_0\left(1-e^{-\gamma{\phi_0}}\right)^2,  
\end{align}  
where $\gamma$ is the potential parameter (or shape factor). Generalized Starobinsky parameters must satisfy observational constraints: slow-roll conditions, scalar power spectrum amplitude, and spectral index consistency. We derive parameter relations and constraints following established methodologies \cite{Domcke:2016bkh,Niu:2022quw}.  

At CMB scales, the spin-$s$ field contributions can be neglected and the slow-roll parameter can be parametrized as the power of e-folds 
\begin{align}
    \epsilon_V=\frac{1}{2}\left(\frac{\dot{\phi_0}}{H}\right)^2=\frac{1}{2}\left(\frac{V'(\phi_0)}{V(\phi_0)}\right)^2\simeq\frac{1}{2\gamma^2N_\star^2},
\end{align}
where $N_\star\simeq 50$ is the e-folds when modes of CMB scales left the horizon and its value is set by the observed value of the spectral index
\begin{align}
    n_s\approx 0.96\approx 1-\frac{2}{N_\star}.
\end{align}
Since the slow-roll parameter and the tensor-to-scalar ratio have the relation
\begin{align}
    r\approx 16\epsilon_V,
\end{align}
and the value of tensor to scalar ratio has been bounded by the CMB observations \cite{Meerburg:2015zua,Planck:2018jri,Planck:2019kim,Galloni:2022mok,BICEP:2021xfz}, that is, $r<0.056$, thus there exists a lower bound of $\gamma$, which yields
\begin{align}
    \gamma^2\simeq\frac{1}{2\epsilon N_\star^2}\gtrsim \frac{2}{35}.
\end{align}
In fact, the parameters in the potential are not independent of each other. At CMB scales, we have already measured the value of the scalar power spectrum. Therefore, in order to obtain the observed value of the scalar power spectrum, the value of $V_0$ is determined by $\gamma$,
\begin{align}
    \mathcal{P}_\zeta=\frac{H^2}{8\pi^2\epsilon_V}\simeq\frac{1}{24\pi^2}\frac{V(\phi_0)}{\epsilon_V}\approx \frac{V_0\gamma^2N_\star^2}{12\pi^2}\Rightarrow V_0= \frac{12\pi^2\mathcal{P}_\zeta}{\gamma^2N_\star^2},
\end{align}
where $ \mathcal{P}_\zeta\simeq 2\times10^{-9}$.  
If we require that the evolution of the background field remains balanced between the potential and the frictional force generated by the source, the background field's evolution rate is entirely determined by the shape of the potential and its initial position. Moreover, inflation must last for $N = 50 \sim 60$ e-folds, with the universe ultimately exiting inflation, i.e., when the slow-roll parameter satisfies $\epsilon_V = 1$. The exit point of inflation is entirely determined by the potential, that is,  
\begin{align}\label{phiend}
   & \epsilon_V=\frac{1}{2}\left(\frac{V'(\phi_0)}{V(\phi_0)}\right)^2=2\gamma^2\left(e^{\gamma\phi_0}-1\right)^{-2}\simeq 1\notag\\
    \Rightarrow&\phi_{\text{end}}=\frac{1}{\gamma}\mathrm{log}\left(\sqrt{2}\gamma+1\right).
\end{align}
Therefore, based on the above discussion, we find that if the mass and spin are fixed, the evolution equation actually has only one free parameter, $\kappa_s$, while the cut-off scale $\Lambda_{c,s}$ is determined by $\kappa_s$ and the shape of the potential whose value can be obtained by solving the equilibrium equation. 
If we neglect the contribution of spin-$s$ fields, the e-folds can be calculated using the formula
\begin{align}\label{efold}
N(\phi_{\text{end}},\phi_{\text{initial}})\simeq \int^{\phi_{\text{initial}}}_{\phi_{\text{end}}}d\phi_0\frac{V(\phi_0)}{V'(\phi_0)}=\frac{1}{2\gamma}\left(\frac{1}{\gamma}e^{\gamma\phi_{\text{end}}}-\frac{1}{\gamma}e^{\gamma\phi_{\text{initial}}}-\left(\phi_{\text{end}}-\phi_{\text{initial}}\right)\right).
\end{align}
 From Eq.~(\ref{phiend}), we obtain the value of $\phi$ corresponding to the end of inflation. Substituting $N=50\sim 60$, we can obtain the value of $\phi_0$ corresponding to the start of slow roll from Eq.~(\ref{efold}).
 It is worth noting that considering the effect of spin-$s$ fields, the evolution rate of the background field $d\phi_0/dN$ slows down due to the additional source, causing the initial position of the background field to be shifted. However, this does not change the total number of e-folds $N=50\sim60$.

\begin{figure}[htp!]
	\centering
	\includegraphics[width=\textwidth]{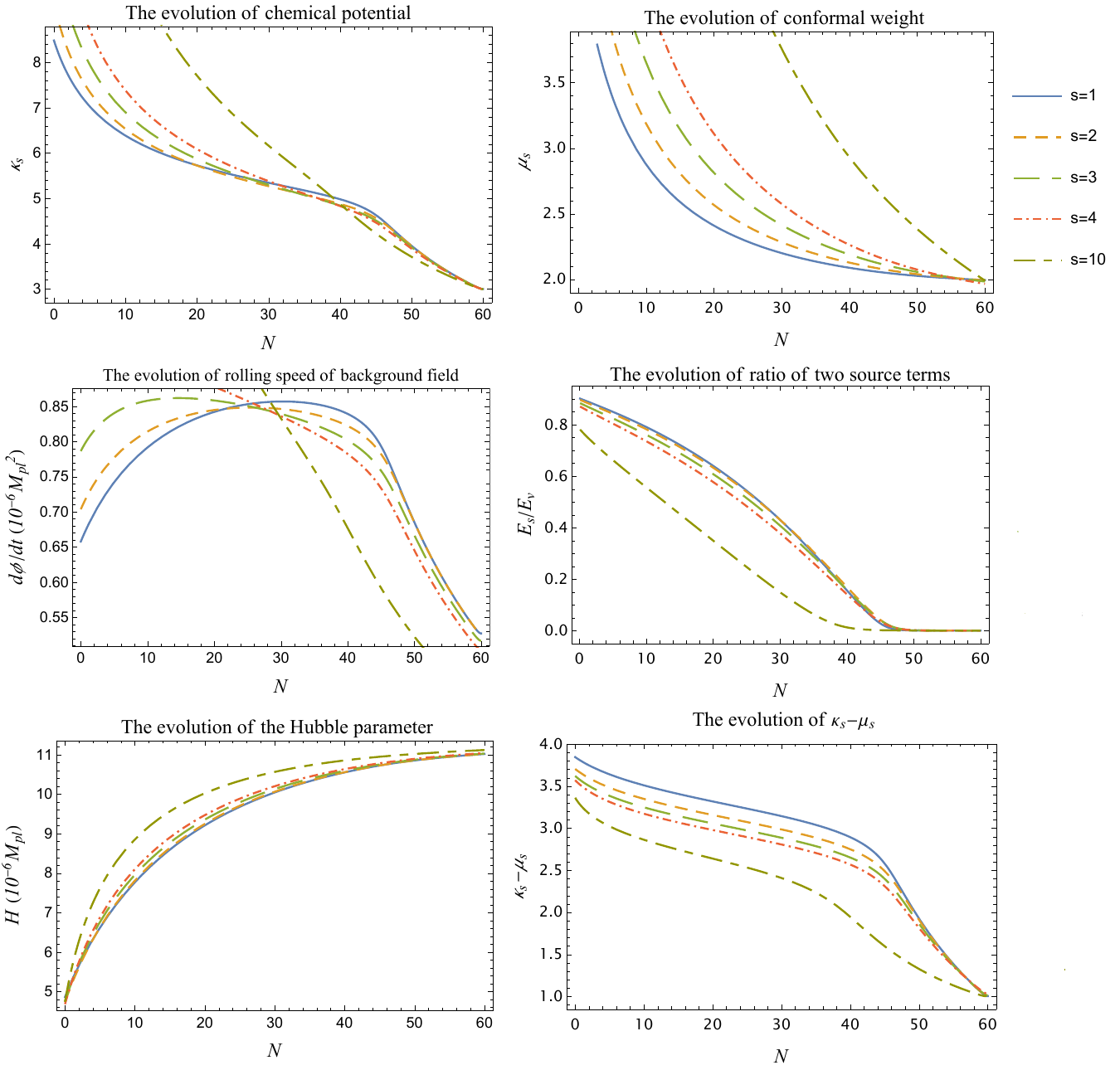}
	\caption{Evolution of the chemical potential \( \kappa_s \), the conformal weight \( \mu_s \), the Hubble parameter \( H \), and the difference \( \kappa_s - \mu_s \), considering the backreaction of fields with different spins on the spacetime background. The figure illustrates results for spin values \( s=1,2,3,4,10 \), with the Hubble parameter at CMB scales given by \( H_{\text{CMB}}\simeq 1.1\times10^{-5}M_{\rm pl} \), and the chemical potential and conformal weight evaluated as \( \kappa_{\text{CMB}}=3 \) and \( \mu_{\text{CMB}}=2 \), respectively. The shape factor of the potential is set to \( \gamma=0.5 \). The background field value at the end of inflation is \( \phi_{\text{end}}=1.07~M_{\rm pl} \). The initial background field values at CMB scales depend on spin: \( \phi^{s=1}_{N=60}\approx 6.17~M_{\rm pl} \), \( \phi^{s=2}_{N=60}\approx 6.17~M_{\rm pl} \), \( \phi^{s=3}_{N=60}\approx 6.20~M_{\rm pl} \), \( \phi^{s=4}_{N=60}\approx 6.25~M_{\rm pl} \), and \( \phi^{s=10}_{N=60}\approx 6.60~M_{\rm pl} \).}\label{backreation_Figure}
 \end{figure}

Next, we analyze the e-fold evolution of physical quantities (Fig.\;(\ref{backreation_Figure})), demonstrating compliance with the slow-roll condition $\epsilon_V \ll 1$ throughout. The $\kappa_s$ evolution for massive spin-$s$ fields follows three distinct stages \cite{Niu:2022quw}:  

In the first stage, spin-$s$ field contributions are subdominant to vacuum fluctuations, rendering cosmic expansion and the tensor power spectrum vacuum-driven. In this period, vacuum energy drives the universe to rapidly expand, while the background field’s rolling velocity $\dot{\phi}_0$ increases markedly. The potential’s relative flatness maintains the slow-roll regime, despite a gradual Hubble parameter decline. The interplay of $\dot{\phi}_0$ and $H$ leads to a rapid increase in the chemical potential $\kappa_s$, which can be seen in Fig.\;(\ref{backreation_Figure}) corresponding to $N\sim46-60$ (taking the spin-$2$ field as an example). 

As $\kappa_s$ gradually increases, the contribution from the spin-$s$ field to Eq.~(\ref{br2}) (the term in the second line) increases progressively, while the vacuum's dominant effects diminish. Consequently, the universe's evolution gradually transitions to a mixed regime where vacuum and spin-$s$ fields jointly govern background dynamics. At this stage, spin-$s$ effects counteract vacuum-driven expansion, resulting in a deceleration of the background field's rolling velocity. Since the Hubble parameter remains approximately constant over time, the $\kappa_s$ growth rate diminishes, leading to a flattening of the curve that corresponds to the stage with e-folds  $N\sim26-46$ in Fig.\;(\ref{backreation_Figure}) (spin-2 example).

When the source term of the spin-$s$ field in Eq.~(\ref{br2}) becomes closer to the vacuum, cosmic evolution transitions into the third stage. The increase in friction generated by the spin-$s$ field at this stage causes the background field’s rolling velocity to peak and subsequently decelerate. Concurrently, as the field gradually rolls away from the potential’s plateau,the rate of decrease of the Hubble parameter increases. These competing effects collectively enhance $\kappa_s$’s growth rate. Concurrently, $\mu_s$ exhibits a monotonically increasing growth rate. Collectively, $\kappa_s - \mu_s$ evolves through two dynamical regimes: an initial rapid increase phase followed by a slower growth phase (Fig.\;(\ref{backreation_Figure})).  

To better understand the backreaction induced by higher-spin fields, we examine the impact of different spins on the evolution of the background field. Fixing the initial conditions $\kappa_s(0)$ and $\mu_s(0)$, the fields with higher spins generate non-negligible contributions, leading to larger initial source terms compared to their lower-spin counterparts. This larger backreaction reduces the background field’s initial rolling acceleration more significantly, resulting in lower rolling velocities ($\dot{\phi}_0$) of the background field for the cases of higher spins in the early stages of inflation. As $\dot{\phi}_0$ is embedded within the source term’s exponential factor (Eq.~(\ref{br2})), it follows that after a certain time, the resistance exerted by the source term of fields with higher spins on the background field will become smaller than that of fields with lower spins. This triggers a relative increase in rolling acceleration for fields with higher spins. A crossover epoch arises where the rolling acceleration of the background field influenced by fields with higher spins surpasses that of lower spins. Subsequently, the rolling velocity of the background field will also exceed that of the lower spins after a certain period.  

On the other hand, due to the suppressed initial rolling velocity under backreaction from fields with higher spins, the background field evolves more slowly within the potential. Consequently, higher spins exhibit diminished early-time Hubble parameter ($H$) decay relative to lower spins. For a fixed initial $\mu_s$, the mass hierarchy enhances the growth of $\mu_s$ for fields with higher spins under the same $H$ decay. Since the decay rate of $H$ varies only slightly with spin, the rate of change of $\mu_s$ for the fields with higher spins consistently exceeds that of the lower spins as the universe evolves.  

Overall, early-stage disparities between $\kappa_s$ and $\mu_s$ are less pronounced for fields with higher spins than for fields with lower spins due to their suppressed $\kappa_s$ and enhanced $\mu_s$. Although $\kappa_s$ eventually grows faster for higher spins, the concurrent acceleration of $\mu_s$ ensures that $\kappa_s - \mu_s$ remains consistently smaller for higher spins, as illustrated in Fig.\;(\ref{backreation_Figure}).

From a physical perspective, chemical potential-driven particle production originates from energy transfer: the background field converts kinetic and potential energy into spin-$s$ field excitations, and the number density of produced particles is governed by the temporal evolution of the background and the particle mass, quantified by $\kappa_s - \mu_s$. If the initial values of $\kappa_s$ and $\mu_s$ are identical across spins, the initial particle number densities are nearly the same. However, higher-spin particles, possessing greater energy density, induce stronger backreaction, increasing resistance to the background field’s motion. This suppression of rolling acceleration temporarily reduces the particle production rate. Once production surpasses a critical threshold, backreaction weakens, allowing acceleration and particle production to resume—explaining why $\kappa_s$ starts smaller but later exceeds that of fields with lower spins. However, the growth of $\mu_s$ for higher spins significantly increases the energy cost of production, ultimately suppressing their number density.

    \subsection{The theoretical and experimental constraints}\label{phenoconstraints}
 In previous subsections, we constructed a model that enhances gravitational wave signals via the chemical potential. However, to satisfy experimental constraints and maintain theoretical consistency, the chemical potential $\kappa_s$ and conformal weight $\mu_s$ must adhere to specific restrictions. 
\paragraph{Non-Gaussianity} The experimental constraints primarily arise from limitations on non-Gaussianity imposed by current Cosmic Microwave Background (CMB) observations. To compute the non-Gaussianity, we employ the Schwinger-Keldysh formalism to evaluate the three-point correlation function of curvature perturbations. In App.\;\ref{FeynmanRuleAppendix} and App.\;\ref{Prop_estim}, we present the vertices and propagators used in the model and provide the estimated results for these propagators. We will now utilize these results to calculate the non-Gaussianity. Since the interaction between the inflaton and the spin-$s$ field arises solely from the chemical potential term, the lowest-order contribution to the three-point correlation function is given by Eq.~(\ref{verticeIIh}).  The most stringent experimental constraint on the curvature bispectrum is $f_{\rm NL}^{\text{equil}}=-26\pm47$ from the Planck collaboration 2018 \cite{Planck:2018jri,Planck:2019kim}. The non-Gaussianity can be calculated as
\begin{align}
\nonumber&\begin{gathered}
			\includegraphics[width=3cm]{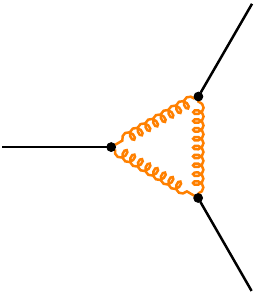}
		\end{gathered}
	=
    f^{\text{equil}}_{\rm NL}=\langle\zeta\zeta\zeta \rangle_{\text{equil}}<21\\
    &\Rightarrow \frac{1}{16\pi^2}\frac{1}{2\pi P_\zeta^{\frac{1}{2}}}\left(\frac{H}{\Lambda_{c,s}}\right)^3\left(\frac{1}{\mu_s^2}e^{2\pi(\kappa_s-\mu_s)}\right)^3
    =\frac{1}{4} P_\zeta\left(\frac{\kappa_s}{\mu_s}\right)^3\mu_s^{-3}e^{6\pi(\kappa_s-\mu_s)}<21,
\end{align}
where $P_{\zeta}\simeq {H^4}/\left({(2\pi)^2\dot{\phi}_0^2}\right)$ is the amplitude of the scalar power spectrum and the relation $\kappa_s={\dot{\phi}_0}/({\Lambda_{c,s}H})$ is applied. In our calculations, we do not include the factor of $1/2^{s-1}$ in Eq.~(\ref{proestimate}) since this factor in the propagator cancels out the spin-dependent contribution from the angular part of the loop integral. This effectively serves as a normalization of the polarization tensor. Subsequent calculations adhere to the same pattern and similarly do not involve this factor.
We also calculate the constraints from the curvature trispectrum, which yields
\begin{align}
\nonumber&\begin{gathered}
			\includegraphics[width=3cm]{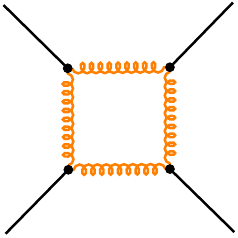}
		\end{gathered}
	=
    g^{\text{loc}}_{\rm NL}=\langle\zeta\zeta\zeta \zeta\rangle_{\text{loc}}<6.5\times 10^4\\
    &\Rightarrow \frac{1}{16\pi^2}\frac{1}{(2\pi)^2 P_\zeta}\left(\frac{H}{\Lambda_{c,s}}\right)^4\left(\frac{1}{\mu_s^2}e^{2\pi(\kappa_s-\mu_s)}\right)^4=\frac{1}{4}P_\zeta\left(\frac{\kappa_s}{\mu_s}\right)^4\mu_s^{-4}e^{8\pi(\kappa_s-\mu_s)}<6.5\times 10^4,
\end{align}
where $P_{\zeta}\simeq {H^4}/\left({(2\pi)^2\dot{\phi}_0^2}\right)$ is the amplitude of the scalar power spectrum and the relation $\kappa_s={\dot{\phi}_0}/({\Lambda_{c,s}H})$ is applied. 

\paragraph{Perturbativity}From a theoretical perspective, as we employ perturbation theory in our calculations, we must ensure the validity of the perturbative expansion.\footnote{For small scales, the perturbativity should be analyzed separately; however, our calculated results for small-scale GW enhancement should remain qualitatively valid.} The inclusion of the chemical potential introduces exponential factors that correspond to chemical potential enhancement in the vertices. Consequently, the chemical potential must have an upper limit to ensure that this exponential enhancement does not cause higher-order contributions to exceed leading-order contributions. Specifically, it is required that self-energy corrections remain subdominant. First, we examine the 1-loop correction of the inflaton propagator mediated by the spin-$s$ field.~\footnote{One noteworthy observation is that despite our imposition of perturbativity in the calculations—specifically, ensuring weak couplings between the graviton and the spin-$s$ field, as well as between the inflaton and the spin-$s$ field—we are still equipped to address this system, which may exhibit strong coupling, by effectively resuming all chain diagrams, more specifically, resolving the linear EoMs and conducting diagonalization. However, for simplicity, we focus on the weak-coupling scenario \cite{Tong:2022cdz}.}.
\begin{align}
		\nonumber&\begin{gathered}
			\includegraphics[width=3cm]{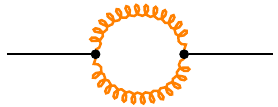}
		\end{gathered}\quad<\quad\begin{gathered}
			\includegraphics[width=3cm]{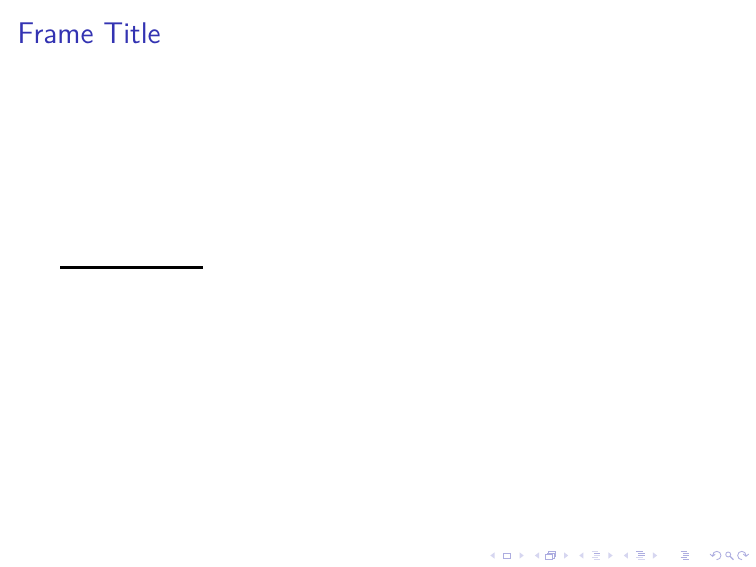}
		\end{gathered}\\
	&\Rightarrow\frac{1}{16\pi^2}\left(\frac{H}{\Lambda_{c,s}}\right)^2\left(\frac{1}{\mu_s^2}e^{2\pi(\kappa_s-\mu_s)}\right)^2=\frac{1}{4} P_\zeta\left(\frac{\kappa_s}{\mu_s}\right)^2\mu_s^{-2}e^{4\pi(\kappa_s-\mu_s)}<1~.
	\end{align}
Next, we consider one-loop correction to the graviton propagator. Similarly, we require it to be smaller than the tree-level propagator. We only consider the minimal coupling between the graviton and the spin-$s$ field (see Eq.~(\ref{verticeIIh})):
\begin{align}
\nonumber&\begin{gathered}
			\includegraphics[width=3cm]{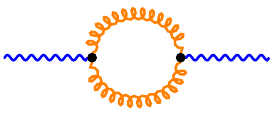}
		\end{gathered}\quad<\quad\begin{gathered}
			\includegraphics[width=3cm]{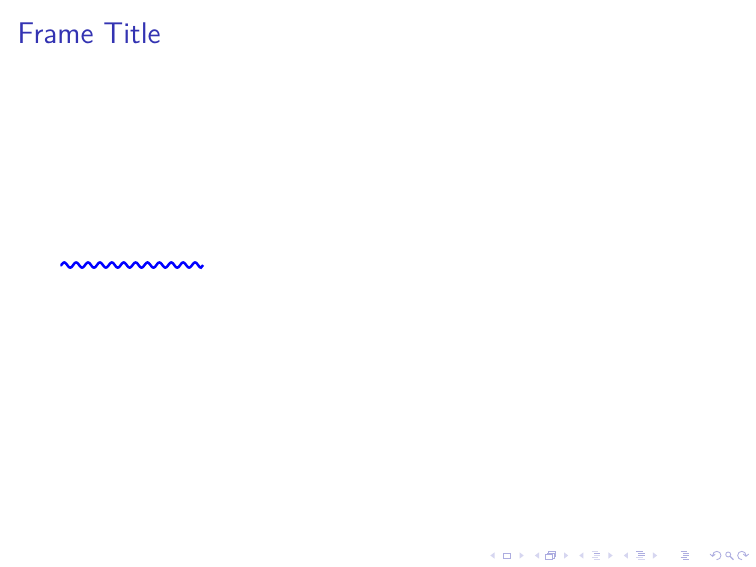}
		\end{gathered}\\
	&\Rightarrow
    \frac{s^2}{16\pi^2}\frac{H^2(\mu_s^2+\frac{(2s-3)^2}{4})^2}{M_{\rm pl}^2}\left(\frac{1}{\mu_s^2}e^{2\pi(\kappa_s-\mu_s)}\right)^2=\frac{s^2}{16 \pi^2}\left(\frac{H}{M_{\rm pl}}\right)^2\left(1+\frac{(2s-3)^2}{4\mu_s^2}\right)^2e^{4\pi(\kappa_s-\mu_s)}<1~.
\end{align}
The final calculation needed is the one-loop self-energy correction of the spin-$s$ field, where both graviton and inflaton contribute. The constraint becomes
\begin{align}
\nonumber&\begin{gathered}
			\includegraphics[width=3cm]{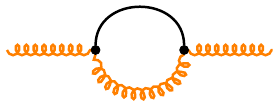}
		\end{gathered}\quad<\quad\begin{gathered}
			\includegraphics[width=3cm]{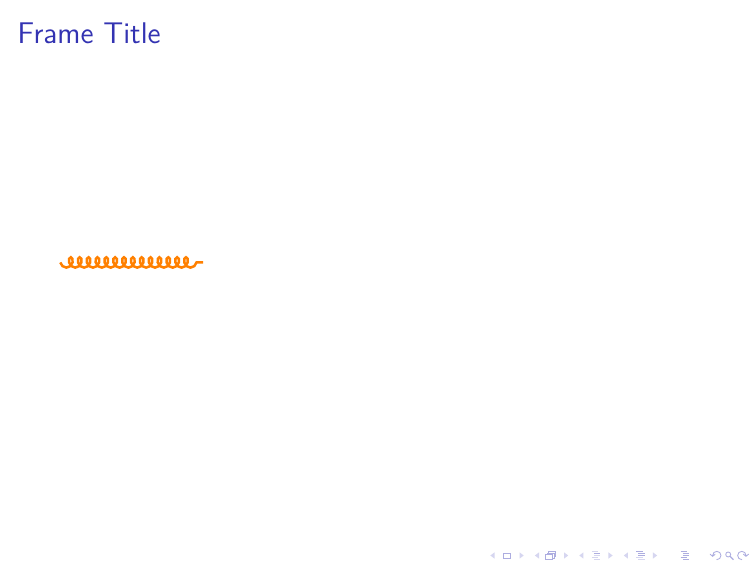}
		\end{gathered}\\
	&\Rightarrow
   \frac{1}{16\pi^2}\frac{H^2}{\Lambda_{c,s}^2}\left(\frac{1}{\mu_s^2}e^{2\pi(\kappa_s-\mu_s)}\right)^2=\frac{1}{4}P_\zeta\left(\frac{\kappa_s}{\mu_s}\right)^2\mu_s^{-2}e^{4\pi(\kappa_s-\mu_s)}<1,
\end{align}
and
\begin{align}
\nonumber&\begin{gathered}
			\includegraphics[width=3cm]{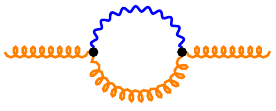}
		\end{gathered}\quad<\quad\begin{gathered}
			\includegraphics[width=3cm]{Fig/new_propgaspins}
		\end{gathered}\\
	&\Rightarrow
    \frac{s^2}{16\pi^2}\frac{H^2(\mu_s^2+\frac{(2s-3)^2}{4})^2}{M_{\rm pl}^2}\left(\frac{1}{\mu_s^2}e^{2\pi(\kappa_s-\mu_s)}\right)^2=\frac{s^2}{16 \pi^2}\left(\frac{H}{M_{\rm pl}}\right)^2\left(1+\frac{(2s-3)^2}{4\mu_s^2}\right)^2e^{4\pi(\kappa_s-\mu_s)}<1~.
\end{align}
\paragraph{EFT constraint} In addition, Since the chemical potential term is not invariant under a linear gauge transformation and the gauge invariance of the spin-$s$ field is realized non-linearly, the theory should be viewed as an EFT under the gauge symmetry breaking scale $\tilde{\Lambda}_s\sim \Lambda_{c,s}$. In the theory, there are three important energy scales: the characteristic physical momentum associated with particle production $\tilde{p}$, the mass of spin-$s$ field $m_s$, and the cutoff scale of effective theories $\Lambda_{s}$. The validity of EFT
should be under control which roughly imposes constraints on the three energy scales:
\begin{align}
    \Lambda_{s}\gtrsim\tilde{p},\quad  \Lambda_{s}\gtrsim m_s.
\end{align}
Thus EFT also sets the upper bound of the chemical potential ($\dot{\phi}_0\simeq (60H)^2$)
\begin{align}
  \kappa_s=\frac{\dot{\phi}_0}{H\Lambda_{c,s}}\simeq \frac{60^2H^2}{{\Lambda_{c,s}} H} \lesssim  \frac{60^2H}{m_s}.
\end{align}
\paragraph{Tachyonic condition} To derive additional constraints on $\kappa_s$ based on energy scale relationships, we first analyze the characteristic physical momentum at which particles are produced. The physical energy scale at which particles are produced is determined by the effective mass derived from the EoMs Eq.~(\ref{EoMhs}). When the square of the effective mass becomes negative, a tachyonic instability arises, resulting in significant particle production. To identify the tachyonic point, we must first transform the EoM Eq.~(\ref{EoMhs}) into a form that includes only a second-order derivative and the mass term. By redefining the mode function as $\chi_s\equiv a^{-(s-1)}I_s$, where we select the mode with helicity $-s$ and disregard the helicity label for simplicity, the EoM becomes
 \begin{align}
     \partial_\tau^2\chi_s^2+\left(k^2+\frac{m_s^2}{H^2}\frac{1}{\tau^2}-\frac{s(s-1)}{\tau^2}+\frac{2k}{\tau}\kappa_s\right)\chi_s=0,
 \end{align}
 where $s(s-1)H^2$ is exactly the Higuchi bound of massive spin-$s$ field in dS. Then we rewrite the above equation in terms of $x\equiv -k\tau$ and it becomes
 \begin{align}\label{chiEoM}
    \partial_x^2\chi_s^2+\left(1+\frac{m_s^2}{H^2}\frac{1}{x^2}-\frac{s(s-1)}{x^2}-\frac{2}{x}\kappa_s\right)\chi_s=0.
 \end{align}
Thus, the tachyonic point can be solved as the root of the effective mass in Eq.~(\ref{chiEoM})    
\begin{align}\label{charamom}
	x_{\text{tach}}=-(k\tau)_{\text{tach}}&\simeq\kappa_s+\sqrt{\kappa_s^2-\left(\frac{m_s^2}{H^2}-s(s-1)\right)}.
\end{align}
 To obtain a real solution, the chemical potential of the massive spin-$s$ field should satisfy  
 \begin{align}\label{kappagtrcd}
	\kappa_s^2 \geq \mu_s^2+\frac{1}{4}=\frac{m_s^2}{H^2}-s(s-1)\equiv M^2(m,s).
\end{align}
It is worth noticing that the above equation Eq.~(\ref{kappagtrcd}) actually represents a condition imposed artificially, rather than a constraint. Since when $\kappa_s<\sqrt{{m_s^2}/{H^2}-s(s-1)}$, particles can still be produced. We can use the method of Stokes lines to obtain the energy scale of particle production \cite{Sou:2021juh}, that is
\begin{align}
    \frac{\tilde{p}}{H}\simeq 0.6627M(m,s)+0.3435\kappa_s-0.0102\frac{\kappa^2_s}{M(m,s)}+0.0064\frac{\kappa_s^3}{M(m,s)^2}+\cdots,
\end{align}
where $\tilde{p}$ is the physical scale of particle production. The physical scenario can be analogized to the simple potential barrier scattering problem in quantum mechanics: even if the barrier is lower than the particle's kinetic energy, the probability of reflection of the particle is still not zero. Therefore, solving the tachyonic point is actually a classical approximation. However, when $\kappa_s<M(m,s)$, the production rate of particles is much lower relative to $\kappa_s>M(m,s)$. In order to obtain a larger gravitational wave signal, we hope that the chemical potential satisfies the condition $\kappa_s>M(m,s)$. 
From this, we can see the necessity of the massive spin-$s$ field satisfying the Higuchi bound: if $m_s^2 < s(s-1)H^2$, even if the chemical potential $\kappa_s$ tends to zero, the tachyonic point still exists, and particles are produced in large quantities, exhibiting instability. It is important to note that when $\kappa_s > M(m, s)$, the time of particle production and the variation of particle number with time cannot be directly solved using the Stokes line method. In this case, the optimal truncation breaks down \cite{Sou:2021juh}. Therefore, in our work, we consider the tachyonic point as the approximate energy scale for particle production
\begin{align}\label{pscaleparpro}
	\frac{\tilde{p}}{H}\simeq x_{\text{tach}}&=\kappa_s+\sqrt{\kappa_s^2-\left(\frac{m_s^2}{H^2}-s(s-1)\right)}= \kappa_s+\sqrt{\kappa_s^2-\left(\mu_s^2+\frac{1}{4}\right)}.
\end{align}

Moreover, the tachyonic point can serve as a natural cutoff for the integral of $x \equiv -k\tau$ in Eq.~(\ref{Eq:GWspins}). From a physical perspective, the meaning of $x$ becomes clearer when we take the tachyonic point as $x_{\rm max}$. The cutoff of the integrand in Eq.~(\ref{Eq:GWspins}) can be understood in two ways: first, with a fixed $k$, $x$ corresponds to the negative of conformal time, representing the universe's evolution time. We focus primarily on contributions after the conformal time $\tau_{\rm min}=-x_{\rm max}/k$, as this period corresponds to a tachyonic instability, indicating that the mode $k$ particles are abundantly produced from this moment onward. Alternatively, $x$ can be seen as a dimensionless physical scale, and $x_{\rm max}$ denotes the scale at which particles are primarily produced. Modes with energies exceeding this scale are less frequently produced, so we focus on modes below this energy scale. 

\paragraph{Backreaction constraint} As discussed in Sec.\;\ref{slowback}, at the scale of the Cosmic Microwave Background (CMB), experimental constraints require this period to be dominated by vacuum energy. Thus, the backreaction should be negligible; hence, the energy density of the spin-$s$ field must be less than the energy density during inflation at CMB scales. According to Eq.~(\ref{enegydensity}), the constraint we obtain is
\begin{align}
   & \rho_s\ll3M_{\rm pl}^2H^2\notag\\
	&\Rightarrow\frac{m_s^4}{16\pi^2}\left[\left(\frac{\kappa_s H}{m_s}+2\left(\frac{\kappa_s H}{m_s}\right)^3\right)\sqrt{1+\left(\frac{\kappa_s H}{m_s}\right)^2}-\mathrm{log}\left[\frac{\kappa_s H}{m_s}+\sqrt{1+\left(\frac{\kappa_s H}{m_s}\right)^2}\right]\right]e^{2\pi(\kappa_s-\mu_s)}\ll 3M_{\rm pl}^2H^2
\end{align}
and we notice that this bound is less constraining than the other bounds.

\begin{figure}[htp!]
	\centering
	\includegraphics[width=\textwidth]{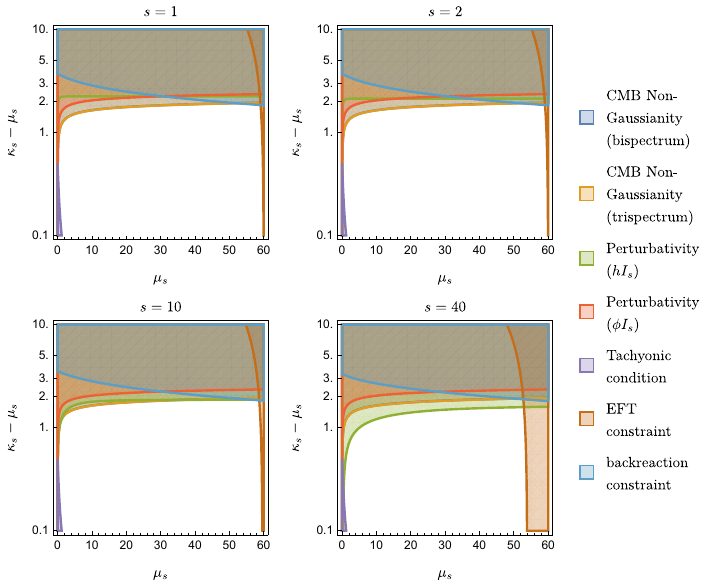}
	\caption{Parameter space satisfying experimental and theoretical constraints in the \( \kappa_s - \mu_s \) vs. \( \mu_s \) plane. The spin parameter takes values \( s = 1, 2, 10, 40 \).}\label{constraint_area_totallargespin}
\end{figure}

\paragraph{Conclusions} Using above constraints, we plot the feasible parameter space of the chemical potential $\kappa_s$ and the conformal weight $\mu_s$ in Fig.\;(\ref{constraint_area_totallargespin}).

From Fig.\;(\ref{constraint_area_totallargespin}), we observe that the perturbative constraints on the interaction term \(hI_sI_s\) become more stringent as the spin increases, and consequently, the feasible parameter range for \(\kappa_s-\mu_s\) diminishes. This is because the interaction, as described by Eq.~(\ref{intera}), is proportional to a positive power of the spin. This phenomenon can be understood as follows: when other parameters are held fixed, a larger spin results in a higher energy density of the gravitons that the field excites. On the other hand, the constraints arising from the bispectrum, trispectrum, and perturbativity of the interaction \(\phi I_sI_s\) show little variation with the spin, since the inflaton is a scalar whose coupling with the massive spin-$s$ fields does not explicitly involve spin.

Furthermore, the tachyonic condition applied imposes a constraint on the feasible parameter space, restricting the values to the lower-left region, which is nearly spin-independent, as the effective mass in Eq.~(\ref{pscaleparpro}) depends explicitly only on the chemical potential and the conformal weight. It is crucial to reiterate that the parameter space that does not satisfy the tachyonic condition is not excluded, as this condition is imposed to enhance the feasibility of experimental observations, serving merely as a phenomenological requirement.

Another important aspect is the constraint imposed by the effective theory on the chemical potential. This constraint is relatively loose for small spins. Specifically, it only becomes apparent at the boundary of \(\mu_s \sim \mathcal{O}(100)\) when \(s \sim \mathcal{O}(1)\). The exclusion line excludes the region with large $\mu_s$, and gradually moves towards the direction of small $\mu_s$ with increasing spin. Moreover, from calculations we observe that it will intersect with the perturbative constraint line of the interaction  $hI_sI_s$ when spin $s\sim 390$, covering the entire parameter space. This indicates that in our model, to ensure the perturbativity of the theory and comply with the constraint of the EFT, the spin cannot be arbitrarily large.

Finally, there is a constraint on the energy density of particles. The size of the energy density of the massive spin-$s$ field is mainly determined by the exponential factor $e^{2\pi(\kappa_s-\mu_s)}$, and is almost independent of spin. In the region where $\mu_s$ is less than $10$, the upper limit of $\kappa_s-\mu_s$ is roughly between $3.5$ and $3.2$. As $\mu_s$ decreases, the constraint becomes looser. From the analysis in Sec.\;\ref{slowback} and Fig.\;(\ref{backreation_Figure}), it can be observed that with increasing spin, the final $\kappa_s-\mu_s$ becomes smaller. For the case of spin-$1$, which has the maximum $\kappa_s-\mu_s$, it can be seen that the corresponding $\kappa_s-\mu_s$ reaches the energy density limit at approximately e-folds $N \sim 10$, corresponding to the late stage of inflation. Since the number density exponentially decreases with $\kappa_s-\mu_s$, the energy of massive fields relative to the vacuum energy can be neglected before $N \sim10$, which means that it will not change the dS spacetime background over a large range of e-folds. And for higher spins $s > 1$, the number of e-folds required to reach the energy density limit is even smaller and remains lower than the vacuum energy until the end of inflation. In regions with larger $\mu_s$, the energy density limit is stronger. Therefore, without changing the spacetime background of inflation, we expect to explore the parameter space of chemical potential enhancement in regions with smaller $\mu_s$.

\subsection{Numerical results}\label{numresult}

\begin{table}[htp!]
	\centering
	\begin{tabular}{ccccccccccccc}
		\hline\hline
        $s$& $\kappa_{(1)}$& $\mu_{(1)}$&$\kappa_{(2)}$ &$\mu_{(2)}$& $\kappa_{(3)}$& $\mu_{(3)}$& $\kappa_{(4)}$& $\mu_{(4)}$& $\kappa_{(5)}$& $\mu_{(5)}$& $\kappa_{(6)}$& $\mu_{(6)}$ \\
		\hline
		1 &2.00 &1.00&3.00 &2.00&5.50&5.00 & 6.00&5.00&6.50 &5.00&8.00 &7.00   \\
		2 &2.00 &1.00&3.00 &2.00&5.50&5.00 & 6.00&5.00&6.50 &5.00&8.00 &7.00   \\
		3&2.00 &1.00&3.00 &2.00&5.50&5.00 & 6.00&5.00&6.50 &5.00&8.00 &7.00  \\
        4&2.00 &1.00&3.00 &2.00&5.50&5.00 & 6.00&5.00&6.50 &5.00&8.00 &7.00    \\
        10&2.00 &1.00&3.00 &2.00&5.50&5.00 & 6.00&5.00&6.50 &5.00&8.00 &7.00    \\
        \hline
	\end{tabular}
	\caption{Benchmark points for gravitational wave signals. We select spins $s=1,2,3,4,10$, corresponding to six sets of different $\kappa_s$ and $\mu_s$ values. In the second, fourth and sixth sets, the difference between $\kappa_s$ and $\mu_s$ remains the same as in the first set, while the overall values decrease. In the third, fourth and fifth sets, the value of $\mu_s$ remains the same as the first set, while the difference between $\kappa_s$ and $\mu_s$ decreases. All selected parameter points satisfy the phenomenological constraints estimated in Sec.\;\ref{phenoconstraints}.}\label{tab:benchmarks}
\end{table}

\begin{table}[htp!]
	\centering
	\begin{tabular}{cccccc}
		\hline\hline
        $(\kappa,\mu)$& $\Omega_{GW}h^2~(s=1)$& $\Omega_{GW}h^2~(s=2)$&$\Omega_{GW}h^2~(s=3)$ &$\Omega_{GW}h^2~(s=4)$& $\Omega_{GW}h^2~(s=10)$ \\
		\hline
		(2.00,1.00) &$6.28\times10^{-26}$&$1.46\times10^{-24}$&$9.09\times10^{-24}$ &$3.26\times10^{-23}$&$1.76\times10^{-21}$   \\(3.00,2.00)&$2.08\times10^{-24}$&$3.25\times10^{-23}$&$2.04\times10^{-22}$ &$7.44\times10^{-22}$&$4.35\times10^{-20}$   \\(5.50,5.00)&$1.37\times10^{-24}$&$6.42\times10^{-24}$&$4.13\times10^{-23}$ &$1.85\times10^{-22}$&$2.30\times10^{-20}$  \\(6.00,5.00)&$4.65\times10^{-22}$&$1.94\times10^{-21}$&$1.02\times10^{-20}$ &$3.98\times10^{-20}$&$3.64\times10^{-18}$    \\(6.50,5.00)&$1.79\times10^{-19}$&$6.52\times10^{-19}$&$3.07\times10^{-18}$ &$1.12\times10^{-17}$&$8.51\times10^{-16}$    \\(8.00,7.00)&$3.15\times10^{-21}$&$1.02\times10^{-20}$&$3.96\times10^{-20}$ &$1.38\times10^{-19}$&$1.39\times10^{-17}$    \\
        \hline
	\end{tabular}
	\caption{Gravitational wave signals produced by massive spin-$s$ fields of different benchmark points at CMB scales. Benchmark points are listed in Tab.\;(\ref{tab:benchmarks}).}\label{tab:GWcalresult}
\end{table}
\begin{figure}[htp!]
\vspace{-0.8cm}
\begin{center}
\hspace*{-0.65cm} 
\begin{minipage}{0.5\linewidth}
\begin{center}
	\includegraphics[width=\linewidth]{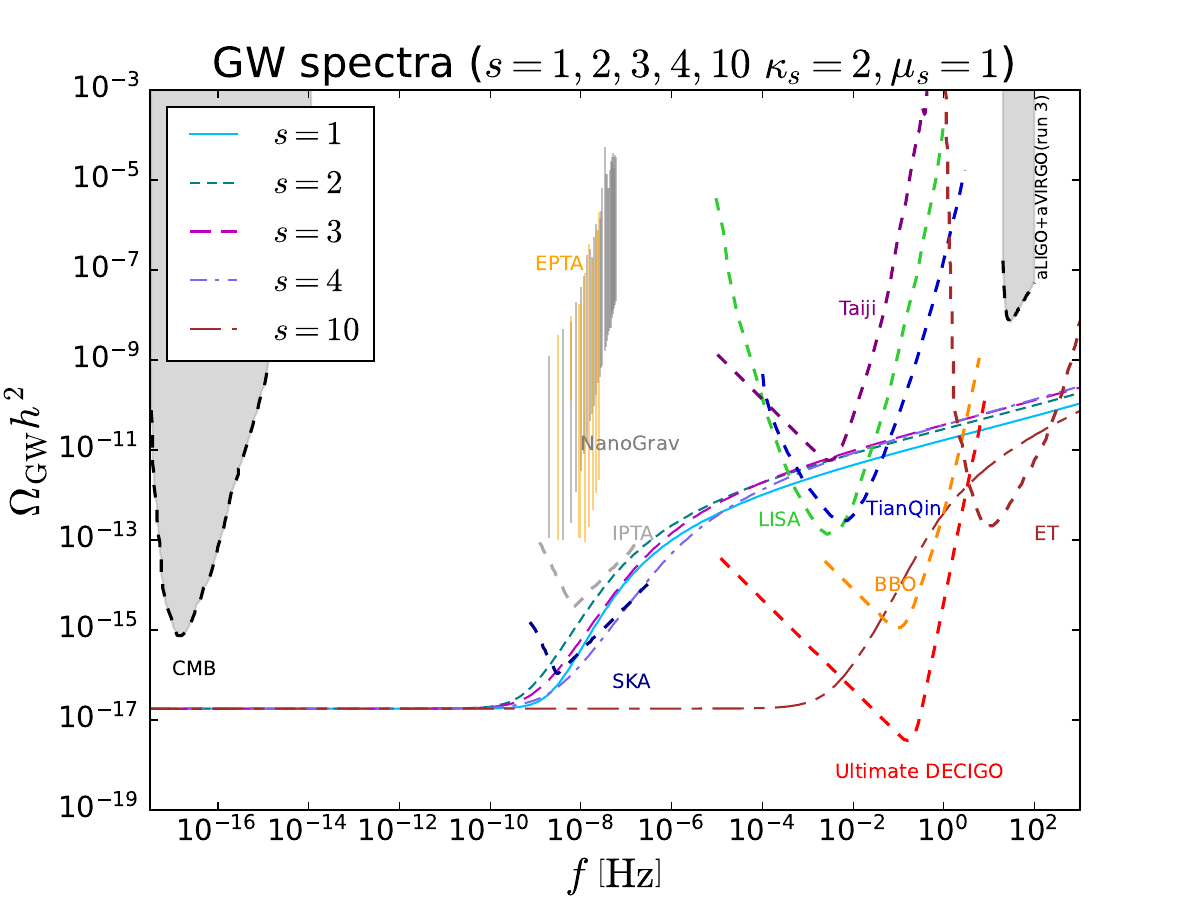}
\end{center}
\end{minipage}
\begin{minipage}{0.5\linewidth}
\begin{center}
	\includegraphics[width=\linewidth]{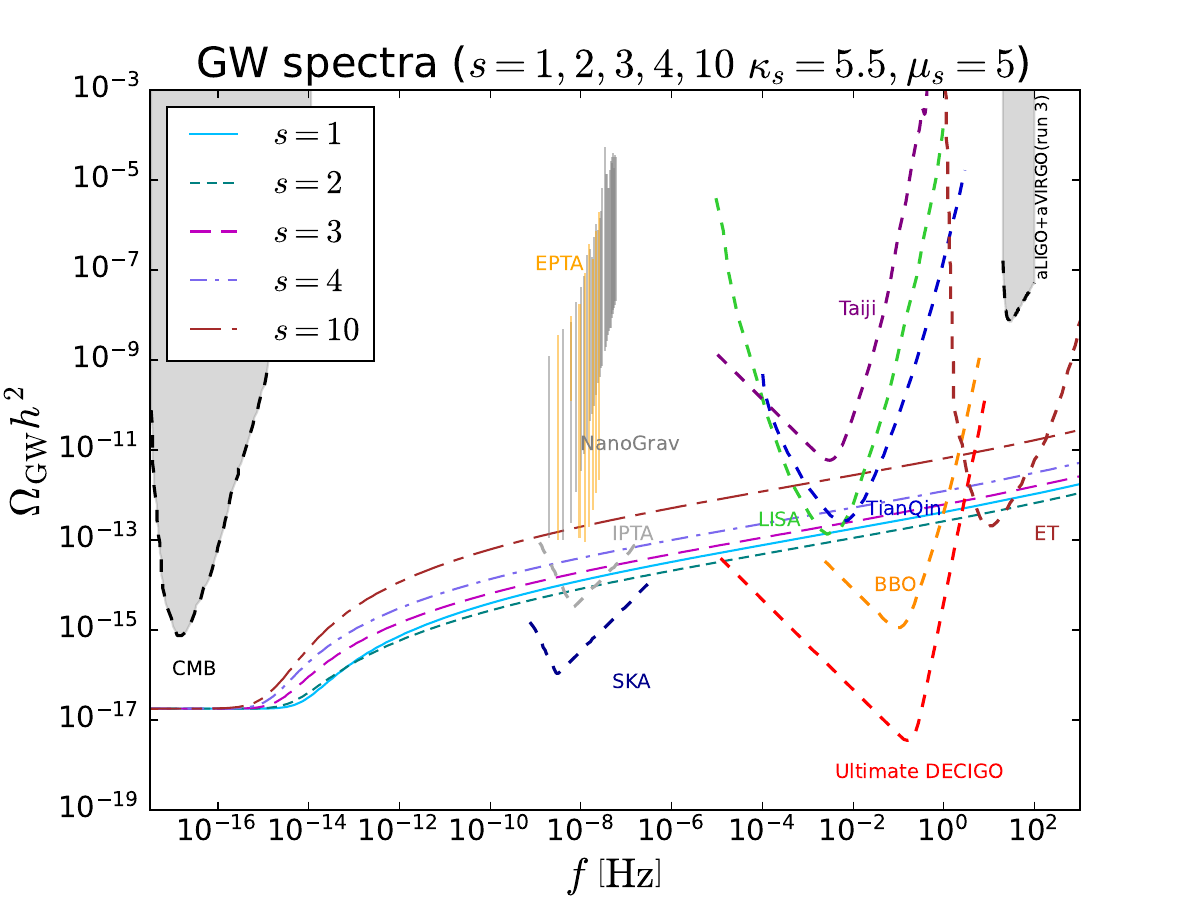}
\end{center}
\end{minipage}
\\[0.1cm]
\hspace*{-0.65cm} 
\begin{minipage}{0.5\linewidth}
\begin{center}
	\includegraphics[width=\linewidth]{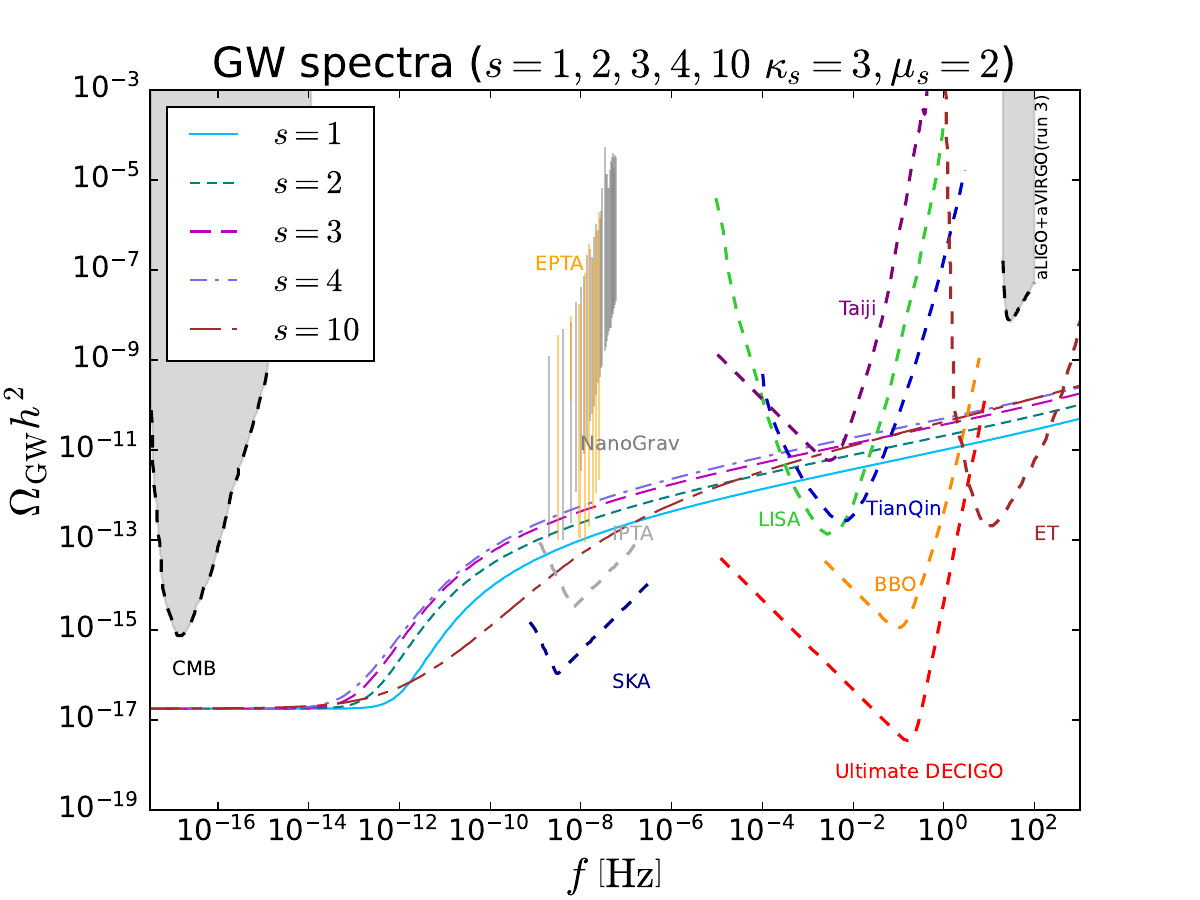}
\end{center}
\end{minipage}
\begin{minipage}{0.5\linewidth}
\begin{center}

\includegraphics[width=\linewidth]{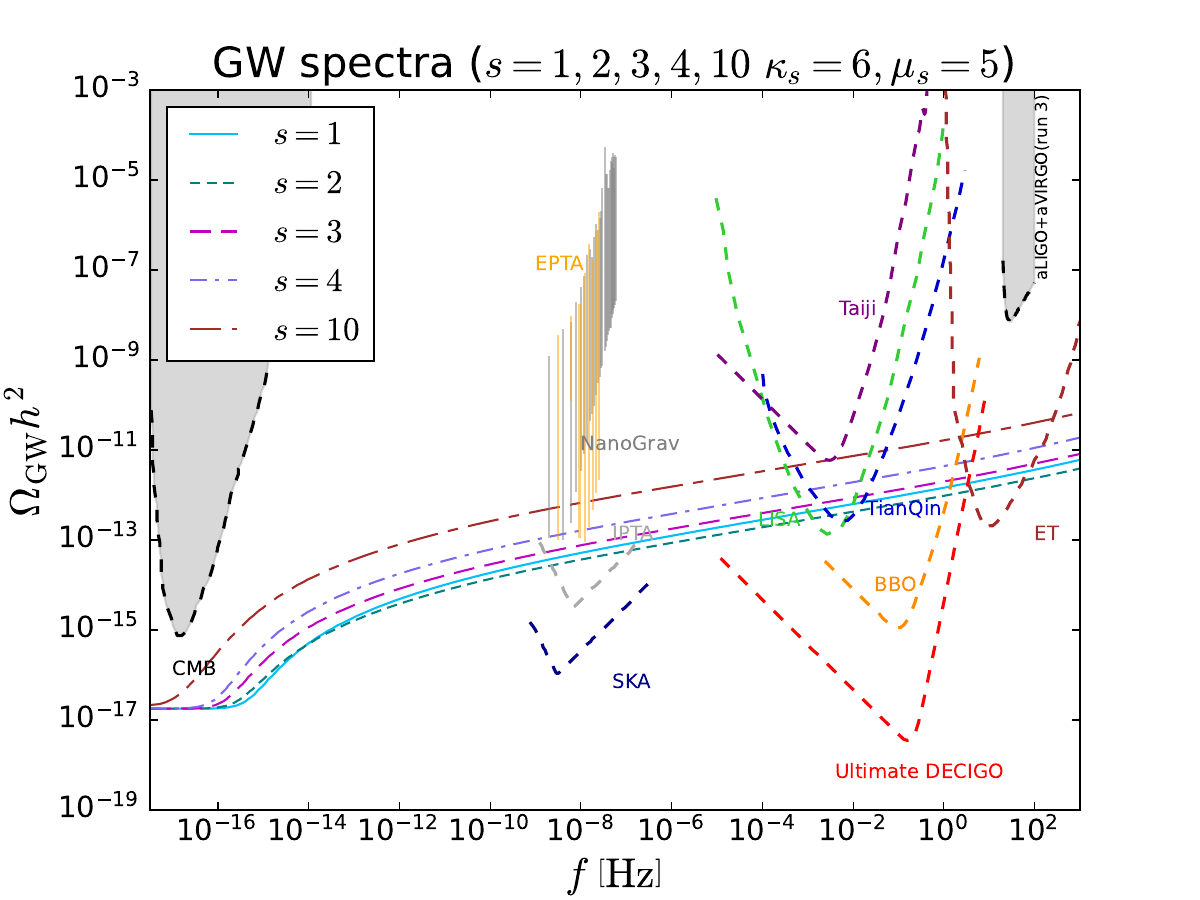}
\end{center}
\end{minipage}
\\[0.1cm]
\hspace*{-0.65cm} 
\begin{minipage}{0.5\linewidth}
\begin{center}
	\includegraphics[width=\linewidth]{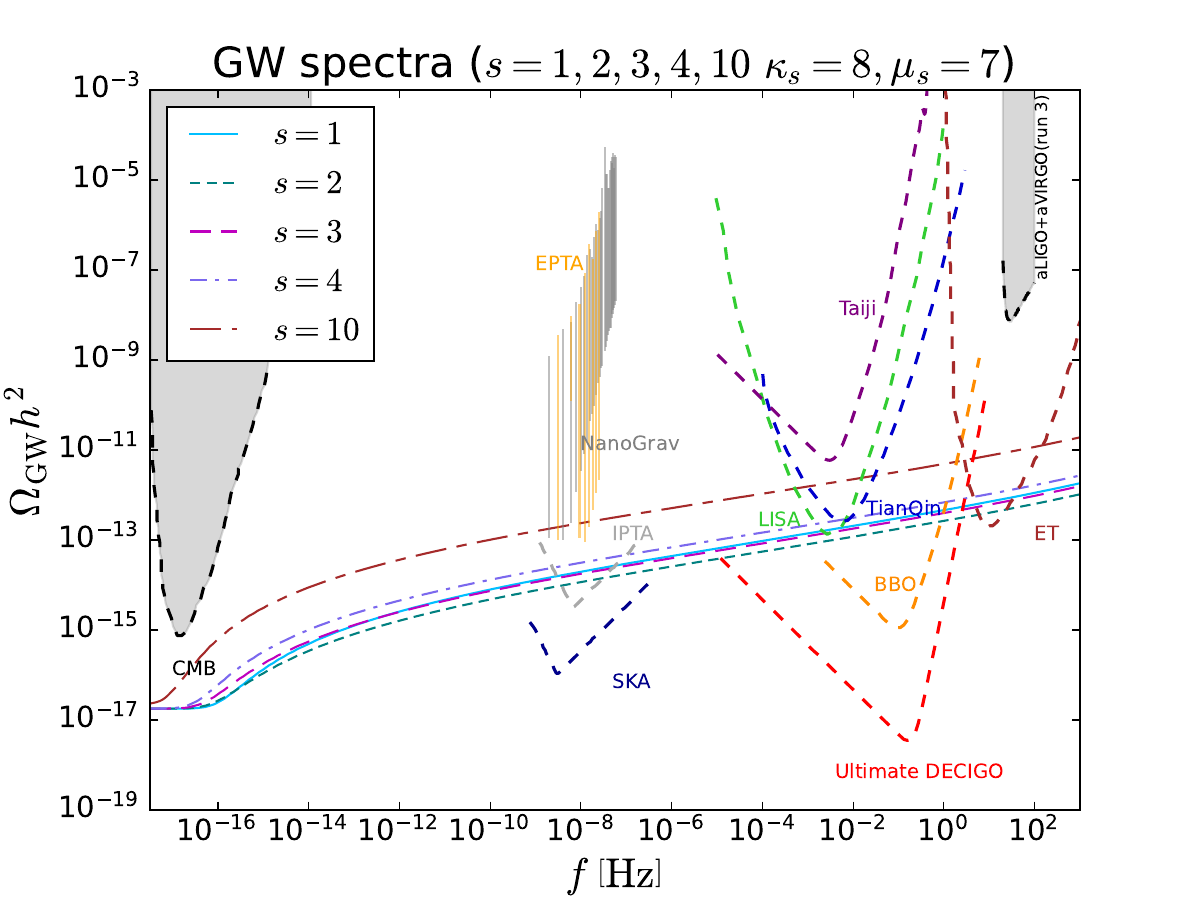}
\end{center}
\end{minipage}
\begin{minipage}{0.5\linewidth}
\begin{center}
	\includegraphics[width=\linewidth]{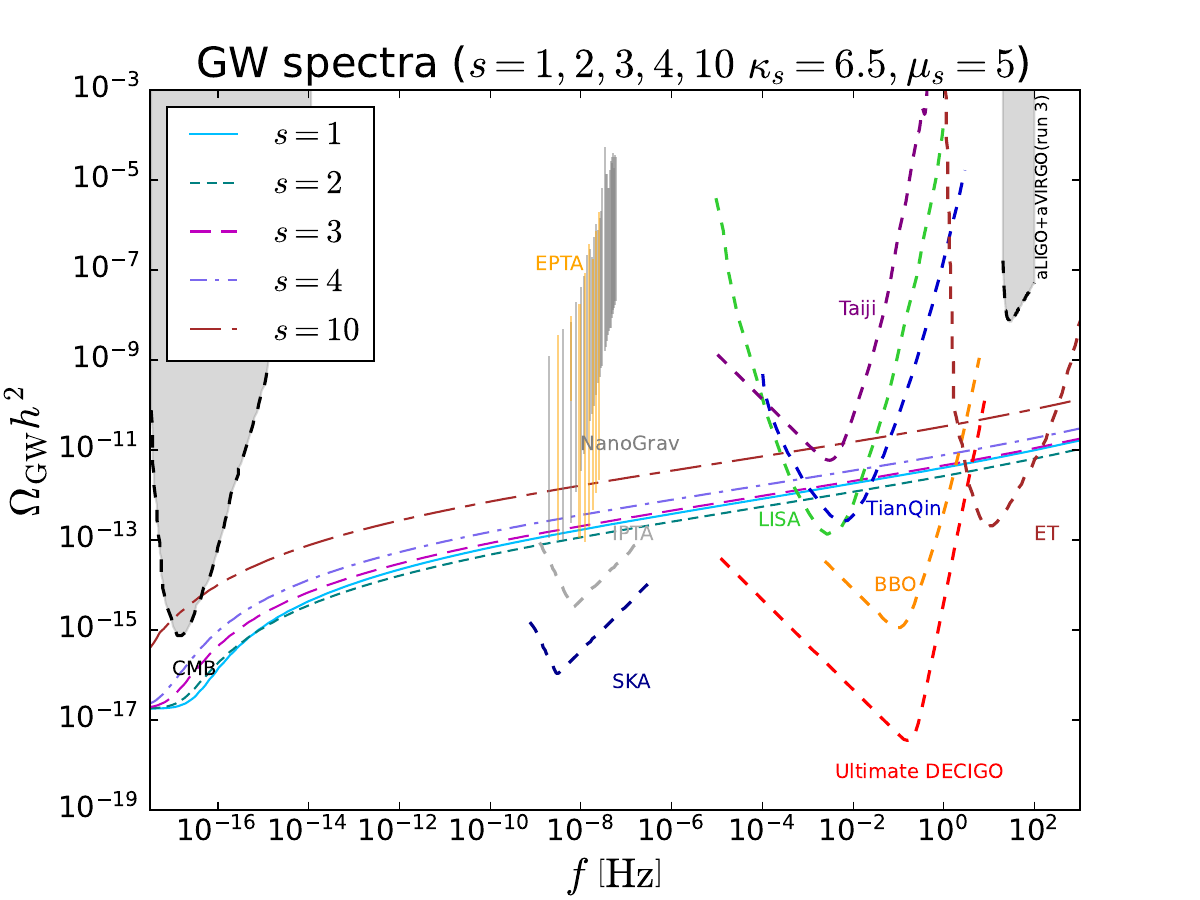}
\end{center}
\end{minipage}
\end{center}
\vspace*{-0.2cm}
\caption{The gravitational wave spectra generated by fields with different spin values. We consider spin values \( s = 1,2,3,4,10 \) with a fixed parameter \( \gamma = 0.5 \). The values of \( \kappa_s \) and \( \mu_s \) correspond to the six benchmark points listed in Tab.\;\ref{tab:benchmarks}. In the left column and the second plot of the right column, we fix \( \kappa_s - \mu_s = 1 \) while varying \( \mu_s \) as \( 1, 2, 5, 7 \). In the plots of the right column, we keep \( \mu_s = 5 \) fixed and progressively increase the difference \( \kappa_s - \mu_s \) to \( 0.5, 1, \) and \( 1.5 \). Additionally, we overlay the constraints from ongoing and proposed gravitational wave experiments. 
}\label{fig:CombinedPlot}

\end{figure}

 \begin{figure}[t!]
	\centering
	\includegraphics[width=\textwidth]{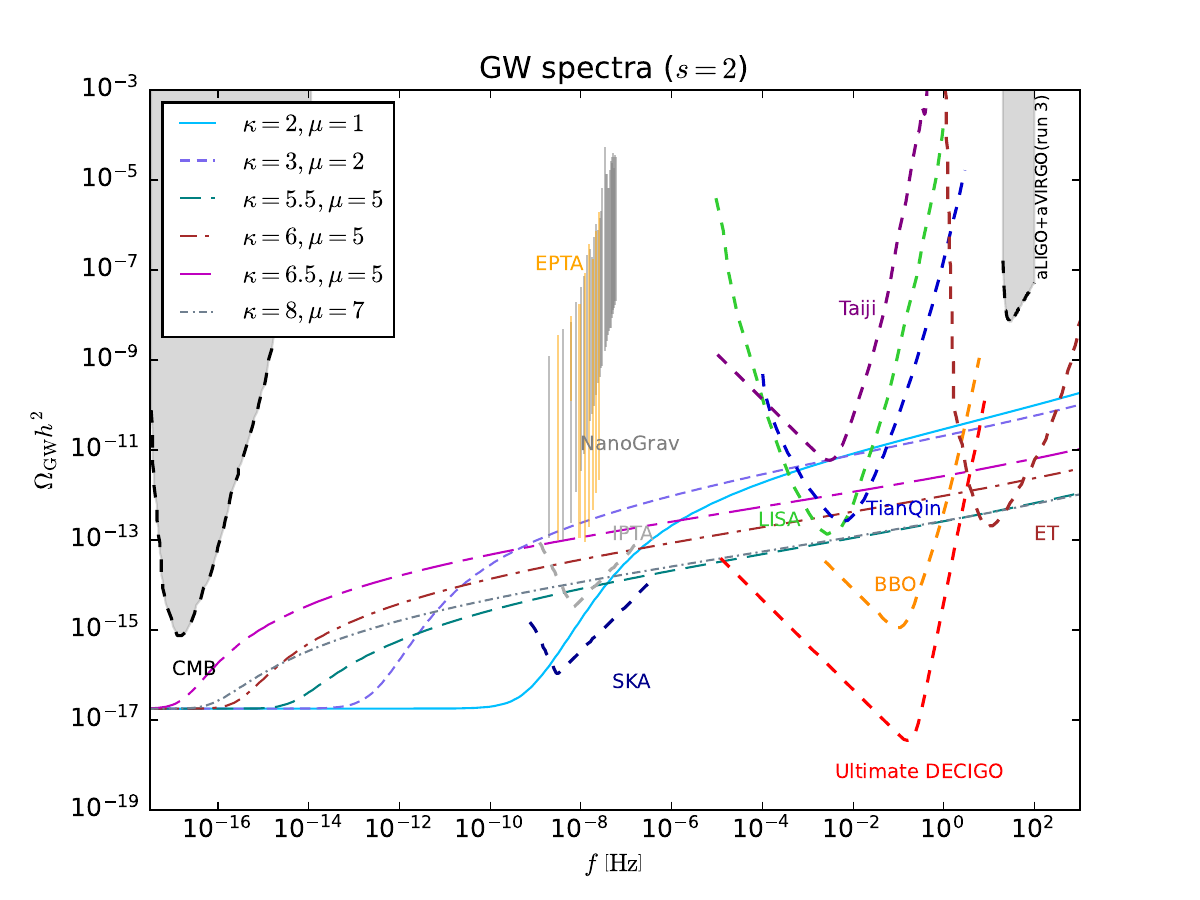}
	\caption{The gravitational wave spectra generated by the massive spin-$2$ field with $\gamma=0.5$. $\kappa$ and $\mu$ are set to be the 6 benchmark points in Tab.\;(\ref{tab:benchmarks}). We also present the constraints of ongoing and proposed gravitational wave experiments.}\label{spin-$2$_GW}
\end{figure}

 \begin{figure}[t!]
	\centering
	\includegraphics[width=\textwidth]{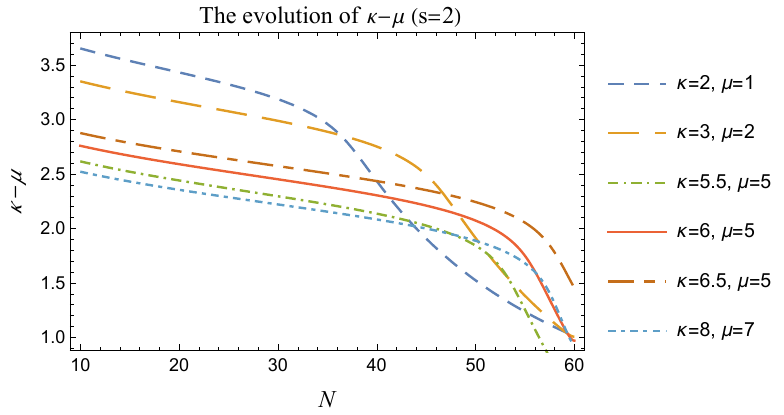}
	\caption{The difference between chemical potential $\kappa$ and conformal weight $\mu$ of massive spin-$2$ field with potential parameter $\gamma=0.5$. $\kappa$ and $\mu$ are set to be the 6 benchmark points in Tab.\;(\ref{tab:benchmarks}).}\label{spin2kappammu}
\end{figure}

\begin{figure}[t!]
	\centering
	\includegraphics[width=\textwidth]{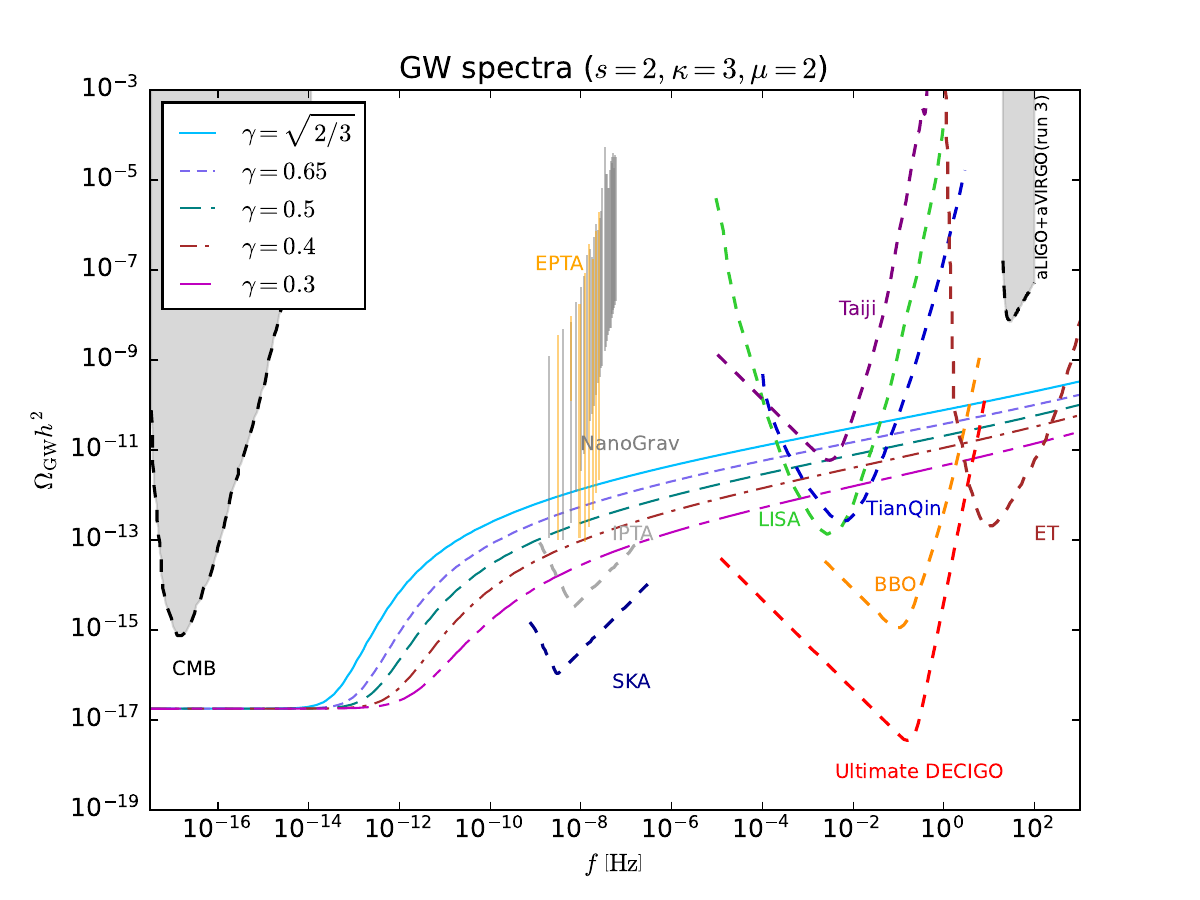}
	\caption{The gravitational wave spectra generated by fields of different types of generalized Starobinsky models. We have chosen $\gamma=\sqrt{2/3},0.65,0.5,0.4,0.3$, with fixed parameters $s=2$, $\mu=2$ and $\kappa=3$. We also present the constraints of ongoing and proposed gravitational wave experiments}\label{constraint_area_totdiff_gamma}
 \end{figure}
In this subsection, we perform a numerical computation of the gravitational wave signals generated by massive spin-$s$ fields, aiming to obtain the characteristics of these signals for various spins under differing parameters. 

At CMB scales, calculations in Sec.\;\ref{slowCMB} indicate that spin-dependent contributions to the spectral tilt arise only at first order, while the GW intensity remains significantly smaller than the leading-order signal from quantum fluctuations. Moreover, even if first-order GW tilt were detectable, the information regarding spin, chemical potential, and conformal weight is intertwined, obscuring spin-specific signatures. At smaller scales, backreaction dominates the GW signal; thus, our numerical calculations prioritize backreaction effects at small scales rather than slow-roll corrections at CMB scales.

We analyze spins $s = 1, 2, 3, 4, 10$ within the generalized Starobinsky model ($\gamma = 0.5$). Six benchmark points (Tab(\ref{tab:benchmarks})) comply with phenomenological constraints from Section~\ref{phenoconstraints}. Tab.\;(\ref{tab:GWcalresult}) summarizes CMB-scale gravitational wave energy densities, and Fig.\;(\ref{fig:CombinedPlot}) presents full-band spectra incorporating backreaction. We systematically analyze these results. In Tab.\;(\ref{tab:GWcalresult}), for the first, second, fourth, and sixth rows, as well as the three plots in the left column and the third plot in the right column of Fig.\;(\ref{fig:CombinedPlot}), we fix $\kappa_s - \mu_s = 1$ and gradually increase $\mu_s$. For the third, fourth, and fifth rows of Tab.\;(\ref{tab:GWcalresult}), along with the three plots in the right column of Fig.\;(\ref{fig:CombinedPlot}), we fix $\mu_s = 5$ and vary $\kappa_s$ across different values.  

From Tab.\;(\ref{tab:GWcalresult}), we observe that for a fixed chemical potential and conformal weight, GW intensities produced by fields with higher spins are larger. This enhancement arises because particles with higher spins excite gravitons with greater energy densities, consistent with the $s^2$-proportional GW power spectrum in Eq.~(\ref{Eq:GWspins}). Furthermore, the $x$-integration coefficients for $\partial_x W_{-i\kappa_s,i\mu_s}(-2ilx)$ scale linearly with $s-1$, reinforcing this trend. 

For a fixed spin and mass, the GW signal strength increases with $\kappa_s$ (Tab.\;(\ref{tab:GWcalresult})). This correlation stems from the fact that a larger $\kappa_s$ enhances the particle number density via the chemical potential, thereby amplifying GW emission. Additionally, for a fixed $\kappa_s - \mu_s$, the GW strength increases with increasing $\mu_s$. Physically, this can be understood as follows: while the number density of particles is governed by the factor $e^{2\pi(\kappa_s - \mu_s)}$ (Eq.~(\ref{num_density})), particles with larger $\mu_s$ (or mass) possess greater energy, leading to an elevated total energy density that facilitates stronger GW excitation. 

A more detailed analysis of the gravitational wave spectra across different frequencies is required to extract parameter information. The $\kappa_s$ and $\mu_s$ evolve with e-folds $N$, as detailed in Sec.\;\ref{slowback}. To map spectral features to frequencies, we convert e-folds $N$ to GW frequencies $f$ via \cite{Domcke:2016bkh,Niu:2022quw}:  
\begin{align}  
    N = N_{\text{CMB}} - 44.9 + \log\frac{k_{\text{CMB}}}{0.002\,\text{Mpc}^{-1}} - \log\frac{f}{100\,\text{Hz}},  
    \label{eq:Ntof}  
\end{align}  
with e-folds $N_{\text{CMB}} \sim 50\text{--}60$ and the comoving momentum $k_{\text{CMB}} = 0.002\,\text{Mpc}^{-1}$ benchmarked at CMB scales.

 The spectral morphology of gravitational waves is primarily governed by $\kappa_s - \mu_s$. Initially, tensor perturbations generated by vacuum fluctuations dominate over those produced by massive spin-$s$ fields. Neglecting slow-roll corrections, this yields a scale-invariant GW spectrum. As $\kappa_s - \mu_s$ increases, the tensor perturbations generated by massive spin-$s$ fields grow, eventually exceeding vacuum-fluctuation dominance. This transition marks the onset of a rapid spectral amplification phase at higher frequencies. Subsequent slow growth of $\kappa_s - \mu_s$ causes GW intensity to a flatter spectral shape. The spectrum is thus characterized by: (i) rapid-phase onset frequency, (ii) rapid-phase slope, (iii) slow-phase onset frequency, and (iv) slow-phase slope.  
 
Spectral morphology variations arise from the interplay between two competing factors: (i) initial amplitudes at CMB scales and (ii) inflationary parameter evolution (backreaction). We first analyze spin dependencies: while higher spins enhance CMB-scale GW intensities at fixed $\kappa_s$, $\mu_s$, they also induce stronger backreaction at early times. This suppresses initial spectral growth rates but amplifies subsequent growth as the backreaction relaxes. 

Fig.\;(\ref{fig:CombinedPlot}) categorizes different cases based on the difference in power spectrum intensities at CMB scales and the gravitational wave intensity generated by vacuum fluctuations, which determines the exit time from the scale-invariant phase. The quantitative analysis focuses only on spin-$10$ and spin-$1$ cases, as smaller spin differences make power spectra trends difficult to determine through qualitative comparison, necessitating numerical resolution.

The first case occurs when the difference is relatively small, corresponding to $(\kappa_s, \mu_s) = (2,1)$. In this case, the field with higher spin initially dominates over its low-spin counterpart but experiences a delayed exit from the scale-invariant stage due to slower growth. This delay allows the low-spin GWs to overtake the high-spin signal before becoming comparable with the vacuum-fluctuation contribution. As a result, the higher-spin signal exits the scale-invariant phase later, and afterward, the backreaction becomes relatively smaller compared to the lower-spin counterpart, leading to a steeper slope in the rapid rise phase of $f$. 

In the second case, such as $\kappa_s=3$, $\mu_s=2$, the reduced CMB-scale disparity between massive spin-$s$ fields and vacuum fluctuations shortens transition times. The GW spectrum produced by the field with higher spin exits scale invariance earlier, but its growth rate remains lower, and it only surpasses the lower-spin counterpart at higher frequencies.

The final case corresponds to other parameter choices, where the exit time is even shorter compared to the previous cases, enabling the higher-spin signal to surpass the lower-spin signal at sufficiently low frequencies. Thus, the growth rate of GWs from the field with higher spin surpasses that of the lower-spin counterpart before the latter exits the scale-invariant phase.

For fixed spin, variations in chemical potential ($\kappa_s$) and conformal weight ($\mu_s$) produce distinct GW spectral features, as shown for spin-2 fields in Fig.\;(\ref{fig:CombinedPlot}). To isolate $\kappa_s$- and $\mu_s$-dependent effects, Fig.\;(\ref{spin-$2$_GW}) contrasts GW spectra across parameter pairs, while Fig.\;(\ref{spin2kappammu}) tracks the time-evolution of $\kappa_s - \mu_s$. At fixed $\kappa_s - \mu_s$ (CMB scales), larger $\mu_s$ enhances GWs and accelerates exit from the scale-invariant regime. However, increased $\mu_s$ amplifies the source term in Eq.~(\ref{br2}), intensifying backreaction on the background field. This feedback suppresses the field's rolling acceleration, throttling the growth rate of $\kappa_s - \mu_s$. This effect is reflected in the gravitational wave spectrum as a smaller slope in the growth stage of GWs. Fixing $\mu$ while increasing $\kappa$ leads to a similar trend in the variation of the gravitational wave spectrum. Thus, both $\kappa_s$ and $\mu_s$ modulate GW morphology through competitive source-amplification and growth-suppression mechanisms.

We further compute GW spectra for various generalized Starobinsky models, focusing on the case of $s=2$ with $(\kappa_s, \mu_s)=(3,2)$ (Fig.\;(\ref{constraint_area_totdiff_gamma})). Key findings reveal that for larger values of the potential parameter $\gamma$, GWs leave the scale-invariant stage and enter the rapid-growth stage earlier, while steepening the spectral slope in this phase. However, after entering the slow increase stage, differences in the spectral slopes for different $\gamma$ values diminish. This behavior originates from $\gamma$-dependent potential dynamics: larger $\gamma$ increases the potential's steepness within the range of $\phi_0$ we consider, accelerating the background field's rolling velocity and enhancing $\kappa_s$ growth. Thus, this allows spin-$s$ tensor perturbations to dominate vacuum fluctuations earlier, hastens the scale-invariant exit and intensifies the rapid-growth slope. However, due to stronger backreaction, the growth rate of $\kappa_s - \mu_s$ decreases over time, leading to a reduced GW enhancement in later stages.

A key feature of the primordial gravitational waves generated by massive spin-$s$ particles during inflation is that the power spectrum maintains a consistent positive slope over a broad frequency range. Moreover, the power spectrum exhibits a steeper slope at lower frequencies and flattens at higher frequencies. This characteristic distinguishes it from gravitational waves produced by phase transitions, cosmic strings, and other mechanisms. 

Since it covers a wide range of frequencies, correlation analysis across multiple detectors in different frequency bands serves as an effective method for detecting gravitational wave signals and extracting parameter information. Observations from EPTA, NanoGrav, IPTA, and SKA in the nanohertz range, along with planned future laser interferometers such as LISA, Taiji, Tianqin, BBO, and DECIGO in the millihertz to hertz range, provide crucial data. In the 10–100 Hz frequency range, gravitational wave signals can be detected by ET, as well as the aLIGO and AdVirgo network. 

Furthermore, a detailed analysis of the power spectrum—such as template matching—could provide insights into the spin, chemical potential, and mass of the sources.

	\section{Conclusion and outlook}\label{ConclusionsSect}
	In this paper, we investigate primordial gravitational waves produced by massive spin-\(s\) bosons, which are enhanced by a chemical potential. The analysis begins with the calculation of a massive spin-$2$ field. An effective chemical potential operator with the lowest mass dimension is introduced based on the requirements of linearity, symmetry, and consistency \cite{Tong:2022cdz}. The study is then extended to arbitrary integer spins, deriving the general EoMs and mode functions for massive spin-$s$ fields with the chemical potential. Our investigation demonstrates that, at the linear theory level, the phenomenon of chemical-potential discontinuity is general for integer spins. This phenomenon is discussed and, in conjunction with the literature \cite{Tong:2022cdz}, an explanation for its occurrence is provided. 
	
	The effects of the slow-roll correction and the backreaction effect are also calculated. The gravitational wave spectrum is predominantly influenced by the backreaction effect. Taking into account the backreaction effect and under the constraints of experimental observations and theoretical consistency, the power spectrum of GWs produced by the mode with helicity $\lambda=-s$ of the massive spin-$s$ field ($\kappa_s>0$) is computed. The gravitational wave power spectrum is numerically computed for different spin, chemical potential, and conformal weight parameters (Fig.\;(\ref{fig:CombinedPlot})). And the impact of variations in the shape of the inflaton potential on the gravitational wave power spectrum (Fig.\;(\ref{constraint_area_totdiff_gamma})) is explored. The characteristic feature of the gravitational wave power spectrum produced by the massive spin-$s$ field is that it is approximately scale-invariant at CMB scales, thereby evading the stringent constraints from CMB observations while exhibiting increased signal strength at small scales, covering a wide frequency range. Consequently, it may be detected by both current and future gravitational wave experiments. A semi-quantitative analysis of the gravitational wave spectrum's characteristics under different parameters is performed. The analysis suggests that by employing a combination of various gravitational wave experiments, it may be possible to distinguish the spin, chemical potential, and conformal weight of the signal source.

 Furthermore, our paper identifies several unresolved issues and avenues for future research. Firstly, the phenomenon of the chemical-potential discontinuity observed in the computation of mode functions for the massive spin-$s$ field is not fully understood. The underlying physical reason for this phenomenon remains unclear and warrants deeper investigation, such as extending the theory to nonlinear orders. Secondly, in our phenomenological calculations, we employ approximations to simplify calculations of the constraints and gravitational wave spectra. To facilitate future comparisons with experimental data and to extract effective parameter information, more precise numerical calculations are needed. Specifically, we need to determine the relationship between the slope of the power spectrum at small scales and the relevant parameters. Lastly, in our paper, we derive an effective chemical potential operator using an effective field theory approach,
thus a further comprehensive exploration of the corresponding UV theory for this effective operator is also warranted.

	\acknowledgments
	We thank Chon Man Sou and Xi Tong for useful discussions.
 HA is supported in part by the National Key R\&D
Program of China under Grant No.\ 2023YFA1607104
and 2021YFC2203100, the National Science Foundation of China (NSFC) under Grant No.\ 11975134, and
the Tsinghua University Dushi Program.
 ZX is supported by NSFC under Grants No.\ 12275146 and No.\ 12247103, the National Key R\&D Program of China (2021YFC2203100), and the Dushi Program of Tsinghua University.

	\appendix
	\section{The Feynman rules}\label{FeynmanRuleAppendix}

 In our model, the interactions between the inflaton, the graviton, and the massive spin-$s$ field are restricted to two specific types: the three-vertex coupling between the inflaton and the massive spin-$s$ field generated by the chemical potential term, and the three-vertex coupling between the graviton and the massive spin-$s$ field generated by the minimal coupling\footnote{In fact, the model may also include effective interactions in the forms of $hI_s$, $\phi h I_s$, and $\phi\phi I_s$, particularly in effective theories where the spontaneous breaking of time translation symmetry occurs at a low scale \cite{Lee:2016vti,Dimastrogiovanni:2018uqy}. However, to reduce the additional assumptions required for theoretical construction and to simplify calculations and discussions, we consider only minimal coupling.}, considering only the highest helicity modes with $s$ spatial indices
 \begin{align}\label{interspins}
     \mathcal{L}^{\phi I_sI_s}_{\text{int}}&=-\epsilon^{nrm}a^{-2(s-1)}(\tau)\frac{\partial_\tau\phi}{2\Lambda_{c,s}}I_{ni_1\cdots i_{s-1}}\partial_rI_{mi_1\cdots i_{s-1}}\\
      \mathcal{L}^{h I_sI_s}_{\text{int}}&=-sa^{-2(s-1)}(\tau)h^{TT}_{ij}\partial_\tau I_{ii_1\cdots i_{s-1}}\partial_\tau I_{ji_1\cdots i_{s-1}}.
 \end{align}
 We write down the Feynman rules based on the Schwinger-Keldysh formalism \cite{Chen:2017ryl}. First, we write down the propagators for the inflaton, the graviton, and the massive spin-$s$ field (we only consider the highest helicity modes with $s$ spatial indices)
 \begin{align}\label{intera}
		&~~~~~\begin{gathered}
			\includegraphics[width=3.2cm]{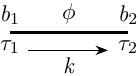}
		\end{gathered}=~P^{b_1b_2}(\tau_1,\tau_2,k)~,\\
		&\begin{gathered}
			\includegraphics[width=4cm]{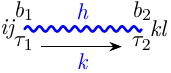}
		\end{gathered}=~G_{ij,kl}^{b_1b_2}(\tau_1,\tau_2,\mathbf{k})~,\\
		&\begin{gathered}
			\includegraphics[width=4cm]{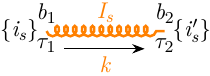}
		\end{gathered}=~I_{\{i_s\},\{i^\prime_s\}}^{b_1b_2}(\tau_1,\tau_2,\mathbf{k})~,
	\end{align}
where $\{i_s\}\equiv i_1i_2\cdots i_s$ and $b=\pm$ denote the time-ordering of the propagators which are all built from the Wightman functions
	\begin{align}
		P^{-+}(\tau_1,\tau_2,k)&=\varphi(\tau_1,k)\varphi^\star(\tau_2,k)~,\\
		G_{ij,kl}^{-+}(\tau_1,\tau_2,\mathbf{k})&=\sum_{\lambda=\pm 2}h^\lambda(\tau_1,k)\epsilon_{ij}^\lambda(\mathbf{\hat{k}})\left[h^\lambda(\tau_2,k)\epsilon_{kl}^\lambda(\mathbf{\hat{k}})\right]^\star~,\\
		 I_{\{i_s\},\{i^\prime_s\}}^{-+}(\tau_1,\tau_2,\mathbf{k})&=\sum_{\lambda=\pm s}I_s^{\lambda}(\tau_1,k)\epsilon_{i_1\cdots i_s}^\lambda(\mathbf{\hat{k}})\left[I_s^\lambda(\tau_2,k)\epsilon_{i^\prime_1\cdots i^\prime_s}^\lambda(\mathbf{\hat{k}})\right]^\star,
	\end{align}
 where the mode functions of the inflaton, the graviton and the spin-$s$ field are
 \begin{align}
     \phi(\tau,k)&=\frac{H}{\sqrt{2k^3}}\left(1+ik\tau\right)e^{-ik\tau}\\
       h^{\lambda}(\tau,k)&=\frac{2H}{M_{\rm pl}\sqrt{2k^3}}(1+ik\tau)e^{-ik\tau}\\
      I^{\pm s}_s(\tau,k)&=\frac{e^{\mp\frac{\pi\kappa_s}{2}}}{2^{\frac{s}{2}}H^{s-1}\sqrt{k}}(-1)^{s-1}\tau^{1-s}W_{\pm\kappa_s,i\mu_s}(2ik\tau).
 \end{align}
 The vertex rules of 3-vertex interactions Eq.~(\ref{interspins}) are given by
\begin{align}
	\begin{gathered}
	\includegraphics[width=5cm]{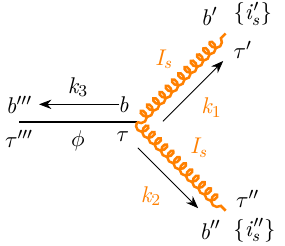}
		\end{gathered}
		&\begin{gathered}
         =\frac{b}{2\Lambda_{c,s}}\epsilon^{nrm}(\mathbf{k}_1-\mathbf{k}_2)_r\int d\tau a^{-2(s-1)}(\tau)\partial_\tau  P^{b'''b}(\tau''',\tau,k_3)\\
         \qquad\qquad\qquad\qquad\qquad\times I_{ni_1\cdots i_{s-1},i^\prime_1\cdots i^\prime_s}^{bb'}(\tau,\tau',\mathbf{k}_1)I_{mi_1\cdots i_{s-1},i^{\prime\prime}_1\cdots i^{\prime\prime}_s}^{bb''}(\tau,\tau'',\mathbf{k}_2)~,
   \end{gathered}
\end{align}
\begin{align}\label{verticeIIh}
		\begin{gathered}
	\includegraphics[width=5cm] {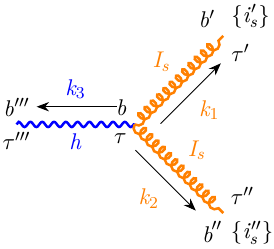}
		\end{gathered}
		&\begin{gathered}
		\\=-i{b}s\int d\tau a^{-2(s-1)}(\tau)  G^{b'''b}_{ij,kl}(\tau''',\tau,\mathbf{k}_3)\\
			\qquad\qquad\times \partial_\tau I_{ki_1\cdots i_{s-1},i^\prime_1\cdots i^\prime_s}^{bb'}(\tau,\tau',\mathbf{k}_1) \partial_\tau I_{li_1\cdots i_{s-1},i^{\prime\prime}_1\cdots i^{\prime\prime}_s}^{bb''}(\tau,\tau'',\mathbf{k}_2)~,
		\end{gathered}
\end{align}

\section{Propagator estimation}\label{Prop_estim}
The complexity of special functions and the breaking of time translation symmetry make the calculation of correlation functions in dS space challenging. Furthermore, our analysis involves loop diagram calculations, which are complex and extremely time-consuming even with numerical methods. Another significant challenge is the introduction of the chemical potential. The chemical potential explicitly breaks dS boost symmetry, rendering many analytical methods such as bootstrap and spectral decomposition \cite{Xianyu:2022jwk} ineffective. However, it can be observed that the dependence of the correlation functions on $\kappa_s$ and $\mu_s$ primarily lies in the exponential factor $e^{\pi(\kappa_s-\mu_s)}$, where the difference between $\kappa_s$ and $\mu_s$ dictates the order of magnitude of the correlation function. Therefore, we can obtain relatively accurate constraints on $\kappa_s$ and $\mu_s$ through some estimation methods following the literature \cite{Wang:2020ioa}. The method of estimating the correlation function can be summarized as
\begin{align}
    \sim \text{time integral}\times\text{vertices}\times \text{propagators}\times\text{loop factors}
\end{align} 
 For the propagator, we initially examine the late-time limit, as this portion of the mode function predominantly contributes to the signa. The late-time expansion of the propagator for the helicity-\(\pm s\) mode is expressed as
\begin{align}
	&I_{s}^{\pm s}(k,\tau_1)I_{s}^{\pm s\star}(k,\tau_2)\notag\\
	&=\left(\frac{e^{\mp\frac{\pi\kappa_s}{2}}}{2^{\frac{s}{2}}H^{(s-1)}}\right)^2(1-i)\left[\frac{(2k)^{i\mu_s}e^{\frac{\pi\mu_s}{2}}\Gamma\left[-2i\mu_s\right]}{\Gamma[\frac{1}{2}\mp i\kappa_s-i\mu_s]}(-\tau)^{\frac{3}{2}-s+i\mu_s}+\frac{(2k)^{-i\mu_s}e^{-\frac{\pi\mu_s}{2}}\Gamma[2i\mu_s]}{\Gamma[\frac{1}{2}\mp i\kappa_s+i\mu_s]}(-\tau)^{\frac{3}{2}-s-i\mu_s}\right]\notag\\
	&\times(1+i)\left[\frac{(2k)^{-i\mu_s}e^{\frac{\pi\mu_s}{2}}\Gamma\left[+2i\mu_s\right]}{\Gamma[\frac{1}{2}\pm i\kappa_s+i\mu_s]}(-\tau)^{\frac{3}{2}-s-i\mu_s}+\frac{(2k)^{i\mu_s}e^{-\frac{\pi\mu_s}{2}}\Gamma[-2i\mu_s]}{\Gamma[\frac{1}{2}\pm i\kappa_s-i\mu_s]}(-\tau)^{\frac{3}{2}-s+i\mu_s}\right]\notag\\
	&=\frac{e^{\mp{\pi\kappa_s}}}{(2H^2)^{(s-1)}}\left[\frac{e^{\pi\mu_s}\Gamma\left[-2i\mu_s\right]\Gamma\left[2i\mu_s\right]}{\Gamma[\frac{1}{2}\mp i\kappa_s-i\mu_s]\Gamma[\frac{1}{2}\pm i\kappa_s+i\mu_s]}(\tau_1\tau_2)^{\frac{3}{2}-s}\left(\frac{\tau_1}{\tau_2}\right)^{i\mu_s}\right.\notag\\
	&+\frac{(2k)^{2i\mu_s}\Gamma^2\left[-2i\mu_s\right]}{\Gamma[\frac{1}{2}\mp i\kappa_s-i\mu_s]\Gamma[\frac{1}{2}\pm i\kappa_s-i\mu_s]}(\tau_1\tau_2)^{\frac{3}{2}-s}\left({\tau_1}{\tau_2}\right)^{i\mu_s}\notag\\
	&+\frac{(2k)^{-2i\mu_s}\Gamma^2\left[2i\mu_s\right]}{\Gamma[\frac{1}{2}\mp i\kappa_s+i\mu_s]\Gamma[\frac{1}{2}\pm i\kappa_s+i\mu_s]}(\tau_1\tau_2)^{\frac{3}{2}-s}\left({\tau_1}{\tau_2}\right)^{-i\mu_s}\notag\\
	&\left.+\frac{e^{-\pi\mu_s}\Gamma\left[-2i\mu_s\right]\Gamma\left[2i\mu_s\right]}{\Gamma[\frac{1}{2}\mp i\kappa_s+i\mu_s]\Gamma[\frac{1}{2}\pm i\kappa_s-i\mu_s]}(\tau_1\tau_2)^{\frac{3}{2}-s}\left(\frac{\tau_1}{\tau_2}\right)^{-i\mu_s}\right],
\end{align}
which can be separated into non-local and local parts
\begin{align}
&D^{\pm s}|_{\text{non-local}}=\frac{(\tau_1\tau_2)^{\frac{3}{2}-s}}{(2H^2)^{(s-1)}}\left[\frac{e^{\mp{\pi\kappa_s}}(2k)^{2i\mu_s}\Gamma^2\left[-2i\mu_s\right]}{\Gamma[\frac{1}{2}\mp i\kappa_s-i\mu_s]\Gamma[\frac{1}{2}\pm i\kappa_s-i\mu_s]}\left({\tau_1}{\tau_2}\right)^{i\mu_s}+(\mu_s\rightarrow-\mu_s)\right]\notag\\
&D^{\pm s}|_{\text{local}}=\frac{(\tau_1\tau_2)^{\frac{3}{2}-s}}{(2H^2)^{(s-1)}}\left[\frac{e^{\pi(\mu_s\mp\kappa_s)}\Gamma\left[-2i\mu_s\right]\Gamma\left[2i\mu_s\right]}{\Gamma[\frac{1}{2}\mp i\kappa_s-i\mu_s]\Gamma[\frac{1}{2}\pm i\kappa_s+i\mu_s]}\left(\frac{\tau_1}{\tau_2}\right)^{i\mu_s}+\frac{e^{\pi(-\mu_s\pm\kappa_s)}\Gamma\left[-2i\mu_s\right]\Gamma\left[2i\mu_s\right]}{\Gamma[\frac{1}{2}\mp i\kappa_s+i\mu_s]\Gamma[\frac{1}{2}\pm i\kappa_s-i\mu_s]}\left(\frac{\tau_1}{\tau_2}\right)^{-i\mu_s}\right].
\end{align}
The local part can be written as
\begin{align}
	D^{\pm s}|_{\text{local}}=\frac{(\tau_1\tau_2)^{\frac{3}{2}-s}}{(2H^2)^{(s-1)}2\mu_s}\left[\frac{1+e^{2\pi(\mp\kappa_s-\mu_s)}}{1-e^{-4\pi\mu_s}}\left(\frac{\tau_1}{\tau_2}\right)^{i\mu_s}-\frac{1+e^{2\pi(\mp\kappa_s+\mu_s)}}{1-e^{4\pi\mu_s}}\left(\frac{\tau_1}{\tau_2}\right)^{-i\mu_s}\right].
\end{align}
And in the large mass limit $\mu_s\gg1$, the local part yields
\begin{align}
	D^{\pm s}|_{\text{local}}=-\frac{(\tau_1\tau_2)^{\frac{3}{2}-s}}{(2H^2)^{(s-1)}2\mu_s}e^{2\pi(\mp\kappa_s-\mu_s)}2\mathrm{cos}\left(\mu_s\mathrm{log}\left(\frac{\tau_1}{\tau_2}\right)\right).
\end{align}
Subsequently, for the non-local part, we consider the conditions $\mu_s \gg 1$ and $\kappa_s > \mu_s$, employing the asymptotic expansion of the $\Gamma$ function. The part containing the $\Gamma$ functions yields
\begin{align}
	&\frac{e^{{\mp\pi\kappa_s}}\Gamma^2\left[-2i\mu_s\right]}{\Gamma[\frac{1}{2}+ i\mp\kappa_s-i\mu_s]\Gamma[\frac{1}{2}- i\mp\kappa_s-i\mu_s]}\left({4k^2\tau_1}{\tau_2}\right)^{i\mu_s}+(\mu_s\rightarrow-\mu_s)\notag\\
	&=\frac{e^{2\pi(\mp\kappa_s-\mu_s)}}{2\mu_s}e^{i\left[-(\mp\kappa_s-\mu_s)\mathrm{log}(\mp\kappa_s-\mu_s)+(\mp\kappa_s+\mu_s)\mathrm{log}(\mp\kappa_s+\mu_s)-2\mu_s\mathrm{log}(2\mu_s)\right]}\left(4k^2\tau_1\tau_2\right)^{i\mu_s}\notag\\
	&-\frac{e^{2\pi(\mp\kappa_s-\mu_s)}}{2\mu_s}e^{i\left[-(\mp\kappa_s+\mu_s)\mathrm{log}(\mp\kappa_s+\mu_s)+(\mp\kappa_s-\mu_s)\mathrm{log}(\mp\kappa_s-\mu_s)+2\mu_s\mathrm{log}(2\mu_s)\right]}\left(4k^2\tau_1\tau_2\right)^{-i\mu_s}\notag\\
	&=\frac{e^{2\pi(\mp\kappa_s-\mu_s)}}{2\mu_s}2\mathrm{Re}\left[e^{i\left[-(\mp\kappa_s-\mu_s)\mathrm{log}(\mp\kappa_s-\mu_s)+(\mp\kappa_s+\mu_s)\mathrm{log}(\mp\kappa_s+\mu_s)-2\mu_s\mathrm{log}(2\mu_s)\right]}\left(4k^2\tau_1\tau_2\right)^{i\mu_s}\right],
\end{align}
thus the non-local part yields
\begin{align}
	D^{\pm s}|_{\text{non-local}}=\frac{(\tau_1\tau_2)^{\frac{3}{2}-s}}{(2H^2)^{(s-1)}}\frac{e^{2\pi(\mp\kappa_s-\mu_s)}}{2\mu_s}2\mathrm{Re}\left[e^{i\left[-(\mp\kappa_s-\mu_s)\mathrm{log}(\mp\kappa_s-\mu_s)+(\mp\kappa_s+\mu_s)\mathrm{log}(\mp\kappa_s+\mu_s)-2\mu_s\mathrm{log}(2\mu_s)\right]}\left(4k^2\tau_1\tau_2\right)^{i\mu_s}\right].
\end{align}
The dominant contribution to the enhancement of the gravitational wave signal arises from the component with helicity $\lambda=-s$; therefore, we exclusively extract the contribution from this component. In the large mass limit, and by considering the contribution of the time integral \cite{Wang:2021qez}, we can extract the dominant component of the propagator containing $\kappa_s$ and $\mu_s$
\begin{align}\label{proestimate}
    D^{- s}|_{\text{local}}\sim\frac{1}{2^{s-1}}\frac{1}{\mu_s}e^{2\pi(\kappa_s-\mu_s)},\qquad  D^{- s}|_{\text{non-local}}\sim \frac{1}{2^{s-1}}e^{2\pi(\kappa_s-\mu_s)},
\end{align}
where we do not calculate the correlation functions exactly, but instead employ an approximate method proposed in the literature  \cite{Wang:2020ioa}. The factor $\mu_s^{-1}$ is dropped in the nonlocal part of the propagator as the in-in integral always yields a positive power of $\mu_s$ \cite{Wang:2020ioa}. For a hard propagator,  the local part can be approximated as $\propto1/\mu_s^2e^{2\pi(\kappa_s-\mu_s)}$ \cite{Wang:2020ioa}. What's more, the other two approximation conditions are also employed: firstly, certain non-exponential factors and coefficients in the propagator are neglected; secondly, the late-time limit is used when estimating the time integral. Regarding the first condition, the contribution of non-exponential factors is relatively small compared to the exponential part and does not significantly affect the order of magnitude. As for the second condition,  adopting the late-time limit results in a certain deviation in the powers of $\mu_s$ in the propagator compared to the actual results. This deviation is an error introduced by the approximate calculation method. In addition, $1/2^{s-1}$ is the exponential part depending on spin, which arises from the normalization of polarization tensors.

\section{Polarization tensors}\label{polar_tensors}
In this section, I will derive the explicit expressions for polarization tensors of arbitrary spin and helicity, following the standard approach in the literature \cite{Lee:2016vti}.
The polarization tensor $\varepsilon^{\lambda}_{i_1\cdots i_n}(\hat{\bm{k}},\epsilon^{\pm})$ is totally symmetric and traceless, which can be further decomposed into transverse and longitudinal parts:
 \begin{align}\label{polarization_tensor}
	\varepsilon^{\lambda}_{i_1\cdots i_n}(\hat{\bm{k}},\epsilon^{\pm})=\epsilon^{\lambda}_{(i_1\cdots i_{|\lambda|}}(\epsilon^\pm)f_{i_{|\lambda|+1}\cdots i_n)}(\hat{\bm{k}}),
 \end{align}
 and $\epsilon^\lambda_{i_1\cdots i_{|\lambda|}}$ is the purely transverse polarization tensor: 
 \begin{align}
\epsilon^\lambda_{i_1\cdots  i_{|\lambda|}}=\prod_{k=1}^{|\lambda|}\epsilon^\pm_{i_k},
 \end{align} 
Here $\epsilon^\pm_i(\hat{\bm{k}})$ is the rank-1 polarization tensor ($\lambda=\pm1$) satisfying
 \begin{align}
     \hat{\bm{k}}_i\epsilon^\pm_i(\hat{\bm{k}})=0,\quad \epsilon^\pm_i(\hat{\bm{k}}) \epsilon^\pm_i(\hat{\bm{k}})^\star=2.
 \end{align}
Furthermore, the maximally transverse polarization tensor satisfies the following relation when contracted with momentum $ \bm{q}_i $: 
 \begin{align}
     \hat{\bm{q}}_{i_1}\cdots \hat{\bm{q}}_{i_n}\epsilon^{\lambda}(\hat{\bm{k}},\epsilon^{\pm})_{i_1\cdots i_{n}}=\hat{Y}^\lambda_s(\theta,\phi),
 \end{align}
 where $\hat{Y}^\lambda_s(\theta,\phi)\propto e^{i\lambda\phi}P^\lambda_s(\mathrm{cos}\theta) \equiv \varepsilon^\lambda_\lambda P^\lambda_s(\mathrm{cos}\theta)$.
And $f_{i_{|\lambda|+1}\cdots i_{n}}$ is the longitudinal part which satisfies
\begin{align}
	\hat{\bm{q}}_{i_1}\cdots \hat{\bm{q}}_{i_n}f^{\lambda}(\hat{\bm{k}})_{i_1\cdots i_{n}}=\hat{P}^\lambda_s(\mathrm{cos}\theta).
\end{align}
We can utilize the properties of $f_{i_{\lambda+1}\cdots i_{n}}$ to express it in terms of $\hat{\bm{k}}_i$ and $\delta_{ij}$. Consequently, the polarization tensor can be written in a more specific form 
\begin{align}\label{polar_tensor}
  \varepsilon^{\lambda}_{i_1\cdots i_s}(\hat{\bm{k}},\epsilon^\pm)=\frac{1}{(2|\lambda|-1)!!}\sum_{n=0}^{s-|\lambda|}B_n\frac{1}{s}\epsilon^\lambda_{(i_1\cdots i_{|\lambda|}}\hat{\bm{k}}_{i_{|\lambda|+1}}\cdots \hat{\bm{k}}_{i_{|\lambda|+n}}\delta_{i_{|\lambda|+n+1}\cdots i_s)},
\end{align}
where $\delta_{i_1\cdots i_m}\equiv \delta_{i_1i_2}\cdots \delta_{i_{m-1}i_m}$ appears only when $n=\text{even}$ and $\lambda$ denotes the helicity. What we have done is expanding $f_{i_{|\lambda|+1}\cdots i_{n}}$ into a linear combination of different longitudinal modes, and the coefficient \(B_n\) is given by:
\begin{align}
  B_n=\frac{2^s}{n!(s-n-|\lambda|)!}\frac{\Gamma\left[\frac{1}{2}(n+|\lambda|+1+s)\right]}{\Gamma\left[\frac{1}{2}(n+|\lambda|+1-s)\right]}.
\end{align}	
For the case $|\lambda|=s$, the coefficient becomes
\begin{align}
	\frac{B_0}{(2s-1)!!}=\frac{2^s\Gamma\left[s+\frac{1}{2}\right]}{\Gamma\left[\frac{1}{2}\right](2s-1)!!}=\frac{2^s}{\Gamma\left[\frac{1}{2}\right](2s-1)!!}\frac{(2s-1)!!}{2^s}\Gamma\left[\frac{1}{2}\right]=1.
\end{align}
The factor $\frac{1}{s}$ is cancelled by the $s$ coming from the totally symmetrizing
\begin{align}
	\epsilon^\lambda_{(i_1\cdots i_{|\lambda|})}=s\epsilon^\lambda_{i_1\cdots i_{|\lambda|}},
\end{align}
thus for the highest helicity $|\lambda|=s$, the polarization tensor becomes
\begin{align}\label{highestpol}
	\varepsilon^{s}_{i_1\cdots i_s}(\hat{\bm{k}},\epsilon^{\pm})=\epsilon^\lambda_{i_1\cdots i_{|\lambda|}}=\prod_{k=1}^{|\lambda|}\epsilon^\pm_{i_k}.
\end{align}
For helicity $|\lambda|=s-1$, the coefficient yields
\begin{align}
	\frac{B_1}{(2s-3)!!}=\frac{2^s}{(2s-3)!!}\frac{\Gamma\left[s+\frac{1}{2}\right]}{\Gamma\left[\frac{1}{2}\right]}=\frac{2^s}{\Gamma\left[\frac{1}{2}\right](2s-1)!!}\frac{(2s-3)!!}{2^s}\Gamma\left[\frac{1}{2}\right]=(2s-1).
\end{align}
Combine all factors into one coefficient, we obtain
\begin{align}
	C_{s-1}^s=\frac{B_1}{(2s-3)!!}\frac{1}{s}=\frac{(2s-1)}{s}  
\end{align}
And the self-contraction of polarization tensors becomes
\begin{align}
	\varepsilon^\lambda_{i_1\cdots i_s}\varepsilon^{\lambda\star}_{i_1\cdots i_s}=\frac{(2s-1)!!(s+|\lambda|)!}{\left[(2|\lambda|-1)!!\right]^2s!(s-|\lambda|)!}.
\end{align}
It is easy and straightforward to check for $|\lambda|=s$, the self-contraction yields
\begin{align}
	\varepsilon^s_{i_1\cdots i_s}\varepsilon^{s\star}_{i_1\cdots i_s}=2^s,
\end{align}
which is consistent with Eq.~(\ref{highestpol}) where $\epsilon^\pm_i\epsilon^{\pm\star}_i=2$. For $|\lambda|=s-1$
\begin{align}
	\varepsilon^{s-1}_{i_1\cdots i_s}\varepsilon^{s-1\star}_{i_1\cdots i_s}=\frac{(2s-1)!!(2s-1)!}{\left[(2s-3)!!\right]^2s!},
\end{align}
where $(2s-1)!=(2s-1)!!2^{s-1}(s-1)!$, thus the above equation can also be written as
\begin{align}
	\varepsilon^{s-1}_{i_1\cdots i_s}\varepsilon^{s-1\star}_{i_1\cdots i_s}=\frac{\left[(2s-1)!!\right]^22^{s-1}}{\left[(2s-3)!!\right]^2s}.
\end{align}

\section{Energy-momentum tensor of the massive spin-$2$ field}\label{spin2EMT}
To compute GWs, we need to derive the energy-momentum tensor corresponding to the massive spin-$2$ field, which can be obtained by taking the variation of the metric~\footnote{There are two primary methods for deriving the energy-momentum tensor. The first is to derive the canonical energy-momentum tensor using Noether's first theorem, which is then modified into the Belinfante tensor by accounting for its uncertainty (which can differ by a total derivative) \cite{belinfante1940current,Weinberg:1995mt,Forger:2003ut,Baker:2021hly}. However, this approach can be subtle, particularly when dealing with spin \cite{Hehl:1976vr}. The alternative method, known as Hilbert's method, involves calculating the variation of the metric with respect to the action and then reverting the metric back to the spacetime background. These two methods are generally considered equivalent and both are physically valid. However, some research suggests that these methods are not completely equivalent in certain models, such as the linearized Gauss-Bonnet gravity model \cite{Baker:2020eqs}. In our study, the definition of the physical energy-momentum tensor is based on the EoMs of tensor perturbations, which naturally aligns with the second derivation method.}. Notably, by applying the constraints of the massive spin-$2$ field and setting the field on shell after calculating the energy-momentum tensor, many terms in the Lagrangian Eq.~(\ref{effect_Lag_spin2}) do not contribute to the result. Terms that include traces and terms where fields contract with covariant derivatives both vanish. Therefore, we only need to perform a variation of the following terms
\begin{align}
    S_{\text{spin-$2$}}&\supset \int d^4x\sqrt{-g}\left[-\frac{1}{4}\nabla_{\mu}\Sigma_{\nu\lambda}\nabla^\mu\Sigma^{\nu\lambda}-\frac{1}{4}(2H^2+m^2)\Sigma_{\mu\nu}\Sigma^{\mu\nu}-\frac{\nabla_{\mu}\phi}{2\Lambda_{c,2}}\varepsilon^{\mu\nu\rho\sigma}\Sigma_{\nu\lambda}\nabla_{\rho}\Sigma_{\sigma}^{~\lambda}\right].
\end{align}
The energy-momentum tensor for the massive spin-$2$ field yields
\begin{align}
	T_{\alpha\beta}&=\frac{-2}{\sqrt{-g}}\frac{\delta S_{\text{spin-$2$}}}{\delta g^{\mu\nu}}\notag\\
 &=\frac{-2}{\sqrt{-g}}\times\left\{\sqrt{-g}\left(-\frac{1}{4}\nabla_\alpha\Sigma_{\nu\lambda}\nabla_\beta\Sigma^{\nu\lambda}-\frac{1}{2}\nabla_\mu\Sigma_{\alpha\lambda}\nabla^\mu\Sigma_\beta^{~\lambda}\right)\right.\notag\\
	&+\frac{\sqrt{-g}}{4}\left[\left(\partial_{\nu}g_{\beta\mu}+\partial_{\mu}g_{\beta\nu}-\partial_{\beta}g_{\mu\nu}\right)\Sigma_{\alpha\lambda}+(\alpha\leftrightarrow\beta)\right]\nabla^\mu\Sigma^{\nu\lambda}\notag\\
	&-\sqrt{-g}\left(H^2+\frac{1}{2}m^2\right)\Sigma_{\alpha\eta}\Sigma_{\beta}^{~\eta}\notag\\
	&-\frac{\sqrt{-g}}{2}g_{\alpha\beta}\left[-\frac{1}{4}\nabla_{\mu}\Sigma_{\nu\lambda}\nabla^\mu\Sigma^{\nu\lambda}-\frac{1}{2}\left(H^2+\frac{1}{2}m^2\right)\Sigma_{\mu\nu}\Sigma^{\mu\nu}\right]\notag\\
	&-\frac{1}{4}\left[g_{\kappa^\prime\alpha}g_{\mu\beta}\partial_\nu+g_{\kappa^\prime\alpha}g_{\nu\beta}\partial_\mu-g_{\nu\alpha}g_{\mu\beta}\partial_{\kappa^\prime}+(\alpha\leftrightarrow\beta)\right]\left[\sqrt{-g}g^{\delta\kappa^\prime}\Sigma_{\delta\kappa}\nabla^\mu\Sigma^{\nu\kappa}\right]\notag\\
	&-\frac{1}{2}\frac{\partial_\mu\phi}{2\Lambda_{c,2}}\epsilon^{\mu\nu\rho\sigma}\left[\Sigma_{\nu\alpha}\nabla_\rho\Sigma_{\sigma\beta}+\Sigma_{\nu\beta}\nabla_\rho\Sigma_{\sigma\alpha}\right]\notag\\
	&+\frac{1}{4}\left[\left(\partial_{\kappa}g_{\beta\rho}+\partial_{\rho}g_{\beta\kappa}-\partial_{\beta}g_{\kappa\rho}\right)\Sigma_{\sigma\alpha}+(\alpha\leftrightarrow\beta)\right]\left[\frac{\partial_\mu\phi}{2\Lambda_{c,2}}\epsilon^{\mu\nu\rho\sigma}\Sigma_\nu^{~\kappa}\right]\notag\\
	&\left.-\frac{1}{4}\left[2g_{\alpha\delta}g_{\beta\rho}\partial_\kappa+g_{\alpha\delta}g_{\beta\kappa}\partial_\rho+(\alpha\leftrightarrow\beta)\right]\left[\frac{\partial_\mu\phi}{2\Lambda_{c,2}}\epsilon^{\mu\nu\rho\sigma}g^{\lambda\kappa}\Sigma_{\nu\lambda}g^{\eta\delta}\Sigma_{\sigma\eta}\right]\right\},
\end{align}
Here, we expand certain covariant derivatives, explicitly express the metric tensor $g_{\mu\nu}$, and omit the backreaction of gravitons on the source. The calculation is notably lengthy and complicated; therefore, we present the result directly. This highlights the substantial challenge in computing the energy-momentum tensor for higher-spin cases, even when we can formulate the Lagrangian. Since only the spatial components of the energy-momentum tensor are necessary for computing GWs, we substitute $(\alpha, \beta)$ with $(i, j)$. The spatial components of the energy-momentum tensor are given by
\begin{align}
	T_{ij}&=\frac{-2}{\sqrt{-g}}\times\left\{\sqrt{-g}\left(-\frac{1}{4}\nabla_i\Sigma_{\nu\lambda}\nabla_j\Sigma^{\nu\lambda}-\frac{1}{2}\nabla_\mu\Sigma_{i\lambda}\nabla^\mu\Sigma_j^{~\lambda}\right)\right.\notag\\
	&+\frac{\sqrt{-g}}{4}\left[\left(\partial_{\nu}g_{j\mu}+\partial_{\mu}g_{j\nu}-\partial_{j}g_{\mu\nu}\right)\Sigma_{i\lambda}+(i\leftrightarrow j)\right]\nabla^\mu\Sigma^{\nu\lambda}\notag\\
	&-\sqrt{-g}\left(H^2+\frac{1}{2}m^2\right)\Sigma_{i\eta}\Sigma_{j}^{~\eta}\notag\\
	&-\frac{\sqrt{-g}}{2}g_{ij}\left[-\frac{1}{4}\nabla_{\mu}\Sigma_{\nu\lambda}\nabla^\mu\Sigma^{\nu\lambda}-\frac{1}{2}\left(H^2+\frac{1}{2}m^2\right)\Sigma_{\mu\nu}\Sigma^{\mu\nu}\right]\notag\\
	&-\frac{1}{4}\left[g_{\kappa^\prime i}g_{\mu j}\partial_\nu+g_{\kappa^\prime i}g_{\nu j}\partial_\mu-g_{\nu i}g_{\mu j}\partial_{\kappa^\prime}+(i\leftrightarrow j)\right]\left[\sqrt{-g}g^{\delta\kappa^\prime}\Sigma_{\delta\kappa}\nabla^\mu\Sigma^{\nu\kappa}\right]\notag\\
	&-\frac{1}{2}\frac{\partial_\mu\phi}{2\Lambda_{c,2}}\epsilon^{\mu\nu\rho\sigma}\left[\Sigma_{\nu i}\nabla_\rho\Sigma_{\sigma j}+\Sigma_{\nu j}\nabla_\rho\Sigma_{\sigma i}\right]\notag\\
	&+\frac{1}{4}\left[\left(\partial_{\kappa}g_{j\rho}+\partial_{\rho}g_{j\kappa}-\partial_{j}g_{\kappa\rho}\right)\Sigma_{\sigma i}+(i\leftrightarrow j)\right]\left[\frac{\partial_\mu\phi}{2\Lambda_{c,2}}\epsilon^{\mu\nu\rho\sigma}\Sigma_\nu^{~\kappa}\right]\notag\\
	&\left.-\frac{1}{4}\left[2g_{i\delta}g_{j\rho}\partial_\kappa+g_{i\delta}g_{j\kappa}\partial_\rho+(i\leftrightarrow j)\right]\left[\frac{\partial_\mu\phi}{2\Lambda_{c,2}}\epsilon^{\mu\nu\rho\sigma}g^{\lambda\kappa}\Sigma_{\nu\lambda}g^{\eta\delta}\Sigma_{\sigma\eta}\right]\right\}.
\end{align}

We can calculate the energy-momentum tensor directly using metric and Christoffel symbol in dS spacetime
\begin{align}
    g_{\mu\nu}=\left(\begin{array}{llll} 
	-1&0&0&0\\
	0&a^2(t)&0&0\\
	0&0&a^2(t)&0\\
	0&0&0&a^2(t)
	\end{array}\right),\quad \Gamma^{0}_{ij}=a^2H\delta_{ij},\quad \Gamma^{i}_{j0}=H\delta_{ij}.
\end{align}
Here, we first use physical time instead of conformal time since the corresponding connection form for physical time is simpler. However, in the calculations of GWs in the previous sections, we will use conformal time. These two are equivalent, differing only by a coordinate transformation. Based on the position of the spatial indices $i,j$ that couple to the graviton $h^{TT}_{ij}$, the energy-momentum tensor can be decomposed into orbital angular momentum and spin angular momentum components. The orbital angular momentum component of the energy-momentum tensor yields
\begin{align}\label{spin2EMTorbit}
	&T_{ij}^{\rm orbit}\notag\\
	&=\frac{1}{2}\partial_{i}\Sigma_{\nu\lambda}\partial_{j}\Sigma^{\nu\lambda}+2g_{ij}g^{\lambda\delta}a^4H^2\Sigma_{0\lambda}\Sigma_{0\delta}\notag\\
     &+\left(\frac{7}{2}H\Sigma_{00}\partial_{i}\Sigma_{j0}-4a^{-2}H\Sigma_{0l}\partial_{i}\Sigma_{jl}+(i\leftrightarrow j)\right)-\left(\frac{1}{2}H\Sigma_{0j}\partial_{i}\Sigma_{00}-\frac{5}{2}a^{-2}H\Sigma_{lj}\partial_{i}\Sigma_{l0}+(i\leftrightarrow j)\right)\notag\\
	 &+\left(\frac{1}{2}\partial_j\Sigma_{00}\partial_0\Sigma_{i0}-\frac{1}{2}a^{-2}\partial_j\Sigma_{0l}\partial_0\Sigma_{il}+(i\leftrightarrow j)\right)+\left(\frac{1}{2}\Sigma_{i0}\partial_0\partial_j\Sigma_{00}-\frac{1}{2}a^{-2}\Sigma_{il}\partial_0\partial_j\Sigma_{0l}+(i\leftrightarrow j)\right)\notag\\
	 &-\left(\frac{1}{2}\partial_j\Sigma_{i0}\partial_0\Sigma_{00}-\frac{1}{2}a^{-2}\partial_j\Sigma_{il}\partial_0\Sigma_{0l}+(i\leftrightarrow j)\right)-\left(\frac{1}{2}\Sigma_{00}\partial_0\partial_j\Sigma_{i0}-\frac{1}{2}a^{-2}\Sigma_{0l}\partial_0\partial_j\Sigma_{il}+(i\leftrightarrow j)\right)\notag\\
	 &-\left(\frac{1}{2}a^{-2}\partial_{j}\Sigma_{k0}\partial_k\Sigma_{i0}+\frac{1}{2}a^{-2}\Sigma_{i0}\partial_k\partial_j\Sigma_{k0}-\frac{1}{2}a^{-2}\partial_{j}\Sigma_{i0}\partial_k\Sigma_{k0}-\frac{1}{2}a^{-2}\Sigma_{k0}\partial_k\partial_j\Sigma_{i0}+(i\leftrightarrow j)\right).
\end{align}
And the spin angular momentum part of the energy-momentum tensor yields
\begin{align}\label{spin2EMTspin}
	&T_{ij}^{\rm spin}\notag\\
	 &=\left(2\partial_0\Sigma_{i0}\partial_0 \Sigma_{j0}-2a^{-2}\partial_0\Sigma_{il}\partial_0 \Sigma_{jl}\right)\notag\\
  &-\partial_m\Sigma_{i0}\partial_m \Sigma_{j0}-\left(a^{-2}H\Sigma_{j0}\partial_{m}\Sigma_{im}+(i\leftrightarrow j)\right)\notag\\
	 &-\left(\frac{13}{2}H\Sigma_{i0}\partial_0 \Sigma_{j0}-\frac{13}{2}a^{-2}H\Sigma_{il}\partial_0 \Sigma_{jl}+(i\leftrightarrow j)\right)\notag\\
    &+\left(-4\frac{\ddot{a}}{a}+18H^2-m^2\right)\Sigma_{i0}\Sigma_{j0}+\left(\frac{\ddot{a}}{a}-20H^2+m^2\right)a^{-2}\Sigma_{il}\Sigma_{jl}\notag\\
	 &+\left(-8H^2+2\frac{\ddot{a}}{a}\right)\Sigma_{ij}\Sigma_{00}\notag\\
	 &+2H\Sigma_{00}\partial_0\Sigma_{ij}+2H\partial_0\Sigma_{00}\Sigma_{ij}-a^{-2}\partial_{l}\Sigma_{i0}\partial_{l}\Sigma_{j0}+\left(a^{-2}H\Sigma_{il}\partial_l \Sigma_{j0}+(i\leftrightarrow j)\right)\notag\\
	 &+\left(\frac{1}{2}\Sigma_{i0}\partial_0^2\Sigma_{j0}-\frac{1}{2}a^{-2}\Sigma_{il}\partial_0^2\Sigma_{jl}+(i\leftrightarrow j)\right)-\left({\frac{1}{2}a^{-2}\Sigma_{i0}\partial^2_l\Sigma_{j0}+\frac{1}{2}a^{-2}\Sigma_{j0}\partial^2_l\Sigma_{i0}}+(i\leftrightarrow j)\right)\notag\\
	 &+\left(a^{-2}H\Sigma_{jl}\partial_{l}\Sigma_{i0}+a^{-2}H\partial_{l}\Sigma_{jl}\Sigma_{i0}+(i\leftrightarrow j)\right)-\left(2a^{-2}H\Sigma_{0l}\partial_{l}\Sigma_{ij}+2a^{-2}H\partial_{l}\Sigma_{0l}\Sigma_{ij}\right)\notag\\
   &+a^{-3}\frac{\dot{\phi}}{2\Lambda_{c,2}}\epsilon^{nrs}\Sigma_{ni}\partial_{r}\Sigma_{sj}+a^{-3}\frac{\dot{\phi}}{2\Lambda_{c,2}}\epsilon^{njs}\left(\Sigma_{si}\partial_k\Sigma_{kn}+\partial_k\Sigma_{si}\Sigma_{kn}\right)+(i\leftrightarrow j)\notag\\
   &+4a^{-1}H\frac{\dot{\phi}}{2\Lambda_{c,2}}\epsilon^{njs}\Sigma_{si}\Sigma_{n0}-a^{-1}\frac{\dot{\phi}}{2\Lambda_{c,2}}\epsilon^{njs}\left(\Sigma_{si}\partial_0\Sigma_{n0}+\partial_0\Sigma_{si}\Sigma_{0n}\right)+(i\leftrightarrow j).
 \end{align}
 
\section{Energy-momentum tensor of massive spin-$s$ fields}
\label{AppendixspinsEMT}
Although it is challenging to explicitly write down the full Lagrangian for higher-spin fields, notably, terms involving the trace $I^{\mu}_{~\mu\cdots }$ and divergence $\nabla^{\mu} I_{\mu\cdots }$ vanish on-shell due to the enforcement of tracelessness and transversality constraints. For spins $s > 5/2$, auxiliary fields are required to maintain consistency in the Lagrangian formulation \cite{Deser:2003gw}. These auxiliary fields, however, are non-dynamical and do not contribute to the production of GWs.  

Furthermore, from the study of the massive spin-$2$ case, we understand that phenomenologically, the dominant contribution to gravitational wave generation originates from the term: $T_{ij} \supset \partial_{\tau} I_{ii_1\cdots i_{s-1}} \partial_{\tau} I_{ji_1\cdots i_{s-1}}$, which arises from the kinetic term of the energy-momentum tensor. While this structure generalizes to higher spins, the spin-dependent coefficients in the term remain undetermined and require explicit derivation. The derivation proceeds as follows.
 The kinetic term for a spin-$s$ field can be expressed as:\footnote{The coefficient of the kinetic term here is $1/4$ because our convention for the Wronskian Eq.~(\ref{Wronskian_spins}) differs by a factor of 2 from other literature. This convention is consistent with the action Eq.~(\ref{effect_Lag_spin2}) for the massive spin-$2$ case in Sec.\;\ref{FreeSpin2ChemPtlSect}.}
  \begin{align}\label{kineticspins}
	\frac{1}{4}\sqrt{-g}g^{\mu\nu}g^{\mu_1\nu_1}\cdots g^{\mu_s\nu_s}\nabla_{\mu}I_{\mu_1\cdots \mu_s}\nabla_{\nu}I_{\nu_1\cdots \nu_s},
  \end{align}
where the covariant derivative can be expanded as
\begin{align}
	\nabla_{\mu}I_{\mu_1\cdots \mu_s}=\partial_{\mu}I_{\mu_1\cdots \mu_s}-\sum^s_{k=1}\Gamma^\sigma_{\mu_k\mu}I_{\mu_1\cdots \sigma\cdots \mu_s},
\end{align}
and the kinetic term yields
\begin{align}
	&\nabla_{\mu}I_{\mu_1\cdots \mu_s}\nabla_{\nu}I_{\nu_1\cdots \nu_s}\notag\\
	&=\left(\partial_{\mu}I_{\mu_1\cdots \mu_s}-\sum^s_{k=1}\Gamma^\sigma_{\mu_k\mu}I_{\mu_1\cdots \sigma\cdots \mu_s}\right)\left(\partial_{\nu}I_{\nu_1\cdots \nu_s}-\sum^s_{k=1}\Gamma^\sigma_{\nu_k\nu}I_{\nu_1\cdots  \sigma\cdots \nu_s}\right)\notag\\
	&=\partial_{\mu}I_{\mu_1\cdots \mu_s}\partial_{\nu}I_{\nu_1\cdots \nu_s}+\sum^s_{k,p=1}\Gamma^\sigma_{\mu_k\mu}\Gamma^\lambda_{\nu_p\nu}I_{\mu_1\cdots \sigma\cdots \mu_s}I_{\nu_1\cdots \lambda\cdots \nu_s}\notag\\
	&-\partial_{\mu}I_{\mu_1\cdots \mu_s}\sum^s_{k=1}\Gamma^\sigma_{\nu_k\nu}I_{\nu_1\cdots  \sigma\cdots  \nu_s}-\partial_{\nu}I_{\nu_1\cdots \nu_s}\sum^s_{k=1}\Gamma^\sigma_{\mu_k\mu}I_{\mu_1\cdots \sigma\cdots \mu_s}.
\end{align}
To obtain the energy-momentum tensor, we need to take the variation of this term with respect to the metric. First, let us write down the variation of the metric and the variation of the connection. The variation of the metric can be written as 
\begin{align}
	\frac{\delta g^{\mu\nu}}{\delta g^{\alpha\beta}}=\frac{1}{2}\left(\delta_{\mu\alpha}\delta_{\nu\beta}+\delta_{\nu\alpha}\delta_{\mu\beta}\right),,
\end{align}
and the variation of the connection with respect to metric yields
\begin{align}
	\frac{\delta\Gamma^\gamma_{\nu\mu}}{\delta g^{\alpha\beta}}(\cdots)&=\frac{1}{2}\frac{\delta g^{\gamma\delta}}{\delta g^{\alpha\beta}}\left(\partial_\nu g_{\delta\mu}+\partial_\mu g_{\delta\nu}-\partial_\delta g_{\mu\nu}\right)(\cdots )\notag\\
	&-\frac{1}{2}\frac{\delta g^{\eta\kappa}}{\delta g^{\alpha\beta}}\left[\partial_{\nu}\left(g_{\delta\eta}g_{\mu\kappa}g^{\gamma\delta}(\cdots )\right)+\partial_{\mu}\left(g_{\delta\eta}g_{\nu\kappa}g^{\gamma\delta}(\cdots )\right)-\partial_{\delta}\left(g_{\mu\eta}g_{\nu\kappa}g^{\gamma\delta}(\cdots )\right)\right]\notag\\
	&=\frac{1}{4}\left[\left(\partial_\nu g_{\alpha\mu}+\partial_\mu g_{\alpha\nu}-\partial_\alpha g_{\mu\nu}\right)\delta_{\gamma\beta}+(\alpha\leftrightarrow\beta)\right]\notag\\
	&-\frac{1}{4}\left[\partial_{\nu}\left(\delta_{\gamma\alpha}g^{\mu\beta}(\cdots )\right)+\partial_{\mu}\left(\delta_{\gamma\alpha}g^{\nu\beta}(\cdots )\right)-\partial_{\delta}\left(g_{\mu\eta}g_{\nu\kappa}g^{\gamma\delta}(\cdots )\right)+(\alpha\leftrightarrow\beta)\right],
\end{align}
where the ellipses correspond to terms related to the connection. Since the connection involves derivatives of the metric, the method of integration by parts is employed in the derivation above. Utilizing the results derived above, the variation of the action with respect to the metric becomes
\begin{align}\label{varEMTkinetic}
	&\frac{\delta\left(\sqrt{-g}g^{\mu\nu}g^{\mu_1\nu_1}\cdots g^{\mu_s\nu_s}\nabla_{\mu}I_{\mu_1\cdots \mu_s}\nabla_{\nu}I_{\nu_1\cdots \nu_s}\right)}{\sqrt{-g}\delta g^{\alpha\beta}}\notag\\
	&=g^{\mu_1\nu_1}\cdots g^{\mu_s\nu_s}\partial_{\alpha}I_{\mu_1\cdots \mu_s}\partial_{\beta}I_{\nu_1\cdots \nu_s}+\underline{\bm{s} g^{\mu\nu}g^{\mu_1\nu_1}\cdots g^{\mu_s\nu_s}\partial_{\mu}I_{\alpha\mu_1\cdots \mu_{s-1}}\partial_{\nu}I_{\beta\nu_1\cdots \nu_{s-1}}}\notag\\
	\notag\\
	&+g^{\mu_1\nu_1}\cdots g^{\mu_s\nu_s}\sum^s_{k,p=1}\Gamma^\sigma_{\mu_k\alpha}\Gamma^\lambda_{\nu_p\beta}I_{\mu_1\cdots \sigma\cdots \mu_{s}}I_{\nu_1\cdots \lambda\cdots \nu_{s}}\notag\\
	&+\bm{s}g^{\mu\nu}g^{\mu_1\nu_1}\cdots g^{\mu_{s-1}\nu_{s-1}}\left(\sum^{s-1}_{k,p=1}\Gamma^\sigma_{\mu_k\mu}\Gamma^\lambda_{\nu_p\nu}I_{\alpha\mu_1\cdots \sigma\cdots \mu_{s-1}}I_{\beta\nu_1\cdots \lambda\cdots \nu_{s-1}}+\Gamma^\sigma_{\alpha\mu}\Gamma^\lambda_{\beta\nu}I_{\sigma\mu_1\cdots \mu_{s-1}}I_{\lambda\nu_1\cdots \nu_{s-1}}\right.\notag\\
	&\left.+\Gamma^\sigma_{\alpha\mu}\sum^{s-1}_{p=1}\Gamma^{\lambda}_{\nu_p\nu}I_{\sigma\mu_1\cdots \mu_{s-1}}I_{\beta\nu_1\cdots \lambda\cdots \nu_{s-1}}+\Gamma^\lambda_{\beta\nu}\sum^{s-1}_{k=1}\Gamma^{\sigma}_{\mu_k\mu}I_{\lambda\nu_1\cdots \nu_{s-1}}I_{\alpha\mu_1\cdots \sigma\cdots \mu_{s-1}}\right)\notag\\
	&+\frac{1}{4}g^{\mu\nu}g^{\mu_1\nu_1}\cdots g^{\mu_s\nu_s}\sum^s_{k,p=1}I_{\mu_1\cdots \sigma\cdots \mu_{s}}I_{\nu_1\cdots \lambda\cdots \nu_{s}}\notag\\
	&\times \left\{\left[\left(\partial_{\mu_k} g_{\alpha\mu}+\partial_\mu g_{\alpha\mu_k}-\partial_\alpha g_{\mu\mu_k}\right)\delta_{\sigma\beta}+(\alpha\leftrightarrow\beta)\right]\Gamma^\lambda_{\nu_p\nu}+\left[\left(\partial_{\nu_p} g_{\alpha\nu}+\partial_\nu g_{\alpha\nu_p}-\partial_\alpha g_{\nu\nu_p}\right)\delta_{\lambda\beta}+(\alpha\leftrightarrow\beta)\right]\Gamma^\sigma_{\mu_k\mu}\right\}\notag\\
	&-\frac{1}{4}\sum^s_{k,p=1}\left\{\left[\partial_{\mu_k}\left(\delta_{\sigma\alpha} g_{\mu\beta}g^{\mu\nu}g^{\mu_1\nu_1}\cdots g^{\mu_s\nu_s}\Gamma^\lambda_{\nu_p\nu}I_{\mu_1\cdots \sigma\cdots \mu_{s}}I_{\nu_1\cdots \lambda\cdots \nu_{s}}\right)\right.\right.\notag\\
	&+\partial_{\mu}\left(\delta_{\sigma\alpha} g_{\mu_k\beta}g^{\mu\nu}g^{\mu_1\nu_1}\cdots g^{\mu_s\nu_s}\Gamma^\lambda_{\nu_p\nu}I_{\mu_1\cdots \sigma\cdots \mu_{s}}I_{\nu_1\cdots \lambda\cdots \nu_{s}}\right)\notag\\
	&\left.\left.-\partial_{\delta}\left(g_{\mu\alpha}g_{\mu_k\beta}g^{\sigma\delta}g^{\mu\nu}g^{\mu_1\nu_1}\cdots g^{\mu_s\nu_s}\Gamma^\lambda_{\nu_p\nu}I_{\mu_1\cdots \sigma\cdots \mu_{s}}I_{\nu_1\cdots \lambda\cdots \nu_{s}}\right)+(\alpha\leftrightarrow\beta)\right]+(\mu\leftrightarrow\nu)\right\}\notag\\
	\notag\\
	&-\left[g^{\mu_1\nu_1}\cdots g^{\mu_s\nu_s}\partial_\alpha I_{\mu_1\cdots \mu_s}\sum^s_{k=1}\Gamma^\lambda_{\nu_k\beta}I_{\nu_1\cdots \lambda\cdots \nu_s}+(\alpha\leftrightarrow\beta)\right]\notag\\
	&-\left[\bm{s}g^{\mu\nu}g^{\mu_1\nu_1}\cdots g^{\mu_{s-1}\nu_{s-1}\partial_{\mu}} I_{\alpha\mu_1\cdots \mu_{s-1}}\left(\sum^{s-1}_{k=1}\Gamma^\lambda_{\nu_k\nu}I_{\beta\nu_1\cdots \lambda\cdots \nu_{s-1}}+\Gamma_{\beta\nu}^\lambda I_{\lambda\nu_1\cdots \nu_{s-1}}\right)+(\alpha\leftrightarrow\beta)\right]\notag\\
	&-\frac{1}{2}g^{\mu\nu}g^{\mu_1\nu_1}\cdots g^{\mu_s\nu_s}\partial_\mu I_{\mu_1\cdots \mu_s}\sum^s_{k=1}\left[\left(\partial_{\nu_k} g_{\alpha\nu}+\partial_\nu g_{\alpha\nu_k}-\partial_\alpha g_{\nu\nu_k}\right)\delta_{\lambda\beta}+(\alpha\leftrightarrow\beta)\right]I_{\nu_1\cdots \lambda\cdots \nu_s}\notag\\
	&+\frac{1}{2}\sum^s_{k=1}\left[\partial_{\nu_k}\left(g^{\mu_1\nu_1}\cdots g^{\mu_s\nu_s}\partial_\beta I_{\mu_1\cdots \mu_s}I_{\nu_1\cdots \alpha\cdots \nu_s}\right)-\partial_{\delta}\left(g^{\mu_1\nu_1}\cdots g^{\mu_k\beta}\cdots g^{\mu_s\nu_s}\partial_\alpha I_{\mu_1\cdots \mu_s}I_{\nu_1\cdots \delta\cdots \nu_s}\right)+(\alpha\leftrightarrow\beta)\right]\notag\\
	&+\left\{\underline{\sum^s_{k=1}g^{\mu\nu}g^{\mu_1\nu_1}g^{\mu_{k-1}\nu_{k-1}}g^{\mu_{k+1}\nu_{k+1}}g^{\mu_s\nu_s}\partial_\mu I_{\mu_1\cdots \beta\cdots \mu_s}\partial_\nu I_{\mu_1\cdots \alpha\cdots \mu_s}}\right.\notag\\
	&+\frac{1}{2}\sum^s_{k=1}\left[g^{\mu\nu}g^{\mu_1\nu_1}g^{\mu_{k-1}\nu_{k-1}}g^{\mu_{k+1}\nu_{k+1}}g^{\mu_s\nu_s}\partial_\mu \partial_\nu I_{\mu_1\cdots \beta\cdots \mu_s} I_{\mu_1\cdots \alpha\cdots \mu_s}+(\alpha\leftrightarrow\beta)\right]\notag\\
	&\left.+\frac{1}{2}\sum^s_{k=1}\left[\partial_\nu\left(g^{\mu\nu}g^{\mu_1\nu_1}g^{\mu_{k-1}\nu_{k-1}}g^{\mu_{k+1}\nu_{k+1}}g^{\mu_s\nu_s}\right)\partial_\mu  I_{\mu_1\cdots \beta\cdots \mu_s} I_{\mu_1\cdots \alpha\cdots \mu_s}+(\alpha\leftrightarrow\beta)\right]\right\}\notag\\
	&+\frac{1}{\sqrt{-g}}\frac{\delta \sqrt{-g}}{\delta g^{{\alpha\beta}}}g^{\mu_1\nu_1}\cdots g^{\mu_s\nu_s}\nabla_{\mu}I_{\mu_1\cdots \mu_s}\nabla_{\nu}I_{\nu_1\cdots \nu_s}.
\end{align}
To obtain the coefficient of the interaction term for the highest helicity modes
\begin{align}\label{internocoe}
	-h^{TT}_{ij}a^{-2(s-1)}\partial_\tau I_{i i_1\cdots i_{s-1}}\partial_\tau I_{j i_1\cdots i_{s-1}},
\end{align}
we should calculate the underlined terms in Eq.~(\ref{varEMTkinetic}) which are
\begin{align}
	&\bm{s} g^{\mu\nu}g^{\mu_1\nu_1}\cdots g^{\mu_s\nu_s}\partial_{\mu}I_{\alpha\mu_1\cdots \mu_{s-1}}\partial_{\nu}I_{\beta\nu_1\cdots \nu_{s-1}}+\sum^s_{k=1}g^{\mu\nu}g^{\mu_1\nu_1}g^{\mu_{k-1}\nu_{k-1}}g^{\mu_{k+1}\nu_{k+1}}g^{\mu_s\nu_s}\partial_\mu I_{\mu_1\cdots \beta\cdots \mu_s}\partial_\nu I_{\mu_1\cdots \alpha\cdots \mu_s}\notag\\
	&=-2\bm{s}a^{-2s}\partial_\tau I_{i i_1\cdots i_{s-1}}\partial_\tau I_{j i_1\cdots i_{s-1}}.
\end{align}
Considering that the kinetic term should be canonical, and given the coefficient in front of it when taking its variation, the final form of the spin-dependent energy-momentum tensor, which significantly contributes to the phenomenology of GWs, is given by
\begin{align}
	T_{ij}&\supset -2\times\left(-\frac{1}{4}\right)\left(-2\bm{s}a^{-2s}\partial_\tau I_{i i_1\cdots i_{s-1}}\partial_\tau I_{j i_1\cdots i_{s-1}}\right)\notag\\
    &=-\bm{s}a^{-2s}\partial_\tau I_{i i_1\cdots i_{s-1}}\partial_\tau I_{j i_1\cdots i_{s-1}},
\end{align}
thus the coefficient of Eq.~(\ref{internocoe}) associated to spin is $\bm{s}$.

\section{The full-order solution of slow-roll corrections}\label{full-order_slowroll}
For the EoMs of the highest helicity modes of a spin-$s$ field with slow-roll corrections, we can expand the solutions in terms of the slow-roll parameters. Accordingly, we can expand the equations to a given order in the slow-roll parameters and solve them order by order. It can be observed that the homogeneous parts of the equations for each solution order are identical, while the non-homogeneous parts include contributions from all lower-order solutions. Given that solutions to the zeroth-order equations are available, solutions for all higher orders can also be derived.
\begin{align}
    (-\tau)^{-(1+\epsilon)}&=(-\tau)^{-1}H_k^{\epsilon}\sum^\infty_{n=0}\frac{(-1)^n}{n!}\epsilon^n(\mathrm{log}(-\tilde{\tau}))^n,\\
      (k)^{-\epsilon}&=H_k^{-\epsilon}\sum^\infty_{n=0}\frac{(-1)^n}{n!}\epsilon^n(\mathrm{log}(\tilde{k}))^n,\\
      I_s&=\sum^\infty_{n=0}\epsilon^n I_n,
\end{align}
where $\tilde{\tau} \equiv H\tau$ and $\tilde{k} \equiv \frac{k}{H}$ are dimensionless, and the subscript of $\epsilon_H$ and the superscript of $I^{\pm s}_n$ are omitted for convenience.
In order to combine terms of the same order, we utilize the following transformation relation of multiple summations:
\begin{align}\label{multisum}
&\sum^{\infty}_{m=0}\sum^{\infty}_{l=m}=\sum^{\infty}_{l=0}\sum^{l}_{m=0}\Rightarrow \sum^{\infty}_{n=0}\sum^{\infty}_{m=0}\sum^{\infty}_{l=m}=\sum^{\infty}_{g=0}\sum^{g}_{n=0}\sum^{g-n}_{m=0}, \quad g=l+n \\
&\quad\text{and}\notag\\
&\sum^{g}_{n=0}\sum^{g}_{k=n}=\sum^{g}_{k=0}\sum^{k}_{n=0},\quad\text{$g$ is a finite value}.
\end{align}
The zeroth-order terms and the first-order terms are separated, and the remaining higher-order terms can be combined together
\begin{align}
   0&= \partial_\tau^2I_0+\epsilon\partial_\tau^2 I_1+\sum_{n=2}^\infty \epsilon^n\partial^2_\tau I_n+  k^2I_0+\epsilon k^2 I_1+\sum_{n=2}^\infty \epsilon^nk^2 I_n\notag\\
    &+\frac{2(s-1)}{\tau}\left[\partial_\tau I_0+\epsilon\left(\partial_\tau I_1+\partial_\tau I_0\right)+\sum_{n=2}^\infty\epsilon^n\left(\partial_\tau I_{n}+\partial_\tau I_{n-1}\right)\right]\notag\\
    &-\frac{2(s-1)}{\tau^2}\left[I_0+\epsilon\left(I_1+2I_0\right)+\sum_{n=2}^\infty\epsilon^n\left( I_{n}+ 2I_{n-1}+I_{n-2}\right)\right]\notag\\
    &+\frac{m_s^2}{H_k^2\tau^2}\left[I_0+\epsilon\left(2I_0+\sum_{n=0}^1\sum_{m=0}^{1-n}\frac{(-1)^{m+n}}{n!m!}\mathrm{log}^n(\tilde{k}^2)\mathrm{log}^m(\tilde{\tau}^2)\right)I_{1-n-m}\right.\notag\\
    &+\sum_{g=2}^\infty\epsilon^g\left[ \sum_{n=0}^g\sum_{m=0}^{g-n}\frac{(-1)^{m+n}}{n!m!}\mathrm{log}^n(\tilde{k}^2)\mathrm{log}^m(\tilde{\tau}^2)I_{g-n-m} \right.\notag\\
    &+2\sum_{n=0}^{g-1}\sum_{m=0}^{g-1-n}\frac{(-1)^{m+n}}{n!m!}\mathrm{log}^n(\tilde{k}^2)\mathrm{log}^m(\tilde{\tau}^2)I_{g-n-m-1}\notag\\
    &\left.\left.+\sum_{n=0}^{g-2}\sum_{m=0}^{g-2-n}\frac{(-1)^{m+n}}{n!m!}\mathrm{log}^n(\tilde{k}^2)\mathrm{log}^m(\tilde{\tau}^2)I_{g-n-m-2}\right]\right]\notag\\
    &\mp\frac{sk\tilde{\kappa}^{(s)}}{H_k\tau}\left[I_0+\epsilon\left(I_0+\sum_{n=0}^1\sum_{m=0}^{1-n}\frac{(-1)^{m+n}}{n!m!}\mathrm{log}^n(\tilde{k})\mathrm{log}^m(-\tilde{\tau})\right)I_{1-n-m}\right.\notag\\
    &+\sum_{g=2}^\infty\epsilon^g\left[ \sum_{n=0}^g\sum_{m=0}^{g-n}\frac{(-1)^{m+n}}{n!m!}\mathrm{log}^n(\tilde{k})\mathrm{log}^m(-\tilde{\tau})I_{g-n-m} \right.\notag\\
    &\left.\left.+\sum_{n=0}^{g-1}\sum_{m=0}^{g-1-n}\frac{(-1)^{m+n}}{n!m!}\mathrm{log}^n(\tilde{k})\mathrm{log}^m(-\tilde{\tau})I_{g-n-m-1}\right]\right].
\end{align}
After simplification, combining terms of the same order, we obtain
\begin{align}
     0&= \partial_\tau^2I_0+\frac{2(s-1)}{\tau}\partial_\tau I_0+k^2I_0+\frac{m_s^2}{H_k^2\tau^2}I_0-\frac{2(s-1)}{\tau^2}I_0\pm\frac{sk\tilde{\kappa}^{(s)}}{H_k\tau}I_0\notag\\
     &+\epsilon\left[\partial_\tau^2 I_1+k^2 I_1+\frac{2(s-1)}{\tau}\left(\partial_\tau I_1+\partial_\tau I_0\right)-\frac{2(s-1)}{\tau^2}\left(I_1+2I_0\right)\right.\notag\\
     &+\frac{m_s^2}{H_k^2\tau^2}\left(2I_0+\sum_{n=0}^1\sum_{m=0}^{1-n}\frac{(-1)^{m+n}}{n!m!}\mathrm{log}^n(\tilde{k}^2)\mathrm{log}^m(\tilde{\tau}^2)\right)I_{1-n-m}\notag\\
     &\left.\pm\frac{sk\tilde{\kappa}^{(s)}}{H_k\tau}\left(I_0+\sum_{n=0}^1\sum_{m=0}^{1-n}\frac{(-1)^{m+n}}{n!m!}\mathrm{log}^n(\tilde{k})\mathrm{log}^m(-\tilde{\tau})\right)I_{1-n-m}\right]\notag\\
     &+\sum_{g=2}^\infty \epsilon^g\left[\partial^2_\tau I_g+k^2 I_g+\frac{2(s-1)}{\tau}\left(\partial_\tau I_{g}+\partial_\tau I_{g-1}\right)-\frac{2(s-1)}{\tau^2}\left( I_{g}+ 2I_{g-1}+I_{g-2}\right)\right.\notag\\
     &+\frac{m_s^2}{H_k^2\tau^2}\left(\sum_{n=0}^g\sum_{m=0}^{g-n}\frac{(-1)^{m+n}}{n!m!}\mathrm{log}^n(\tilde{k}^2)\mathrm{log}^m(\tilde{\tau}^2)I_{g-n-m} +2\sum_{n=0}^{g-1}\sum_{m=0}^{g-1-n}\frac{(-1)^{m+n}}{n!m!}\mathrm{log}^n(\tilde{k}^2)\mathrm{log}^m(\tilde{\tau}^2)I_{g-n-m-1}\right.\notag\\
     &\left.+\sum_{n=0}^{g-2}\sum_{m=0}^{g-2-n}\frac{(-1)^{m+n}}{n!m!}\mathrm{log}^n(\tilde{k}^2)\mathrm{log}^m(\tilde{\tau}^2)I_{g-n-m-2}\right)\notag\\
     &\left.\mp\frac{sk\tilde{\kappa}^{(s)}}{H_k\tau}\left(\sum_{n=0}^g\sum_{m=0}^{g-n}\frac{(-1)^{m+n}}{n!m!}\mathrm{log}^n(\tilde{k})\mathrm{log}^m(-\tilde{\tau})I_{g-n-m}+\sum_{n=0}^{g-1}\sum_{m=0}^{g-1-n}\frac{(-1)^{m+n}}{n!m!}\mathrm{log}^n(\tilde{k})\mathrm{log}^m(-\tilde{\tau})I_{g-n-m-1}\right)\right].
\end{align}
The zeroth-order equation yields
\begin{align}
    0= \partial_\tau^2I_0+\frac{2(s-1)}{\tau}\partial_\tau I_0+k^2I_0+\frac{m_s^2}{H_k^2\tau^2}I_0-\frac{2(s-1)}{\tau^2}I_0\mp\frac{sk\tilde{\kappa}^{(s)}}{H_k\tau}I_0.
\end{align}
The first-order equation yields
\begin{align}
    &\partial_\tau^2I_{1}+\frac{(2s-2)}{\tau}\partial_\tau I_{1}^{\pm(s)}+\left[k^2+\frac{m_s^2}{H^2_k\tau^2}-\frac{2\left(s-1\right)}{\tau^2}\right]I_{1}
    \mp sk\frac{1}{H_k\tau}{\tilde{\kappa}}^{(s)}I_{1}\notag\\
    &=-\frac{2(s-1)}{\tau}\partial_\tau I_0-\frac{1}{\tau^2}\left(2\frac{m_s^2}{H_k^2}-4(s-1)\mp sk\tau\frac{\tilde{\kappa}^{(s)}}{H_k}-\frac{m_s^2}{H_k^2}\mathrm{log}(k^2\tau^2)\pm sk\frac{\tilde{\kappa}^{(s)}}{H_k}\tau\mathrm{log}(-k\tau)\right)I_0
\end{align}
which is the same as Eq.~(\ref{firsteom}).
We can also obtain equations of higher orders ($g\geq 2$)
\begin{align}
    0&=\partial^2_\tau I_g+k^2 I_g+\frac{2(s-1)}{\tau}\partial_\tau I_{g}-\frac{2(s-1)}{\tau^2} I_{g}+\frac{2(s-1)}{\tau}\partial_\tau I_{g-1}-\frac{2(s-1)}{\tau^2}\left(2I_{g-1}+I_{g-2}\right)\notag\\
     &+\frac{m_s^2}{H_k^2\tau^2}\left(\sum_{k=0}^g\sum_{n=0}^{k}\frac{(-1)^{k}}{n!(k-n)!}\mathrm{log}^n(\tilde{k}^2)\mathrm{log}^{k-n}(\tilde{\tau}^2)I_{g-k} +2\sum_{k=0}^{g-1}\sum_{n=0}^{k}\frac{(-1)^{k}}{n!(k-n)!}\mathrm{log}^n(\tilde{k}^2)\mathrm{log}^{k-n}(\tilde{\tau}^2)I_{g-k-1}\right.\notag\\
     &\left.+\sum_{k=0}^{g-2}\sum_{n=0}^{k}\frac{(-1)^{k}}{n!(k-n)!}\mathrm{log}^n(\tilde{k}^2)\mathrm{log}^{k-n}(\tilde{\tau}^2)I_{g-k-2}\right)\notag\\
     &\mp\frac{sk\tilde{\kappa}^{(s)}}{H_k\tau}\left(\sum_{k=0}^g\sum_{n=0}^{k}\frac{(-1)^{k}}{n!(k-n)!}\mathrm{log}^n(\tilde{k})\mathrm{log}^{k-n}(-\tilde{\tau})I_{g-k}+\sum_{k=0}^{g-1}\sum_{n=0}^{k}\frac{(-1)^{k}}{n!(k-n)!}\mathrm{log}^n(\tilde{k})\mathrm{log}^{k-n}(-\tilde{\tau})I_{g-k-1}\right)\notag\\
     &\Downarrow\notag\\
     &\partial^2_\tau I_g+\frac{2(s-1)}{\tau}\partial_\tau I_{g}+k^2 I_g-\frac{2(s-1)}{\tau^2} I_{g}+\frac{m_s^2}{H_k^2\tau^2}I_g\mp\frac{sk\tilde{\kappa}^{(s)}}{H_k\tau}I_g\notag\\
     &=-\frac{2(s-1)}{\tau}\partial_\tau I_{g-1}+\frac{2(s-1)}{\tau^2}\left(2I_{g-1}+I_{g-2}\right)\notag\\
     &-\sum_{k=1}^g\sum_{n=0}^{k}\frac{(-1)^{k}}{n!(k-n)!}\left(\frac{m_s^2}{H_k^2\tau^2}\mathrm{log}^n(\tilde{k}^2)\mathrm{log}^{k-n}(\tilde{\tau}^2)\mp\frac{sk\tilde{\kappa}^{(s)}}{H_k\tau}\mathrm{log}^n(\tilde{k})\mathrm{log}^{k-n}(-\tilde{\tau})\right)I_{g-k}\notag\\
     &-\sum_{k=0}^{g-1}\sum_{n=0}^{k}\frac{(-1)^{k}}{n!(k-n)!}\left(2\frac{m_s^2}{H_k^2\tau^2}\mathrm{log}^n(\tilde{k}^2)\mathrm{log}^{k-n}(\tilde{\tau}^2)\mp\frac{sk\tilde{\kappa}^{(s)}}{H_k\tau}\mathrm{log}^n(\tilde{k})\mathrm{log}^{k-n}(-\tilde{\tau})\right)I_{g-k-1}\notag\\
     &-\sum_{k=0}^{g-2}\sum_{n=0}^{k}\frac{(-1)^{k}}{n!(k-n)!}\frac{m_s^2}{H_k^2\tau^2}\mathrm{log}^n(\tilde{k}^2)\mathrm{log}^{k-n}(\tilde{\tau}^2)I_{g-k-2},
\end{align}
where we redefine $k \equiv m+n$ and utilize the transformation relation for multiple summations, as given in Eq.~(\ref{multisum}). For higher-order solutions, the corresponding equations are non-homogeneous, but their homogeneous parts are identical to those of the zeroth-order solution. The non-homogeneous parts include contributions from lower-order solutions and their first derivatives. Thus, by iterating from the known solution of the zeroth-order equation, we can derive the first-order solution, then the second-order solution, and ultimately obtain solutions of any desired order. Consequently, for the slow-roll modified equation, expanding the solution in terms of the slow-roll parameter allows us to obtain solutions of all orders. Thus, our calculations can achieve any desired level of accuracy.

\bibliographystyle{JHEP}
\normalem
\bibliography{reference}

\providecommand{\href}[2]{#2}\begingroup\raggedright\begin{thebibliography}{100}

\bibitem{Weinberg:1964ew}
S.~Weinberg, \emph{{Photons and Gravitons in $S$-Matrix Theory: Derivation of
  Charge Conservation and Equality of Gravitational and Inertial Mass}},
  \href{https://doi.org/10.1103/PhysRev.135.B1049}{\emph{Phys. Rev.} {\bfseries
  135} (1964) B1049}.

\bibitem{Coleman:1967ad}
S.R.~Coleman and J.~Mandula, \emph{{All Possible Symmetries of the S Matrix}},
  \href{https://doi.org/10.1103/PhysRev.159.1251}{\emph{Phys. Rev.} {\bfseries
  159} (1967) 1251}.

\bibitem{Grisaru:1976vm}
M.T.~Grisaru, H.N.~Pendleton and P.~van Nieuwenhuizen, \emph{{Supergravity and
  the S Matrix}}, \href{https://doi.org/10.1103/PhysRevD.15.996}{\emph{Phys.
  Rev. D} {\bfseries 15} (1977) 996}.

\bibitem{Grisaru:1977kk}
M.T.~Grisaru and H.N.~Pendleton, \emph{{Soft Spin 3/2 Fermions Require Gravity
  and Supersymmetry}},
  \href{https://doi.org/10.1016/0370-2693(77)90383-5}{\emph{Phys. Lett. B}
  {\bfseries 67} (1977) 323}.

\bibitem{Aragone:1979hx}
C.~Aragone and S.~Deser, \emph{{Consistency Problems of Hypergravity}},
  \href{https://doi.org/10.1016/0370-2693(79)90808-6}{\emph{Phys. Lett. B}
  {\bfseries 86} (1979) 161}.

\bibitem{Weinberg:1980kq}
S.~Weinberg and E.~Witten, \emph{{Limits on Massless Particles}},
  \href{https://doi.org/10.1016/0370-2693(80)90212-9}{\emph{Phys. Lett. B}
  {\bfseries 96} (1980) 59}.

\bibitem{Porrati:2008rm}
M.~Porrati, \emph{{Universal Limits on Massless High-Spin Particles}},
  \href{https://doi.org/10.1103/PhysRevD.78.065016}{\emph{Phys. Rev. D}
  {\bfseries 78} (2008) 065016}
  [\href{https://arxiv.org/abs/0804.4672}{{\ttfamily 0804.4672}}].

\bibitem{Porrati:2012rd}
M.~Porrati, \emph{{Old and New No Go Theorems on Interacting Massless Particles
  in Flat Space}},  in \emph{{17th International Seminar on High Energy
  Physics}}, 9, 2012 [\href{https://arxiv.org/abs/1209.4876}{{\ttfamily
  1209.4876}}].

\bibitem{Bekaert:2006py}
X.~Bekaert and N.~Boulanger, \emph{{The unitary representations of the
  Poincar'e group in any spacetime dimension}},
  \href{https://doi.org/10.21468/SciPostPhysLectNotes.30}{\emph{SciPost Phys.
  Lect. Notes} {\bfseries 30} (2021) 1}
  [\href{https://arxiv.org/abs/hep-th/0611263}{{\ttfamily hep-th/0611263}}].

\bibitem{Rahman:2012thy}
R.~Rahman, \emph{{Higher Spin Theory - Part I}},
  \href{https://doi.org/10.22323/1.195.0004}{\emph{PoS} {\bfseries ModaveVIII}
  (2012) 004} [\href{https://arxiv.org/abs/1307.3199}{{\ttfamily 1307.3199}}].

\bibitem{Rahman:2015pzl}
R.~Rahman and M.~Taronna, \emph{{From Higher Spins to Strings: A Primer}},
  \href{https://doi.org/10.1007/978-3-031-59656-8_1}{\emph{Lect. Notes Phys.}
  {\bfseries 1028} (2024) 1}
  [\href{https://arxiv.org/abs/1512.07932}{{\ttfamily 1512.07932}}].

\bibitem{Ponomarev:2022vjb}
D.~Ponomarev, \emph{{Basic Introduction to Higher-Spin Theories}},
  \href{https://doi.org/10.1007/s10773-023-05399-5}{\emph{Int. J. Theor. Phys.}
  {\bfseries 62} (2023) 146}
  [\href{https://arxiv.org/abs/2206.15385}{{\ttfamily 2206.15385}}].

\bibitem{Fierz:1939ix}
M.~Fierz and W.~Pauli, \emph{{On relativistic wave equations for particles of
  arbitrary spin in an electromagnetic field}},
  \href{https://doi.org/10.1098/rspa.1939.0140}{\emph{Proc. Roy. Soc. Lond. A}
  {\bfseries 173} (1939) 211}.

\bibitem{fronsdal_theory_1958}
C.~Fronsdal, \emph{On the theory of higher spin fields},
  \href{https://doi.org/10.1007/BF02747684}{\emph{Nuovo Cim} {\bfseries 9}
  (1958) 416}.

\bibitem{Chang:1967zzc}
S.-J.~Chang, \emph{{Lagrange Formulation for Systems with Higher Spin}},
  \href{https://doi.org/10.1103/PhysRev.161.1308}{\emph{Phys. Rev.} {\bfseries
  161} (1967) 1308}.

\bibitem{Singh:1974qz}
L.P.S.~Singh and C.R.~Hagen, \emph{{Lagrangian formulation for arbitrary spin.
  1. The boson case}},
  \href{https://doi.org/10.1103/PhysRevD.9.898}{\emph{Phys. Rev. D} {\bfseries
  9} (1974) 898}.

\bibitem{deWit:1979sib}
B.~de~Wit and D.Z.~Freedman, \emph{{Systematics of Higher Spin Gauge Fields}},
  \href{https://doi.org/10.1103/PhysRevD.21.358}{\emph{Phys. Rev. D} {\bfseries
  21} (1980) 358}.

\bibitem{Arkani-Hamed:2002bjr}
N.~Arkani-Hamed, H.~Georgi and M.D.~Schwartz, \emph{{Effective field theory for
  massive gravitons and gravity in theory space}},
  \href{https://doi.org/10.1016/S0003-4916(03)00068-X}{\emph{Annals Phys.}
  {\bfseries 305} (2003) 96}
  [\href{https://arxiv.org/abs/hep-th/0210184}{{\ttfamily hep-th/0210184}}].

\bibitem{Hinterbichler:2011tt}
K.~Hinterbichler, \emph{{Theoretical Aspects of Massive Gravity}},
  \href{https://doi.org/10.1103/RevModPhys.84.671}{\emph{Rev. Mod. Phys.}
  {\bfseries 84} (2012) 671} [\href{https://arxiv.org/abs/1105.3735}{{\ttfamily
  1105.3735}}].

\bibitem{Hassan:2011zd}
S.F.~Hassan and R.A.~Rosen, \emph{{Bimetric Gravity from Ghost-free Massive
  Gravity}}, \href{https://doi.org/10.1007/JHEP02(2012)126}{\emph{JHEP}
  {\bfseries 02} (2012) 126} [\href{https://arxiv.org/abs/1109.3515}{{\ttfamily
  1109.3515}}].

\bibitem{Volkov:2011an}
M.S.~Volkov, \emph{{Cosmological solutions with massive gravitons in the
  bigravity theory}},
  \href{https://doi.org/10.1007/JHEP01(2012)035}{\emph{JHEP} {\bfseries 01}
  (2012) 035} [\href{https://arxiv.org/abs/1110.6153}{{\ttfamily 1110.6153}}].

\bibitem{Hassan:2012wr}
S.F.~Hassan, A.~Schmidt-May and M.~von Strauss, \emph{{On Consistent Theories
  of Massive Spin-2 Fields Coupled to Gravity}},
  \href{https://doi.org/10.1007/JHEP05(2013)086}{\emph{JHEP} {\bfseries 05}
  (2013) 086} [\href{https://arxiv.org/abs/1208.1515}{{\ttfamily 1208.1515}}].

\bibitem{Dalmazi:2013sva}
D.~Dalmazi, \emph{{Massive spin-2 particle from a rank-2 tensor}},
  \href{https://doi.org/10.1103/PhysRevD.87.125027}{\emph{Phys. Rev. D}
  {\bfseries 87} (2013) 125027}
  [\href{https://arxiv.org/abs/1305.1513}{{\ttfamily 1305.1513}}].

\bibitem{Kluson:2013jlo}
J.~Kluson, \emph{{Is Bimetric Gravity Really Ghost Free?}},
  \href{https://doi.org/10.1142/S0217751X13501431}{\emph{Int. J. Mod. Phys. A}
  {\bfseries 28} (2013) 1350143}
  [\href{https://arxiv.org/abs/1301.3296}{{\ttfamily 1301.3296}}].

\bibitem{deRham:2014zqa}
C.~de~Rham, \emph{{Massive Gravity}},
  \href{https://doi.org/10.12942/lrr-2014-7}{\emph{Living Rev. Rel.} {\bfseries
  17} (2014) 7} [\href{https://arxiv.org/abs/1401.4173}{{\ttfamily
  1401.4173}}].

\bibitem{Koenigstein:2015asa}
A.~Koenigstein, F.~Giacosa and D.H.~Rischke, \emph{{Classical and quantum
  theory of the massive spin-two field}},
  \href{https://doi.org/10.1016/j.aop.2016.01.024}{\emph{Annals Phys.}
  {\bfseries 368} (2016) 16}
  [\href{https://arxiv.org/abs/1508.00110}{{\ttfamily 1508.00110}}].

\bibitem{Goon:2018fyu}
G.~Goon, K.~Hinterbichler, A.~Joyce and M.~Trodden, \emph{{Shapes of gravity:
  Tensor non-Gaussianity and massive spin-2 fields}},
  \href{https://doi.org/10.1007/JHEP10(2019)182}{\emph{JHEP} {\bfseries 10}
  (2019) 182} [\href{https://arxiv.org/abs/1812.07571}{{\ttfamily
  1812.07571}}].

\bibitem{Kaluza:1921tu}
T.~Kaluza, \emph{{Zum Unit\"atsproblem der Physik}},
  \href{https://doi.org/10.1142/S0218271818700017}{\emph{Sitzungsber. Preuss.
  Akad. Wiss. Berlin (Math. Phys. )} {\bfseries 1921} (1921) 966}
  [\href{https://arxiv.org/abs/1803.08616}{{\ttfamily 1803.08616}}].

\bibitem{Klein:1926tv}
O.~Klein, \emph{{Quantum Theory and Five-Dimensional Theory of Relativity. (In
  German and English)}}, \href{https://doi.org/10.1007/BF01397481}{\emph{Z.
  Phys.} {\bfseries 37} (1926) 895}.

\bibitem{Aragone:1987dtt}
C.~Aragone, S.~Deser and Z.~Yang, \emph{{Massive Higher Spin From Dimensional
  Reduction of Gauge Fields}},
  \href{https://doi.org/10.1016/S0003-4916(87)80005-2}{\emph{Annals Phys.}
  {\bfseries 179} (1987) 76}.

\bibitem{Randall:1999ee}
L.~Randall and R.~Sundrum, \emph{{A Large mass hierarchy from a small extra
  dimension}}, \href{https://doi.org/10.1103/PhysRevLett.83.3370}{\emph{Phys.
  Rev. Lett.} {\bfseries 83} (1999) 3370}
  [\href{https://arxiv.org/abs/hep-ph/9905221}{{\ttfamily hep-ph/9905221}}].

\bibitem{Csaki:2018muy}
C.~Cs\'aki, S.~Lombardo and O.~Telem, \emph{{TASI Lectures on
  Non-supersymmetric BSM Models}},  in \emph{{Proceedings, Theoretical Advanced
  Study Institute in Elementary Particle Physics : Anticipating the Next
  Discoveries in Particle Physics (TASI 2016)}: {Boulder, CO, USA, June 6-July
  1, 2016}}, R.~Essig and I.~Low, eds., pp.~501--570, WSP (2018),
  \href{https://doi.org/10.1142/9789813233348_0007}{DOI}
  [\href{https://arxiv.org/abs/1811.04279}{{\ttfamily 1811.04279}}].

\bibitem{Witten:1979kh}
E.~Witten, \emph{{Baryons in the 1/n Expansion}},
  \href{https://doi.org/10.1016/0550-3213(79)90232-3}{\emph{Nucl. Phys. B}
  {\bfseries 160} (1979) 57}.

\bibitem{Sorokin:2004ie}
D.~Sorokin, \emph{{Introduction to the classical theory of higher spins}},
  \href{https://doi.org/10.1063/1.1923335}{\emph{AIP Conf. Proc.} {\bfseries
  767} (2005) 172} [\href{https://arxiv.org/abs/hep-th/0405069}{{\ttfamily
  hep-th/0405069}}].

\bibitem{Sagnotti:2011jdy}
A.~Sagnotti, \emph{{Notes on Strings and Higher Spins}},
  \href{https://doi.org/10.1088/1751-8113/46/21/214006}{\emph{J. Phys. A}
  {\bfseries 46} (2013) 214006}
  [\href{https://arxiv.org/abs/1112.4285}{{\ttfamily 1112.4285}}].

\bibitem{Chang:2012kt}
C.-M.~Chang, S.~Minwalla, T.~Sharma and X.~Yin, \emph{{ABJ Triality: from
  Higher Spin Fields to Strings}},
  \href{https://doi.org/10.1088/1751-8113/46/21/214009}{\emph{J. Phys. A}
  {\bfseries 46} (2013) 214009}
  [\href{https://arxiv.org/abs/1207.4485}{{\ttfamily 1207.4485}}].

\bibitem{Buchbinder:2000ta}
I.L.~Buchbinder and V.D.~Pershin, \emph{{Gravitational interaction of higher
  spin massive fields and string theory}},  in \emph{{Conference on Geometrical
  Aspects of Quantum Fields}}, pp.~11--30, 4, 2000,
  \href{https://doi.org/10.1142/9789812810366_0002}{DOI}
  [\href{https://arxiv.org/abs/hep-th/0009026}{{\ttfamily hep-th/0009026}}].

\bibitem{Camanho:2014apa}
X.O.~Camanho, J.D.~Edelstein, J.~Maldacena and A.~Zhiboedov, \emph{{Causality
  Constraints on Corrections to the Graviton Three-Point Coupling}},
  \href{https://doi.org/10.1007/JHEP02(2016)020}{\emph{JHEP} {\bfseries 02}
  (2016) 020} [\href{https://arxiv.org/abs/1407.5597}{{\ttfamily 1407.5597}}].

\bibitem{Noumi:2019ohm}
T.~Noumi, T.~Takeuchi and S.~Zhou, \emph{{String Regge trajectory in de Sitter
  space and implications for inflation}},
  \href{https://doi.org/10.1103/PhysRevD.102.126012}{\emph{Phys. Rev. D}
  {\bfseries 102} (2020) 126012}
  [\href{https://arxiv.org/abs/1907.02535}{{\ttfamily 1907.02535}}].

\bibitem{Kato:2021rdz}
M.~Kato, K.~Nishii, T.~Noumi, T.~Takeuchi and S.~Zhou, \emph{{Spiky strings in
  de Sitter space}}, \href{https://doi.org/10.1007/JHEP05(2021)047}{\emph{JHEP}
  {\bfseries 05} (2021) 047}
  [\href{https://arxiv.org/abs/2102.09746}{{\ttfamily 2102.09746}}].

\bibitem{Basile:2016aen}
T.~Basile, X.~Bekaert and N.~Boulanger, \emph{{Mixed-symmetry fields in de
  Sitter space: a group theoretical glance}},
  \href{https://doi.org/10.1007/JHEP05(2017)081}{\emph{JHEP} {\bfseries 05}
  (2017) 081} [\href{https://arxiv.org/abs/1612.08166}{{\ttfamily
  1612.08166}}].

\bibitem{Sun:2021thf}
Z.~Sun, \emph{{A note on the representations of SO(1,d + 1)}},
  \href{https://doi.org/10.1142/S0129055X24300073}{\emph{Rev. Math. Phys.}
  {\bfseries 37} (2025) 2430007}
  [\href{https://arxiv.org/abs/2111.04591}{{\ttfamily 2111.04591}}].

\bibitem{Guth:1980zm}
A.H.~Guth, \emph{{The Inflationary Universe: A Possible Solution to the Horizon
  and Flatness Problems}},
  \href{https://doi.org/10.1103/PhysRevD.23.347}{\emph{Phys. Rev. D} {\bfseries
  23} (1981) 347}.

\bibitem{Linde:1981mu}
A.D.~Linde, \emph{{A New Inflationary Universe Scenario: A Possible Solution of
  the Horizon, Flatness, Homogeneity, Isotropy and Primordial Monopole
  Problems}}, \href{https://doi.org/10.1016/0370-2693(82)91219-9}{\emph{Phys.
  Lett. B} {\bfseries 108} (1982) 389}.

\bibitem{Albrecht:1982wi}
A.~Albrecht and P.J.~Steinhardt, \emph{{Cosmology for Grand Unified Theories
  with Radiatively Induced Symmetry Breaking}},
  \href{https://doi.org/10.1103/PhysRevLett.48.1220}{\emph{Phys. Rev. Lett.}
  {\bfseries 48} (1982) 1220}.

\bibitem{Baumann:2009ds}
D.~Baumann, \emph{{Inflation}},  in \emph{{Theoretical Advanced Study Institute
  in Elementary Particle Physics}: {Physics of the Large and the Small}},
  pp.~523--686, 2011, \href{https://doi.org/10.1142/9789814327183_0010}{DOI}
  [\href{https://arxiv.org/abs/0907.5424}{{\ttfamily 0907.5424}}].

\bibitem{Wang:2013zva}
Y.~Wang, \emph{{Inflation, Cosmic Perturbations and Non-Gaussianities}},
  \href{https://doi.org/10.1088/0253-6102/62/1/19}{\emph{Commun. Theor. Phys.}
  {\bfseries 62} (2014) 109} [\href{https://arxiv.org/abs/1303.1523}{{\ttfamily
  1303.1523}}].

\bibitem{Meerburg:2015zua}
P.D.~Meerburg, R.~Hlo\v{z}ek, B.~Hadzhiyska and J.~Meyers,
  \emph{{Multiwavelength constraints on the inflationary consistency
  relation}}, \href{https://doi.org/10.1103/PhysRevD.91.103505}{\emph{Phys.
  Rev. D} {\bfseries 91} (2015) 103505}
  [\href{https://arxiv.org/abs/1502.00302}{{\ttfamily 1502.00302}}].

\bibitem{Planck:2018jri}
{\scshape Planck} collaboration, \emph{{Planck 2018 results. X. Constraints on
  inflation}}, \href{https://doi.org/10.1051/0004-6361/201833887}{\emph{Astron.
  Astrophys.} {\bfseries 641} (2020) A10}
  [\href{https://arxiv.org/abs/1807.06211}{{\ttfamily 1807.06211}}].

\bibitem{Planck:2019kim}
{\scshape Planck} collaboration, \emph{{Planck 2018 results. IX. Constraints on
  primordial non-Gaussianity}},
  \href{https://doi.org/10.1051/0004-6361/201935891}{\emph{Astron. Astrophys.}
  {\bfseries 641} (2020) A9}
  [\href{https://arxiv.org/abs/1905.05697}{{\ttfamily 1905.05697}}].

\bibitem{Galloni:2022mok}
G.~Galloni, N.~Bartolo, S.~Matarrese, M.~Migliaccio, A.~Ricciardone and
  N.~Vittorio, \emph{{Updated constraints on amplitude and tilt of the tensor
  primordial spectrum}},
  \href{https://doi.org/10.1088/1475-7516/2023/04/062}{\emph{JCAP} {\bfseries
  04} (2023) 062} [\href{https://arxiv.org/abs/2208.00188}{{\ttfamily
  2208.00188}}].

\bibitem{BICEP:2021xfz}
{\scshape BICEP, Keck} collaboration, \emph{{Improved Constraints on Primordial
  Gravitational Waves using Planck, WMAP, and BICEP/Keck Observations through
  the 2018 Observing Season}},
  \href{https://doi.org/10.1103/PhysRevLett.127.151301}{\emph{Phys. Rev. Lett.}
  {\bfseries 127} (2021) 151301}
  [\href{https://arxiv.org/abs/2110.00483}{{\ttfamily 2110.00483}}].

\bibitem{Sohn:2024xzd}
W.~Sohn, D.-G.~Wang, J.R.~Fergusson and E.P.S.~Shellard, \emph{{Searching for
  cosmological collider in the Planck CMB data}},
  \href{https://doi.org/10.1088/1475-7516/2024/09/016}{\emph{JCAP} {\bfseries
  09} (2024) 016} [\href{https://arxiv.org/abs/2404.07203}{{\ttfamily
  2404.07203}}].

\bibitem{Dalal:2007cu}
N.~Dalal, O.~Dore, D.~Huterer and A.~Shirokov, \emph{{The imprints of
  primordial non-gaussianities on large-scale structure: scale dependent bias
  and abundance of virialized objects}},
  \href{https://doi.org/10.1103/PhysRevD.77.123514}{\emph{Phys. Rev. D}
  {\bfseries 77} (2008) 123514}
  [\href{https://arxiv.org/abs/0710.4560}{{\ttfamily 0710.4560}}].

\bibitem{Slosar:2008hx}
A.~Slosar, C.~Hirata, U.~Seljak, S.~Ho and N.~Padmanabhan, \emph{{Constraints
  on local primordial non-Gaussianity from large scale structure}},
  \href{https://doi.org/10.1088/1475-7516/2008/08/031}{\emph{JCAP} {\bfseries
  08} (2008) 031} [\href{https://arxiv.org/abs/0805.3580}{{\ttfamily
  0805.3580}}].

\bibitem{Cabass:2024wob}
G.~Cabass, O.H.E.~Philcox, M.M.~Ivanov, K.~Akitsu, S.-F.~Chen, M.~Simonovi\'c
  et~al., \emph{{BOSS constraints on massive particles during inflation: The
  cosmological collider in action}},
  \href{https://doi.org/10.1103/PhysRevD.111.063510}{\emph{Phys. Rev. D}
  {\bfseries 111} (2025) 063510}
  [\href{https://arxiv.org/abs/2404.01894}{{\ttfamily 2404.01894}}].

\bibitem{Chen:2009zp}
X.~Chen and Y.~Wang, \emph{{Quasi-Single Field Inflation and
  Non-Gaussianities}},
  \href{https://doi.org/10.1088/1475-7516/2010/04/027}{\emph{JCAP} {\bfseries
  04} (2010) 027} [\href{https://arxiv.org/abs/0911.3380}{{\ttfamily
  0911.3380}}].

\bibitem{Baumann:2011nk}
D.~Baumann and D.~Green, \emph{{Signatures of Supersymmetry from the Early
  Universe}}, \href{https://doi.org/10.1103/PhysRevD.85.103520}{\emph{Phys.
  Rev. D} {\bfseries 85} (2012) 103520}
  [\href{https://arxiv.org/abs/1109.0292}{{\ttfamily 1109.0292}}].

\bibitem{Noumi:2012vr}
T.~Noumi, M.~Yamaguchi and D.~Yokoyama, \emph{{Effective field theory approach
  to quasi-single field inflation and effects of heavy fields}},
  \href{https://doi.org/10.1007/JHEP06(2013)051}{\emph{JHEP} {\bfseries 06}
  (2013) 051} [\href{https://arxiv.org/abs/1211.1624}{{\ttfamily 1211.1624}}].

\bibitem{Gong:2013sma}
J.-O.~Gong, S.~Pi and M.~Sasaki, \emph{{Equilateral non-Gaussianity from heavy
  fields}}, \href{https://doi.org/10.1088/1475-7516/2013/11/043}{\emph{JCAP}
  {\bfseries 11} (2013) 043} [\href{https://arxiv.org/abs/1306.3691}{{\ttfamily
  1306.3691}}].

\bibitem{Arkani-Hamed:2015bza}
N.~Arkani-Hamed and J.~Maldacena, \emph{{Cosmological Collider Physics}},
  \href{https://arxiv.org/abs/1503.08043}{{\ttfamily 1503.08043}}.

\bibitem{Higuchi:1986py}
A.~Higuchi, \emph{{Forbidden Mass Range for Spin-2 Field Theory in De Sitter
  Space-time}}, \href{https://doi.org/10.1016/0550-3213(87)90691-2}{\emph{Nucl.
  Phys. B} {\bfseries 282} (1987) 397}.

\bibitem{Zinoviev:2001dt}
Y.M.~Zinoviev, \emph{{On massive high spin particles in AdS}},
  \href{https://arxiv.org/abs/hep-th/0108192}{{\ttfamily hep-th/0108192}}.

\bibitem{Deser:2003gw}
S.~Deser and A.~Waldron, \emph{{Arbitrary spin representations in de Sitter
  from dS / CFT with applications to dS supergravity}},
  \href{https://doi.org/10.1016/S0550-3213(03)00348-1}{\emph{Nucl. Phys. B}
  {\bfseries 662} (2003) 379}
  [\href{https://arxiv.org/abs/hep-th/0301068}{{\ttfamily hep-th/0301068}}].

\bibitem{Hinterbichler:2016fgl}
K.~Hinterbichler and A.~Joyce, \emph{{Manifest Duality for Partially Massless
  Higher Spins}}, \href{https://doi.org/10.1007/JHEP09(2016)141}{\emph{JHEP}
  {\bfseries 09} (2016) 141}
  [\href{https://arxiv.org/abs/1608.04385}{{\ttfamily 1608.04385}}].

\bibitem{Lee:2016vti}
H.~Lee, D.~Baumann and G.L.~Pimentel, \emph{{Non-Gaussianity as a Particle
  Detector}}, \href{https://doi.org/10.1007/JHEP12(2016)040}{\emph{JHEP}
  {\bfseries 12} (2016) 040}
  [\href{https://arxiv.org/abs/1607.03735}{{\ttfamily 1607.03735}}].

\bibitem{Baumann:2017jvh}
D.~Baumann, G.~Goon, H.~Lee and G.L.~Pimentel, \emph{{Partially Massless Fields
  During Inflation}},
  \href{https://doi.org/10.1007/JHEP04(2018)140}{\emph{JHEP} {\bfseries 04}
  (2018) 140} [\href{https://arxiv.org/abs/1712.06624}{{\ttfamily
  1712.06624}}].

\bibitem{Turner:1987bw}
M.S.~Turner and L.M.~Widrow, \emph{{Inflation Produced, Large Scale Magnetic
  Fields}}, \href{https://doi.org/10.1103/PhysRevD.37.2743}{\emph{Phys. Rev. D}
  {\bfseries 37} (1988) 2743}.

\bibitem{Garretson:1992vt}
W.D.~Garretson, G.B.~Field and S.M.~Carroll, \emph{{Primordial magnetic fields
  from pseudoGoldstone bosons}},
  \href{https://doi.org/10.1103/PhysRevD.46.5346}{\emph{Phys. Rev. D}
  {\bfseries 46} (1992) 5346}
  [\href{https://arxiv.org/abs/hep-ph/9209238}{{\ttfamily hep-ph/9209238}}].

\bibitem{Barnaby:2010vf}
N.~Barnaby and M.~Peloso, \emph{{Large Nongaussianity in Axion Inflation}},
  \href{https://doi.org/10.1103/PhysRevLett.106.181301}{\emph{Phys. Rev. Lett.}
  {\bfseries 106} (2011) 181301}
  [\href{https://arxiv.org/abs/1011.1500}{{\ttfamily 1011.1500}}].

\bibitem{Barnaby:2011vw}
N.~Barnaby, R.~Namba and M.~Peloso, \emph{{Phenomenology of a Pseudo-Scalar
  Inflaton: Naturally Large Nongaussianity}},
  \href{https://doi.org/10.1088/1475-7516/2011/04/009}{\emph{JCAP} {\bfseries
  04} (2011) 009} [\href{https://arxiv.org/abs/1102.4333}{{\ttfamily
  1102.4333}}].

\bibitem{Adshead:2015kza}
P.~Adshead and E.I.~Sfakianakis, \emph{{Fermion production during and after
  axion inflation}},
  \href{https://doi.org/10.1088/1475-7516/2015/11/021}{\emph{JCAP} {\bfseries
  11} (2015) 021} [\href{https://arxiv.org/abs/1508.00891}{{\ttfamily
  1508.00891}}].

\bibitem{Adshead:2018oaa}
P.~Adshead, L.~Pearce, M.~Peloso, M.A.~Roberts and L.~Sorbo,
  \emph{{Phenomenology of fermion production during axion inflation}},
  \href{https://doi.org/10.1088/1475-7516/2018/06/020}{\emph{JCAP} {\bfseries
  06} (2018) 020} [\href{https://arxiv.org/abs/1803.04501}{{\ttfamily
  1803.04501}}].

\bibitem{Chen:2018xck}
X.~Chen, Y.~Wang and Z.-Z.~Xianyu, \emph{{Neutrino Signatures in Primordial
  Non-Gaussianities}},
  \href{https://doi.org/10.1007/JHEP09(2018)022}{\emph{JHEP} {\bfseries 09}
  (2018) 022} [\href{https://arxiv.org/abs/1805.02656}{{\ttfamily
  1805.02656}}].

\bibitem{Wang:2019gbi}
L.-T.~Wang and Z.-Z.~Xianyu, \emph{{In Search of Large Signals at the
  Cosmological Collider}},
  \href{https://doi.org/10.1007/JHEP02(2020)044}{\emph{JHEP} {\bfseries 02}
  (2020) 044} [\href{https://arxiv.org/abs/1910.12876}{{\ttfamily
  1910.12876}}].

\bibitem{Bodas:2020yho}
A.~Bodas, S.~Kumar and R.~Sundrum, \emph{{The Scalar Chemical Potential in
  Cosmological Collider Physics}},
  \href{https://doi.org/10.1007/JHEP02(2021)079}{\emph{JHEP} {\bfseries 02}
  (2021) 079} [\href{https://arxiv.org/abs/2010.04727}{{\ttfamily
  2010.04727}}].

\bibitem{Wang:2020ioa}
L.-T.~Wang and Z.-Z.~Xianyu, \emph{{Gauge Boson Signals at the Cosmological
  Collider}}, \href{https://doi.org/10.1007/JHEP11(2020)082}{\emph{JHEP}
  {\bfseries 11} (2020) 082}
  [\href{https://arxiv.org/abs/2004.02887}{{\ttfamily 2004.02887}}].

\bibitem{Sou:2021juh}
C.M.~Sou, X.~Tong and Y.~Wang, \emph{{Chemical-potential-assisted particle
  production in FRW spacetimes}},
  \href{https://doi.org/10.1007/JHEP06(2021)129}{\emph{JHEP} {\bfseries 06}
  (2021) 129} [\href{https://arxiv.org/abs/2104.08772}{{\ttfamily
  2104.08772}}].

\bibitem{Wang:2021qez}
L.-T.~Wang, Z.-Z.~Xianyu and Y.-M.~Zhong, \emph{{Precision calculation of
  inflation correlators at one loop}},
  \href{https://doi.org/10.1007/JHEP02(2022)085}{\emph{JHEP} {\bfseries 02}
  (2022) 085} [\href{https://arxiv.org/abs/2109.14635}{{\ttfamily
  2109.14635}}].

\bibitem{Tong:2022cdz}
X.~Tong and Z.-Z.~Xianyu, \emph{{Large spin-2 signals at the cosmological
  collider}}, \href{https://doi.org/10.1007/JHEP10(2022)194}{\emph{JHEP}
  {\bfseries 10} (2022) 194}
  [\href{https://arxiv.org/abs/2203.06349}{{\ttfamily 2203.06349}}].

\bibitem{Qin:2022lva}
Z.~Qin and Z.-Z.~Xianyu, \emph{{Phase information in cosmological collider
  signals}}, \href{https://doi.org/10.1007/JHEP10(2022)192}{\emph{JHEP}
  {\bfseries 10} (2022) 192}
  [\href{https://arxiv.org/abs/2205.01692}{{\ttfamily 2205.01692}}].

\bibitem{Qin:2022fbv}
Z.~Qin and Z.-Z.~Xianyu, \emph{{Helical inflation correlators: partial
  Mellin-Barnes and bootstrap equations}},
  \href{https://doi.org/10.1007/JHEP04(2023)059}{\emph{JHEP} {\bfseries 04}
  (2023) 059} [\href{https://arxiv.org/abs/2208.13790}{{\ttfamily
  2208.13790}}].

\bibitem{Lue:1998mq}
A.~Lue, L.-M.~Wang and M.~Kamionkowski, \emph{{Cosmological signature of new
  parity violating interactions}},
  \href{https://doi.org/10.1103/PhysRevLett.83.1506}{\emph{Phys. Rev. Lett.}
  {\bfseries 83} (1999) 1506}
  [\href{https://arxiv.org/abs/astro-ph/9812088}{{\ttfamily
  astro-ph/9812088}}].

\bibitem{Alexander:2004us}
S.H.-S.~Alexander, M.E.~Peskin and M.M.~Sheikh-Jabbari, \emph{{Leptogenesis
  from gravity waves in models of inflation}},
  \href{https://doi.org/10.1103/PhysRevLett.96.081301}{\emph{Phys. Rev. Lett.}
  {\bfseries 96} (2006) 081301}
  [\href{https://arxiv.org/abs/hep-th/0403069}{{\ttfamily hep-th/0403069}}].

\bibitem{Anber:2012du}
M.M.~Anber and L.~Sorbo, \emph{{Non-Gaussianities and chiral gravitational
  waves in natural steep inflation}},
  \href{https://doi.org/10.1103/PhysRevD.85.123537}{\emph{Phys. Rev. D}
  {\bfseries 85} (2012) 123537}
  [\href{https://arxiv.org/abs/1203.5849}{{\ttfamily 1203.5849}}].

\bibitem{Crowder:2012ik}
S.G.~Crowder, R.~Namba, V.~Mandic, S.~Mukohyama and M.~Peloso,
  \emph{{Measurement of Parity Violation in the Early Universe using
  Gravitational-wave Detectors}},
  \href{https://doi.org/10.1016/j.physletb.2013.08.077}{\emph{Phys. Lett. B}
  {\bfseries 726} (2013) 66} [\href{https://arxiv.org/abs/1212.4165}{{\ttfamily
  1212.4165}}].

\bibitem{Domcke:2016bkh}
V.~Domcke, M.~Pieroni and P.~Bin\'etruy, \emph{{Primordial gravitational waves
  for universality classes of pseudoscalar inflation}},
  \href{https://doi.org/10.1088/1475-7516/2016/06/031}{\emph{JCAP} {\bfseries
  06} (2016) 031} [\href{https://arxiv.org/abs/1603.01287}{{\ttfamily
  1603.01287}}].

\bibitem{Machado:2018nqk}
C.S.~Machado, W.~Ratzinger, P.~Schwaller and B.A.~Stefanek, \emph{{Audible
  Axions}}, \href{https://doi.org/10.1007/JHEP01(2019)053}{\emph{JHEP}
  {\bfseries 01} (2019) 053}
  [\href{https://arxiv.org/abs/1811.01950}{{\ttfamily 1811.01950}}].

\bibitem{Machado:2019xuc}
C.S.~Machado, W.~Ratzinger, P.~Schwaller and B.A.~Stefanek,
  \emph{{Gravitational wave probes of axionlike particles}},
  \href{https://doi.org/10.1103/PhysRevD.102.075033}{\emph{Phys. Rev. D}
  {\bfseries 102} (2020) 075033}
  [\href{https://arxiv.org/abs/1912.01007}{{\ttfamily 1912.01007}}].

\bibitem{Salehian:2020dsf}
B.~Salehian, M.A.~Gorji, S.~Mukohyama and H.~Firouzjahi, \emph{{Analytic study
  of dark photon and gravitational wave production from axion}},
  \href{https://doi.org/10.1007/JHEP05(2021)043}{\emph{JHEP} {\bfseries 05}
  (2021) 043} [\href{https://arxiv.org/abs/2007.08148}{{\ttfamily
  2007.08148}}].

\bibitem{Niu:2022quw}
X.~Niu, M.H.~Rahat, K.~Srinivasan and W.~Xue, \emph{{Gravitational wave probes
  of massive gauge bosons at the cosmological collider}},
  \href{https://doi.org/10.1088/1475-7516/2023/02/013}{\emph{JCAP} {\bfseries
  02} (2023) 013} [\href{https://arxiv.org/abs/2211.14331}{{\ttfamily
  2211.14331}}].

\bibitem{Maleknejad:2018nxz}
A.~Maleknejad and E.~Komatsu, \emph{{Production and Backreaction of Spin-2
  Particles of $SU(2)$ Gauge Field during Inflation}},
  \href{https://doi.org/10.1007/JHEP05(2019)174}{\emph{JHEP} {\bfseries 05}
  (2019) 174} [\href{https://arxiv.org/abs/1808.09076}{{\ttfamily
  1808.09076}}].

\bibitem{NANOGrav:2023hde}
{\scshape NANOGrav} collaboration, \emph{{The NANOGrav 15 yr Data Set:
  Observations and Timing of 68 Millisecond Pulsars}},
  \href{https://doi.org/10.3847/2041-8213/acda9a}{\emph{Astrophys. J. Lett.}
  {\bfseries 951} (2023) L9}
  [\href{https://arxiv.org/abs/2306.16217}{{\ttfamily 2306.16217}}].

\bibitem{NANOGrav:2023hvm}
{\scshape NANOGrav} collaboration, \emph{{The NANOGrav 15 yr Data Set: Search
  for Signals from New Physics}},
  \href{https://doi.org/10.3847/2041-8213/acdc91}{\emph{Astrophys. J. Lett.}
  {\bfseries 951} (2023) L11}
  [\href{https://arxiv.org/abs/2306.16219}{{\ttfamily 2306.16219}}].

\bibitem{NANOGrav:2023gor}
{\scshape NANOGrav} collaboration, \emph{{The NANOGrav 15 yr Data Set: Evidence
  for a Gravitational-wave Background}},
  \href{https://doi.org/10.3847/2041-8213/acdac6}{\emph{Astrophys. J. Lett.}
  {\bfseries 951} (2023) L8}
  [\href{https://arxiv.org/abs/2306.16213}{{\ttfamily 2306.16213}}].

\bibitem{EPTA:2023sfo}
{\scshape EPTA} collaboration, \emph{{The second data release from the European
  Pulsar Timing Array - I. The dataset and timing analysis}},
  \href{https://doi.org/10.1051/0004-6361/202346841}{\emph{Astron. Astrophys.}
  {\bfseries 678} (2023) A48}
  [\href{https://arxiv.org/abs/2306.16224}{{\ttfamily 2306.16224}}].

\bibitem{EPTA:2023fyk}
{\scshape EPTA, InPTA:} collaboration, \emph{{The second data release from the
  European Pulsar Timing Array - III. Search for gravitational wave signals}},
  \href{https://doi.org/10.1051/0004-6361/202346844}{\emph{Astron. Astrophys.}
  {\bfseries 678} (2023) A50}
  [\href{https://arxiv.org/abs/2306.16214}{{\ttfamily 2306.16214}}].

\bibitem{EPTA:2023xxk}
{\scshape EPTA, InPTA} collaboration, \emph{{The second data release from the
  European Pulsar Timing Array - IV. Implications for massive black holes, dark
  matter, and the early Universe}},
  \href{https://doi.org/10.1051/0004-6361/202347433}{\emph{Astron. Astrophys.}
  {\bfseries 685} (2024) A94}
  [\href{https://arxiv.org/abs/2306.16227}{{\ttfamily 2306.16227}}].

\bibitem{Verbiest:2016vem}
J.P.W.~Verbiest et~al., \emph{{The International Pulsar Timing Array: First
  Data Release}}, \href{https://doi.org/10.1093/mnras/stw347}{\emph{Mon. Not.
  Roy. Astron. Soc.} {\bfseries 458} (2016) 1267}
  [\href{https://arxiv.org/abs/1602.03640}{{\ttfamily 1602.03640}}].

\bibitem{Janssen:2014dka}
G.~Janssen et~al., \emph{{Gravitational wave astronomy with the SKA}},
  \href{https://doi.org/10.22323/1.215.0037}{\emph{PoS} {\bfseries AASKA14}
  (2015) 037} [\href{https://arxiv.org/abs/1501.00127}{{\ttfamily
  1501.00127}}].

\bibitem{LISA:2017pwj}
{\scshape LISA} collaboration, \emph{{Laser Interferometer Space Antenna}},
  \href{https://arxiv.org/abs/1702.00786}{{\ttfamily 1702.00786}}.

\bibitem{Ruan:2018tsw}
W.-H.~Ruan, Z.-K.~Guo, R.-G.~Cai and Y.-Z.~Zhang, \emph{{Taiji program:
  Gravitational-wave sources}},
  \href{https://doi.org/10.1142/S0217751X2050075X}{\emph{Int. J. Mod. Phys. A}
  {\bfseries 35} (2020) 2050075}
  [\href{https://arxiv.org/abs/1807.09495}{{\ttfamily 1807.09495}}].

\bibitem{TianQin:2020hid}
{\scshape TianQin} collaboration, \emph{{The TianQin project: current progress
  on science and technology}},
  \href{https://doi.org/10.1093/ptep/ptaa114}{\emph{PTEP} {\bfseries 2021}
  (2021) 05A107} [\href{https://arxiv.org/abs/2008.10332}{{\ttfamily
  2008.10332}}].

\bibitem{Kudoh:2005as}
H.~Kudoh, A.~Taruya, T.~Hiramatsu and Y.~Himemoto, \emph{{Detecting a
  gravitational-wave background with next-generation space interferometers}},
  \href{https://doi.org/10.1103/PhysRevD.73.064006}{\emph{Phys. Rev. D}
  {\bfseries 73} (2006) 064006}
  [\href{https://arxiv.org/abs/gr-qc/0511145}{{\ttfamily gr-qc/0511145}}].

\bibitem{Cutler:2005qq}
C.~Cutler and J.~Harms, \emph{{BBO and the neutron-star-binary subtraction
  problem}}, \href{https://doi.org/10.1103/PhysRevD.73.042001}{\emph{Phys. Rev.
  D} {\bfseries 73} (2006) 042001}
  [\href{https://arxiv.org/abs/gr-qc/0511092}{{\ttfamily gr-qc/0511092}}].

\bibitem{Kawamura:2020pcg}
S.~Kawamura et~al., \emph{{Current status of space gravitational wave antenna
  DECIGO and B-DECIGO}},
  \href{https://doi.org/10.1093/ptep/ptab019}{\emph{PTEP} {\bfseries 2021}
  (2021) 05A105} [\href{https://arxiv.org/abs/2006.13545}{{\ttfamily
  2006.13545}}].

\bibitem{Hild:2008ng}
S.~Hild, S.~Chelkowski and A.~Freise, \emph{{Pushing towards the ET sensitivity
  using 'conventional' technology}},
  \href{https://arxiv.org/abs/0810.0604}{{\ttfamily 0810.0604}}.

\bibitem{Abac:2025saz}
A.~Abac et~al., \emph{{The Science of the Einstein Telescope}},
  \href{https://arxiv.org/abs/2503.12263}{{\ttfamily 2503.12263}}.

\bibitem{KAGRA:2021kbb}
{\scshape KAGRA, Virgo, LIGO Scientific} collaboration, \emph{{Upper limits on
  the isotropic gravitational-wave background from Advanced LIGO and Advanced
  Virgo\textquoteright{}s third observing run}},
  \href{https://doi.org/10.1103/PhysRevD.104.022004}{\emph{Phys. Rev. D}
  {\bfseries 104} (2021) 022004}
  [\href{https://arxiv.org/abs/2101.12130}{{\ttfamily 2101.12130}}].

\bibitem{LIGOScientific:2022sts}
{\scshape LIGO Scientific, Virgo,, KAGRA, VIRGO} collaboration, \emph{{Search
  for Gravitational-wave Transients Associated with Magnetar Bursts in Advanced
  LIGO and Advanced Virgo Data from the Third Observing Run}},
  \href{https://doi.org/10.3847/1538-4357/ad27d3}{\emph{Astrophys. J.}
  {\bfseries 966} (2024) 137}
  [\href{https://arxiv.org/abs/2210.10931}{{\ttfamily 2210.10931}}].

\bibitem{Jiang:2022uxp}
Y.~Jiang and Q.-G.~Huang, \emph{{Upper limits on the polarized isotropic
  stochastic gravitational-wave background from advanced LIGO-Virgo's first
  three observing runs}},
  \href{https://doi.org/10.1088/1475-7516/2023/02/026}{\emph{JCAP} {\bfseries
  02} (2023) 026} [\href{https://arxiv.org/abs/2210.09952}{{\ttfamily
  2210.09952}}].

\bibitem{Lozanov:2018kpk}
K.D.~Lozanov, A.~Maleknejad and E.~Komatsu, \emph{{Schwinger Effect by an
  $SU(2)$ Gauge Field during Inflation}},
  \href{https://doi.org/10.1007/JHEP02(2019)041}{\emph{JHEP} {\bfseries 02}
  (2019) 041} [\href{https://arxiv.org/abs/1805.09318}{{\ttfamily
  1805.09318}}].

\bibitem{Maleknejad:2019hdr}
A.~Maleknejad, \emph{{Dark Fermions and Spontaneous $CP$ violation in
  $SU(2)$-axion Inflation}},
  \href{https://doi.org/10.1007/JHEP07(2020)154}{\emph{JHEP} {\bfseries 07}
  (2020) 154} [\href{https://arxiv.org/abs/1909.11545}{{\ttfamily
  1909.11545}}].

\bibitem{Chua:2018dqh}
W.Z.~Chua, Q.~Ding, Y.~Wang and S.~Zhou, \emph{{Imprints of Schwinger Effect on
  Primordial Spectra}},
  \href{https://doi.org/10.1007/JHEP04(2019)066}{\emph{JHEP} {\bfseries 04}
  (2019) 066} [\href{https://arxiv.org/abs/1810.09815}{{\ttfamily
  1810.09815}}].

\bibitem{Bodas:2024hih}
A.~Bodas, E.~Broadberry and R.~Sundrum, \emph{{Grand unification at the
  cosmological collider with chemical potential}},
  \href{https://doi.org/10.1007/JHEP01(2025)115}{\emph{JHEP} {\bfseries 01}
  (2025) 115} [\href{https://arxiv.org/abs/2409.07524}{{\ttfamily
  2409.07524}}].

\bibitem{maggiore2007gravitational}
M.~Maggiore, \emph{Gravitational waves: Volume 1: Theory and experiments}, OUP
  Oxford (2007).

\bibitem{Cook:2011hg}
J.L.~Cook and L.~Sorbo, \emph{{Particle production during inflation and
  gravitational waves detectable by ground-based interferometers}},
  \href{https://doi.org/10.1103/PhysRevD.85.023534}{\emph{Phys. Rev. D}
  {\bfseries 85} (2012) 023534}
  [\href{https://arxiv.org/abs/1109.0022}{{\ttfamily 1109.0022}}].

\bibitem{Chen:2017ryl}
X.~Chen, Y.~Wang and Z.-Z.~Xianyu, \emph{{Schwinger-Keldysh Diagrammatics for
  Primordial Perturbations}},
  \href{https://doi.org/10.1088/1475-7516/2017/12/006}{\emph{JCAP} {\bfseries
  12} (2017) 006} [\href{https://arxiv.org/abs/1703.10166}{{\ttfamily
  1703.10166}}].

\bibitem{Cespedes:2023aal}
S.~C{\'e}spedes, A.-C.~Davis and D.-G.~Wang, \emph{{On the IR divergences in de
  Sitter space: loops, resummation and the semi-classical wavefunction}},
  \href{https://doi.org/10.1007/JHEP04(2024)004}{\emph{JHEP} {\bfseries 04}
  (2024) 004} [\href{https://arxiv.org/abs/2311.17990}{{\ttfamily
  2311.17990}}].

\bibitem{Wang:2020uic}
Y.~Wang and Y.~Zhu, \emph{{Cosmological Collider Signatures of Massive Vectors
  from Non-Gaussian Gravitational Waves}},
  \href{https://doi.org/10.1088/1475-7516/2020/04/049}{\emph{JCAP} {\bfseries
  04} (2020) 049} [\href{https://arxiv.org/abs/2001.03879}{{\ttfamily
  2001.03879}}].

\bibitem{deser_massive_2004}
S.~Deser, \emph{Massive higher spin fields in ({A}){dS} and in interaction},
  \href{https://doi.org/10.1016/S0920-5632(03)02398-3}{\emph{Nuclear Physics B
  - Proceedings Supplements} {\bfseries 127} (2004) 36}.

\bibitem{Zakharov:1970cc}
V.I.~Zakharov, \emph{{Linearized gravitation theory and the graviton mass}},
  {\emph{JETP Lett.} {\bfseries 12} (1970) 312}.

\bibitem{vanDam:1970vg}
H.~van Dam and M.J.G.~Veltman, \emph{{Massive and massless Yang-Mills and
  gravitational fields}},
  \href{https://doi.org/10.1016/0550-3213(70)90416-5}{\emph{Nucl. Phys. B}
  {\bfseries 22} (1970) 397}.

\bibitem{Vainshtein:1972sx}
A.I.~Vainshtein, \emph{{To the problem of nonvanishing gravitation mass}},
  \href{https://doi.org/10.1016/0370-2693(72)90147-5}{\emph{Phys. Lett. B}
  {\bfseries 39} (1972) 393}.

\bibitem{Deffayet:2001uk}
C.~Deffayet, G.R.~Dvali, G.~Gabadadze and A.I.~Vainshtein,
  \emph{{Nonperturbative continuity in graviton mass versus perturbative
  discontinuity}},
  \href{https://doi.org/10.1103/PhysRevD.65.044026}{\emph{Phys. Rev. D}
  {\bfseries 65} (2002) 044026}
  [\href{https://arxiv.org/abs/hep-th/0106001}{{\ttfamily hep-th/0106001}}].

\bibitem{Babichev:2013usa}
E.~Babichev and C.~Deffayet, \emph{{An introduction to the Vainshtein
  mechanism}},
  \href{https://doi.org/10.1088/0264-9381/30/18/184001}{\emph{Class. Quant.
  Grav.} {\bfseries 30} (2013) 184001}
  [\href{https://arxiv.org/abs/1304.7240}{{\ttfamily 1304.7240}}].

\bibitem{Barry:1989zz}
M.V.~Barry, \emph{{Uniform Asymptotic Smoothing of Stokes's Discontinuities}},
  \href{https://doi.org/10.1098/rspa.1989.0018}{\emph{Proc. Roy. Soc. Lond. A}
  {\bfseries 422} (1989) 7}.

\bibitem{Enomoto:2020xlf}
S.~Enomoto and T.~Matsuda, \emph{{The exact WKB for cosmological particle
  production}}, \href{https://doi.org/10.1007/JHEP03(2021)090}{\emph{JHEP}
  {\bfseries 03} (2021) 090}
  [\href{https://arxiv.org/abs/2010.14835}{{\ttfamily 2010.14835}}].

\bibitem{Hashiba:2021npn}
S.~Hashiba and Y.~Yamada, \emph{{Stokes phenomenon and gravitational particle
  production \textemdash{} How to evaluate it in practice}},
  \href{https://doi.org/10.1088/1475-7516/2021/05/022}{\emph{JCAP} {\bfseries
  05} (2021) 022} [\href{https://arxiv.org/abs/2101.07634}{{\ttfamily
  2101.07634}}].

\bibitem{Starobinsky:1980te}
A.A.~Starobinsky, \emph{{A New Type of Isotropic Cosmological Models Without
  Singularity}},
  \href{https://doi.org/10.1016/0370-2693(80)90670-X}{\emph{Phys. Lett. B}
  {\bfseries 91} (1980) 99}.

\bibitem{Dimastrogiovanni:2018uqy}
E.~Dimastrogiovanni, M.~Fasiello and G.~Tasinato, \emph{{Probing the
  inflationary particle content: extra spin-2 field}},
  \href{https://doi.org/10.1088/1475-7516/2018/08/016}{\emph{JCAP} {\bfseries
  08} (2018) 016} [\href{https://arxiv.org/abs/1806.00850}{{\ttfamily
  1806.00850}}].

\bibitem{Xianyu:2022jwk}
Z.-Z.~Xianyu and H.~Zhang, \emph{{Bootstrapping one-loop inflation correlators
  with the spectral decomposition}},
  \href{https://doi.org/10.1007/JHEP04(2023)103}{\emph{JHEP} {\bfseries 04}
  (2023) 103} [\href{https://arxiv.org/abs/2211.03810}{{\ttfamily
  2211.03810}}].

\bibitem{belinfante1940current}
F.J.~Belinfante, \emph{On the current and the density of the electric charge,
  the energy, the linear momentum and the angular momentum of arbitrary
  fields}, {\emph{Physica} {\bfseries 7} (1940) 449}.

\bibitem{Weinberg:1995mt}
S.~Weinberg, \emph{{The Quantum theory of fields. Vol. 1: Foundations}},
  Cambridge University Press (6, 2005),
  \href{https://doi.org/10.1017/CBO9781139644167}{10.1017/CBO9781139644167}.

\bibitem{Forger:2003ut}
M.~Forger and H.~Romer, \emph{{Currents and the energy momentum tensor in
  classical field theory: A Fresh look at an old problem}},
  \href{https://doi.org/10.1016/j.aop.2003.08.011}{\emph{Annals Phys.}
  {\bfseries 309} (2004) 306}
  [\href{https://arxiv.org/abs/hep-th/0307199}{{\ttfamily hep-th/0307199}}].

\bibitem{Baker:2021hly}
M.R.~Baker, N.~Linnemann and C.~Smeenk, \emph{{Noether's first theorem and the
  energy-momentum tensor ambiguity problem}},
  \href{https://arxiv.org/abs/2107.10329}{{\ttfamily 2107.10329}}.

\bibitem{Hehl:1976vr}
F.W.~Hehl, \emph{{On the Energy Tensor of Spinning Massive Matter in Classical
  Field Theory and General Relativity}},
  \href{https://doi.org/10.1016/0034-4877(76)90016-1}{\emph{Rept. Math. Phys.}
  {\bfseries 9} (1976) 55}.

\bibitem{Baker:2020eqs}
M.R.~Baker, N.~Kiriushcheva and S.~Kuzmin, \emph{{Noether and Hilbert (metric)
  energy-momentum tensors are not, in general, equivalent}},
  \href{https://doi.org/10.1016/j.nuclphysb.2020.115240}{\emph{Nucl. Phys. B}
  {\bfseries 962} (2021) 115240}
  [\href{https://arxiv.org/abs/2011.10611}{{\ttfamily 2011.10611}}].

\end{thebibliography}\endgroup

\end{document}